\documentclass[lettersize, 12pt]{article}

\usepackage{graphics}
\usepackage{float}
\usepackage{theorem}
\usepackage{latexsym}
\usepackage{amsfonts}
\usepackage{amssymb}
\usepackage{amsmath}
\usepackage{graphicx}
\usepackage{natbib}
\usepackage{lscape}
\usepackage{booktabs}
\usepackage{threeparttable}
\usepackage{appendix}
\usepackage{subfig}
\usepackage[colorlinks=true,linkcolor=black,citecolor=black,urlcolor=black,pdfstartview=FitH,bookmarks=false]{hyperref}
\usepackage{setspace}
\usepackage{titlesec}
\usepackage{color}
\usepackage{natbib}
\usepackage{bm}
\usepackage{multicol}
\usepackage{soul}
\usepackage{tabularx} 

\usepackage[utf8]{inputenc}

\newtheorem{theorem}{Theorem}
\newtheorem{assumption}{Assumption}

\newtheorem{corollary}{Corollary}
\newtheorem{lemma}{Lemma}

\newtheorem{remark}{Remark}

\renewcommand{\thesection}{\arabic{section}}
\renewcommand{\theequation}{\arabic{section}.\arabic{equation}}
\renewcommand{\thetheorem}{\arabic{section}.\arabic{theorem}}
\renewcommand{\theassumption}{\arabic{section}.\arabic{assumption}}

\renewcommand{\theremark}{\arabic{section}.\arabic{remark}}

\newcounter{bean}

\usepackage{authblk}

\begin{document}
\title{\Large 
Double Debiased Machine Learning Nonparametric Inference with Continuous Treatments
}


%
%

\author{\large Kyle Colangelo 
\hspace{0.6cm} 
Ying-Ying Lee\footnote{Ying-Ying Lee, Department of economics, University of California Irvine.
E-mail: \href{yingying.lee@uci.edu}{yingying.lee@uci.edu}.
Kyle Colangelo, Amazon.
}\vspace{-15pt}
}
\date{\normalsize September 2023\\[-8pt]
{\footnotesize (first version: February 2019)}
}

\maketitle
\vspace{-25pt}

\begin{center}
{\bf \small Abstract}
\end{center}

\vspace{-10pt}

We propose a doubly robust inference method for causal effects of continuous treatment variables, under unconfoundedness and with nonparametric or high-dimensional nuisance functions.  Our double debiased machine learning (DML) estimators for the average dose-response function (or the average structural function) and the partial effects are asymptotically normal with nonparametric convergence rates.  The first-step estimators for the nuisance conditional expectation function and the conditional density can be nonparametric or ML methods.  Utilizing a kernel-based doubly robust moment function and cross-fitting, we give high-level conditions under which the nuisance function estimators do not affect the first-order large sample distribution of the DML estimators.  We  provide sufficient low-level conditions for kernel, series, and deep neural networks.  We justify the use of kernel to localize the continuous treatment at a given value by the Gateaux derivative.  We implement various ML methods in Monte Carlo simulations and an empirical application on a job training program evaluation. 
\\
\textbf{Keywords}: Average structural function, cross-fitting, dose-response function, doubly robust.
\\
\textbf{JEL Classification}: C14, C21, C55
%

\newpage
\section{Introduction}

We propose a nonparametric inference method for {\it continuous} treatment or structural effects, under the unconfoundedness assumption
and in the presence of nonparametric nuisance functions.
We focus on the heterogenous effect with respect to the continuous treatment or policy variable $T$.
To identify the causal effects, it is plausible to have a large number of the control variables $X$ that $T$ is randomly assigned conditional on.
To achieve doubly robust inference and to employ machine learning (ML) methods, 
we use a double debiased ML approach that combines a doubly robust moment function and cross-fitting.

We consider a nonparametric and model-free outcome equation $Y = g(T, X, \varepsilon)$.
No functional form assumption is imposed on the unobserved disturbances $\varepsilon$, such as restrictions on dimensionality, monotonicity, or separability.
The potential outcome is $Y(t) = g(t,X,\varepsilon)$ indexed by the hypothetical treatment value $t$.
The object of interest is the {\it average dose-response function} as a function of $t$, defined as the expected value of the potential outcome across observations with the observed and unobserved heterogeneity $(X, \varepsilon)$, i.e.\ 
$\beta_t \equiv \mathbb{E}[Y(t)] = \int\int g(t, X, \varepsilon) dF_{X\varepsilon}$.
It is also known as the {\it average structural function} in nonseparable models in \cite{BP03}.
The well-studied average treatment effect of switching from treatment $t$ to $s$ is $\beta_{s} - \beta_{t}$.
We allow the continuous treatment to be multi-dimensional and hence capture the unrestricted heterogenous effects with respect to the multivariate treatment variables.
We define the {\it partial (or marginal) effect} of the first component of the continuous variable $T$ at $t = (t_1,... t_{d_T})'$ to be the partial derivative $\theta_t \equiv \partial \beta_t/ \partial t_1$.
In program evaluation, the average dose response function $\beta_t$ shows how participants' labor market outcomes vary with the length of exposure to a job training program.
In demand analysis when $T$ contains price and income, the average structural function $\beta_t$ can be the Engel curve. 
The partial effect $\theta_t$ reveals the average price elasticity at given values of price and income.
Other examples include the efficacy of political advertisements on campaign contributions in \cite{Fong},
the effect of nurse staffing on hospital readmissions penalties in \cite{Kennedy}, etc.

We are among the first to apply the double debiased ML approach to inference on 
the average structural function $\beta_t$ and the partial effect $\theta_t$ of continuous variables, to our knowledge.
They are {\it non-regular nonparametric} objects that cannot be estimated at a root-$n$ convergence rate.
We propose a kernel-based {\it double debiased machine learning} (DML) estimator 
that utilizes a doubly robust moment function and cross-fitting via sample-splitting.
The DML estimator uses the moment function
\begin{align}
\gamma(t, X_i) + \frac{K_h(T_i - t)}{ f_{T|X}(t|X_i)} \big( Y_i - \gamma(t, X_i)\big),
\label{EDR}
\end{align}
where 
the conditional expectation function $\gamma(t, x) \equiv \mathbb{E}[Y|T=t, X=x]$
and 
the conditional density $f_{T|X}(t|x)$ is also known as the generalized propensity score (GPS).
A kernel $K_h(T_i-t)$ weights observation $i$ with treatment value around $t$ in a distance of $h$. 
The number of such observations shrinks as the bandwidth~$h$ vanishes with the sample size $n$.
A {\it $L$-fold cross-fitting} splits the sample into $L$ subsamples.
 The nuisance function estimators for $\gamma(t, X_i)$ and $f_{T|X}(t|X_i)$ use observations in the other $(L-1)$ subsamples that do not contain the observation $i$.
The DML estimator averages over the subsamples.
Then we estimate the partial effect $\theta_t$ by a numerical differentiation.

The doubly robust moment function in equation (\ref{EDR}) has appeared in \cite{Kallus} without asymptotic theory and has been extensively studied in \cite{SUZ} for Lasso-type estimators.
We utilize cross-fitting and provide high-level and low-level conditions that facilitate a variety of nonparametric and ML methods. 
As each ML method has its strengths and weaknesses depending on the data generating process and applications,
flexible employment of various nuisance function estimators is desirable.
High-dimensional control variables can be accommodated via the nuisance function estimators; for example, Lasso allows the dimension of $X$ to grow with the sample size.
Importantly our inference theory for the proposed DML estimator $\hat \beta_t$ allows one of the nuisance functions to be misspecified.
The doubly robust inference is useful, especially when the nuisance functions are estimated 
under a parametric model or sparsity approximation as Lasso.

We show that the proposed kernel-based DML estimators are asymptotically normal and provide high-level conditions under which the nuisance function estimators for $\gamma(t, X_i)$ and $f_{T|X}(t|X_i)$ do not affect the first-order asymptotic distribution. 
Specifically the high-level conditions on the convergence rates use a {\it partial $L_2$ norm} that fixes the treatment value at $t$, in contrast to the standard $L_2$ norm that integrates over the joint distribution of $(T,X)$, i.e.\ the root-mean-square rate.
We further give low-level conditions for nonparametric kernel, series estimators, 
and the deep neural networks in \cite{Farrell18} that is widely popular in industrial applications.
These results on the convergence rates of the nuisance function estimators are new to the literature.

Furthermore, we propose a Simulated DML (SDML) estimator that replaces $\gamma(t, X_i)$ in (\ref{EDR}) with $\gamma(U_i, X_i)$ where a simulated variable $U_i$ localizes the realized treatment values around the target value $t$.
Introducing such a local variation enables the standard $L_2$ convergence rate of the nuisance function estimator
that is available for nonparametric kernel and series estimators, as well as recent ML estimators, such as Lasso in \citep{BRT09AS}, neural networks in \citep{ChenWhite, SH20AS, Farrell18}, random forests in \citep{SZ20}, and empirical $L_2$ rate for boosting in \cite{LeoSpindler}, as discussed in \cite{CCDDHNR} (CCDDHNR, hereafter) and \cite{CNS21ADML}.\footnote{
We are grateful to Whitney Newey for the idea of simulating $U_i$ from the probability density function 
$f_U(u) = f_{T}\left((u-t)/h\right)/h$.
}
This result is valuable, as the SDML estimator readily allows for a wider class of ML methods.

In addition, we propose generic ML estimators for the {\it reciprocal of the conditional density} $1/f_{T|X}(t|x)$ when $d_T = 1$ and for $f_{T|X}(t|x)$ when $d_T > 1$ respectively, which may be of independent interest.
We also propose a data-driven bandwidth to consistently estimate the optimal bandwidth that minimizes the asymptotic mean squared error.

We aim for a tractable inference procedure that is flexible to employ 
nonparametric or ML nuisance function estimators and delivers a reliable distributional approximation in practice.
Toward that end, the DML method contains two key ingredients: a doubly robust moment function and cross-fitting.
The doubly robust moment function reduces sensitivity in estimating $\beta_t$ with respect to nuisance parameters.\footnote{
The double robustness usually refers to consistency of the estimator even if either one of the two nuisance functions is misspecified.
The rapidly growing ML literature has utilized the double robustness 
to reduce regularization and modeling biases in estimating the nuisance functions by ML or nonparametric methods;
for example, \cite{BCH14RES}, \cite{Farrell15},  \cite{BCFH17}, \cite{Farrell18}, \cite{CEINR}, CCDDHNR, \cite{RotheFirpo}, and references therein.
}
Cross-fitting further removes bias induced by overfitting and achieves stochastic equicontinuity without strong entropy conditions.\footnote{
CCDDHNR  point out that the commonly used results in empirical process theory, such as Donsker properties, could break down in high-dimensional settings.
For example, \cite{BCFH17} show how cross-fitting weakens the entropy condition and hence the sparsity assumption on nuisance Lasso estimator.
The benefit of cross-fitting is further investigated by \cite{WagerAthey} for heterogeneous causal effects, \cite{NeweyRobins} for double cross-fitting, and \cite{CJ19ET} for cross-fitting bootstrap.}
\\[5pt]
\textbf{Related literature:}\
Our work builds on the results for semiparametric models in \cite{IchimuraNewey22QE}, \cite{CEINR}, and CCDDHNR and extends the literature to nonparametric continuous treatment/structural effects.
Note that the doubly robust estimator for a binary/multivalued treatment replaces the kernel $K_h(T_i - t)$ with the indicator function ${\bf 1}\{T_i=t\}$ in equation (\ref{EDR}) and has been widely studied, especially in the recent ML literature.
We show that the advantageous properties of the DML estimator for the binary treatment carry over to the continuous treatments case.
Our DML estimator utilizes the kernel function $K_h(T_i - t)$ for the continuous treatments $T$ of fixed low dimension and averages out the covariates $X$.
While our kernel-based estimator appears to be a simple modification of the binary treatment case in practice, 
we show that one important distinct feature of {\it non-regular nonparametric} parameters is that 
the Gateaux derivative and the Riesz representer are not unique, depending on how we approximate the continuous treatment distribution approaching a point mass.    
And the kernel function is a natural choice for localization at $t$.
Neyman orthogonality holds as $h \rightarrow 0$ (\citep{Neyman}).
Therefore we provide a foundational justification for the proposed kernel-based DML estimator $\hat\beta_t$, relative to alternative approaches, such as \cite{Kennedy} and \cite{SC}; see also \cite{VBL}.
Furthermore, 
our estimator is doubly robust in the sense that our inference theory is valid, i.e.\ the asymptotic distribution is the same, if either one of the nuisance functions $\mathbb{E}[Y|T,X]$ or $f_{T|X}$ is misspecified, as in \cite{Kennedy} and \cite{Westling}.
This is a stronger result than the usual doubly robustness on the consistency of the estimator.
When one nuisance function is misspecified, we require the other nuisance function to be estimated consistently at a convergence rate faster than $\sqrt{nh^{d_T}}$, which is the convergence rate of $\hat\beta_t$.
In contrast, the DML estimators in the semiparametric model of CCDDHNR converge at a regular root-$n$ rate, 
so their inference theory does not allow this doubly robust property.


There is a small yet rapidly growing literature on employing the DML approach for non-regular nonparametric infinite-dimesional objects.
\cite{HHLL} propose a test for monotonicity of $\beta_t$.
\cite{CNS21EJ}, \cite{SC}, \cite{Fan}, and \cite{ZimmertLechner} study the conditional average binary treatment effect $\mathbb{E}[Y(1) - Y(0)|X_1]$ for a low-dimensional subset $X_1$ of $X$.
\cite{BonviniKennedy} use higher-order influence functions to achieve a faster convergence rate.
Despite the advantageous theoretical properties, we note potential drawbacks of the DML approach:
Double robustness often requires additional nuisance function estimation, such as the conditional density in the Riesz representer, which could introduce additional variation and implementation complication. 
Cross-fitting could result in a small effective sample size.
For example, the Monte Carlo simulations in \cite{Fan} show that cross-fitting does not improve the finite-sample performance for Lasso-type estimation.  

Our paper adds to the literature on continuous treatment effects estimation.
In low-dimensional settings, see \cite{Imbens00}, \cite{HI04}, \cite{Flores07}, and \cite{Lee} for examples of a class of regression estimators $n^{-1}\sum_{i=1}^n \hat \gamma(t, X_i)$.
\cite{GW} and \cite{HHLP} study a class of inverse probability weighting estimators.
The empirical applications in \cite{FFGN12ReStat} and \cite{KSUZ12} focus on semiparametric results.
We extend this literature to a DML framework that enables ML methods for nonparametric inference in practice.

A main contribution of this paper is a formal inference theory for the fully nonparametric causal effects of continuous variables.
To uncover the causal effect of the continuous variable $T$ on $Y$, our nonparametric nonseparable model $Y = g(T,X,\varepsilon)$ can be compared to the partially linear model $Y = \theta T + g(X) + \varepsilon$ in \cite{Robinson} that specifies the homogenous effect by $\theta$ and hence is a semiparametric problem.
The important partially linear model has many applications and is one of the leading examples in the recent ML literature, where the nuisance function $g(X)$ can be high-dimensional and estimated by a ML method.\footnote{
See \cite{CCDDHNR} and references therein.
\cite{DSLC}  and \cite{OSW} extend to more general functional forms.
\cite{CJN18ET}, \cite{CJN18JASA}, \cite{CJM19}, and \cite{FLM2} propose different approaches.
} 
Another semiparametric parameter of interest is the weighted average of $\beta_t$ or $\theta_t$ over a range of treatment values $t$,
such as the average derivative that summarizes certain aggregate effects \citep{PSS89ETA}
and the bound of the average welfare effect in \cite{CHN}. 
In contrast, our average structural function $\beta_t$ and the partial effect $\theta_t$ capture the fully nonparametric  heterogenous effects of $T$.

The paper proceeds as follows.  Section~\ref{SecEst} introduces the framework and estimation procedure. 
Section~\ref{SecAsy} presents the asymptotic theory of the DML estimators and low-level conditions for various nuisance function estimators.
Section~\ref{SecSDML} introduces the Simulated DML estimator.  
Section~\ref{SecNEx} demonstrates the usefulness of our DML estimator with various ML methods
in Monte Carlo simulations and an empirical example on the Job Corps program evaluation.
All the proofs are in the Appendix.
Additional results, such as uniform inference theory, are in the online supplementary appendix.

\section{Setup and estimation}
\label{SecEst}
Let $\{Y_i, T_i, X_i\}_{i=1}^n$ be an i.i.d.~sample from 
$Z = (Y, T^\prime, X^\prime)^\prime \in \mathcal{Z} = \mathcal{Y}\times\mathcal{T}_0\times\mathcal{X} 
\subseteq \mathcal{R}^{1+d_T+d_X}$ from a population $P$ with
a cumulative distribution function (CDF) $F_{Z}(Z)$.
Consider a set of treatment values of interest $\mathcal{T}$ to be an interior of the support of $T$.

\begin{assumption}
(a)~{(Conditional independence)}
$T$ and $\varepsilon$ are independent conditional on $X$;
\\
(b)~{(Common support)} 
$\inf_{t\in\mathcal{T}}{\rm ess}\inf_{x\in\mathcal{X}} f_{T|X}(t|x) \geq c$ for some positive constant $c$;
(c)~$f_{Z}(y,t,x)$ and $\mathbb{E}[Y|T=t, X=x]$ are three-times differentiable with respect to $t$ with all three derivatives being bounded uniformly over $(y,t^\prime,x^\prime)^\prime \in \mathcal{Z}$;
(d) $var(Y|T=t, X=x)$ and its derivatives with respect to $t$ are bounded uniformly over $(t^\prime,x^\prime)^\prime \in \mathcal{T}\times\mathcal{X}$.
\label{ACIACS}
\end{assumption}

The commonly used identifying Assumption~\ref{ACIACS}(a) based on observational data (also known as unconfoundedness, selection on observables, or ignorability) assumes that conditional on observables, the treatment variable is as good as randomly assigned, or conditionally exogenous.

Define the product kernel as $K_h(T_i-t) \equiv \Pi_{j=1}^{d_T} k((T_{ji} - t_j)/h)/h^{d_T}$, where $T_{ji}$ is the $j^{th}$ component of $T_i$ and the kernel function $k()$ satisfies Assumption~\ref{Akernel}.
Denote the roughness of $k$ as $R_k \equiv \int_{-\infty}^\infty k(u)^2 du$ and $\kappa \equiv \int_{-\infty}^\infty  u^2 k(u) du$.

\begin{assumption}[Kernel]
The second-order symmetric kernel function $k()$ is bounded differentiable, i.e.\ $\int_{-\infty}^\infty  k(u)du =1$, $\int_{-\infty}^\infty  uk(u) du = 0$, and $0 < \kappa < \infty$.
For some finite positive constants $C, \bar{U}$, and for some $\nu > 1$,  $|dk(u)/du| \leq C |u|^{-\nu}$ for $|u| > \bar{{U}}$.
\label{Akernel}
\end{assumption}

Assumption~\ref{Akernel} is standard in nonparametric kernel estimation and holds for commonly used kernel functions, such as Epanechnikov and Gaussian.
By Assumptions~\ref{ACIACS}-\ref{Akernel} and the same reasoning for the binary treatment, it is straightforward to show the identification,
\begin{align}
\beta_t \equiv \mathbb{E}[Y(t)] & = \int_\mathcal{X} \mathbb{E}[Y|T=t, X] dF_{X}(X) 
= \mathbb{E}\big[\gamma(t, X)\big]
\label{IDreg} \\
&= 
\lim_{h\rightarrow 0}\int_\mathcal{Z}  \frac{K_h(T-t) Y}{f_{T|X}(t|X)} dF_{Z}(Z)
= \lim_{h\rightarrow 0} \mathbb{E}\left[\frac{K_h(T-t) Y}{f_{T|X}(t|X)}  \right], \label{IDIPW}
\end{align}
 for $t\in\mathcal{T}$.\footnote{For identification, we only need $\inf_{t\in\mathcal{T}} {\rm ess}\inf_{x\in\mathcal{X}} f_{T|X}(t|x) > 0$
 for (\ref{IDreg}) and additionally $\gamma(t,x)$ and $f_{T|X}(t,x)$ to be continuous in $t$ for (\ref{IDIPW}).
Such standard identifying conditions are weaker than Assumption~\ref{ACIACS} used for the inference theory.} The expression in equation (\ref{IDreg}) motivates the class of regression-based (or imputation) estimators, while equation~(\ref{IDIPW}) motivates the class of inverse probability weighting estimators; see Section~S2.1 in the online supplementary appendix for further discussion.
 Now we introduce the double debiased machine learning estimator.
\\
\\
\textbf{Estimation procedure:}

       \setcounter{bean}{0}
       \begin{center}
        \begin{list}
         {\textsc{Step} \arabic{bean}.}{\usecounter{bean}}
\item  (Cross-fitting)
For some fixed $L \in \{2,...,n\}$, randomly partition the observation indices into $L$ distinct groups $I_\ell$, $\ell=1,...,L$, 
such that the sample size of each group is the largest integer smaller than $n/L$. 
For each $\ell=1,...,L$, the estimators $\hat \gamma_\ell(t,x)$ for $\gamma(t,x) \equiv \mathbb{E}[Y|T=t, X=x]$ and $\hat f_\ell(t|x)$ for $f_{T|X}(t|x)$ use observations not in $I_\ell$
and satisfy Assumption~\ref{A1st} below. 

\item (Double robustness)
Define the double debiased ML (DML) estimator as
\begin{align}
\hat \beta_t \equiv \frac{1}{n}\sum_{\ell=1}^L \sum_{i \in I_\ell} \left\{ \hat \gamma_\ell (t, X_i) + \frac{K_h(T_i - t)}{ \hat f_\ell(t|X_i)} \left( Y_i - \hat \gamma_\ell(t,X_i) \right)\right\}.
\label{EDML}
\end{align}

\item (Partial effect)
Let $t^+ \equiv (t_1 + \eta/2, t_2,..., t_{d_T})'$ and $t^- \equiv (t_1 - \eta/2, t_2,..., t_{d_T})'$, where $\eta$ is a positive sequence converging to zero as $n \rightarrow \infty$.
We estimate the partial effect of the first component of the continuous treatment $\theta_t \equiv \partial \beta_t/\partial t_1$ 
by $\hat \theta_{t} \equiv (\hat \beta_{t^+} - \hat \beta_{t^-})/\eta$.

        \end{list}
       \end{center}
       
       The number of folds in cross-fitting $L$ is not random and typically small, such as five or ten in practice; see, e.g.\ Section~\ref{SecMC} or CCDDHNR.  
When there is no sample splitting ($L=1$), $\hat \gamma_1$ and $\hat f_1$ use all observations in the full sample. 
Then the DML estimator $\hat\beta_t$ in (\ref{EDML}) is the doubly robust estimator considered in \cite{Kallus} and \cite{SUZ}.


Our inference theory requires the estimators $\hat\gamma_\ell$ and $\hat f_\ell$ in Step~1 to satisfy Assumption~\ref{A1st} below.
We define the {\it partial $L_2(tX)$ norm} for any $t\in \mathcal{T}$ as 
$\|\hat\gamma_\ell-\gamma\|_{F_{tX}} \equiv \|\hat\gamma_\ell(T,X)-\gamma(T,X)\|_{F_{tX}} \equiv 
\big(\int_{\mathcal{X}} (\hat \gamma_\ell(t,x) - \gamma(t,x))^2 f_{TX}(t,x) dx \big)^{1/2}$
and 
$\|\hat f_\ell -f_{T|X}\|_{F_{tX}} \equiv \|\hat f_\ell(T|X)- f_{T|X}(T|X)\|_{F_{tX}} \equiv 
\big(\int_{\mathcal{X}} (\hat f_\ell(t|x) - f_{T|X}(t|x))^2 f_{TX}(t, x) dx \big)^{1/2}$,
where the joint distribution $F_{TX}(t,X)$ is evaluated at a fixed value of $T$ equal to $t$.

\begin{assumption}
There exist functions $\bar \gamma(t,x)$ and $\bar f(t,x)$ that 
are three-times differentiable with respect to $t$ with all
three derivatives being bounded uniformly over $(y,t',x') \in\mathcal{Z}$,
$\inf_{t\in\mathcal{T}}{\rm ess}\inf_{x\in\mathcal{X}}
 $\\$\bar f(t,x) \geq c$ for some positive constant $c$, and satisfy the following: 
for each $\ell= 1,...,L$ and for any $t \in \mathcal{T}$,
 {\it (a)} $\big\|\hat \gamma_\ell - \bar\gamma \big\|_{F_{tX}} = o_p(1)$ and 
$\big\|\hat f_\ell - \bar{f} \big\|_{F_{tX}} = o_p(1)$,
{\it (b)} $\sqrt{nh^{d_T}} \big\| \hat f_\ell - {f}_{T|X} \big\|_{F_{tX}}  \big\| \hat\gamma_\ell - \gamma\big\|_{F_{tX}}  \stackrel{p}{\rightarrow}~0$;
{\it (c)} Either $\bar \gamma = \gamma$ or $\bar{f} = f_{T|X}$.
\label{A1st}
\end{assumption}

The nuisance function estimators $\hat\gamma_\ell$ and $\hat f_\ell$ converge to some fixed functions $\bar\gamma$ and $\bar f$ respectively in the sense of Assumption~\ref{A1st}(a)(b). 
Assumption~\ref{A1st}(c) allows one of the nuisance functions to be misspecified, 
or requires at least one of the nuisance function estimators to be consistent.
Assumption~\ref{A1st}(b) and (c) imply that if one nuisance function is misspecified, then the other needs to be estimated consistently at a convergence rate faster than $\sqrt{nh^{d_T}}$.
This is the cost of our doubly robust inference.
Specifically, 
if $\bar\gamma \neq \gamma$, then $\|\hat\gamma_\ell - \gamma\|_{F_{tX}} = O_p(1)$.  So  Assumption~\ref{A1st}(b) requires $\sqrt{n h^{d_T}} \|\hat f_{\ell} - f_{T|X}\|_{F_{tX}} = o_p(1)$.
On the other hand, if $\bar f \neq f_{T|X}$, then we require $\sqrt{n h^{d_T}} \|\hat \gamma_{\ell} - \gamma\|_{F_{tX}} = o_p(1)$.
\cite{Kennedy} show such double robustness using a stronger uniform norm for a local linear estimation.

In Section~\ref{Sec1stEst}, we provide sufficient low-level conditions for Assumption~\ref{A1st} 
when the nuisance function estimators are kernel estimators, series, and the deep neural networks in \cite{Farrell18}. 
The partial $L_2$ convergence rate also appears in \cite{Kennedy}, the conditional average treatment effect in \cite{Fan}, and 
covariate adjustments in regression discontinuity designs in \cite{RotheRD}.

\begin{remark}{\rm 
Consider the binary treatment effect in CCDDHNR as a heuristic example of a regular parameter that can be estimated at the regular $\sqrt{n}$-rate.
When $\bar\gamma \neq \gamma$,
Assumption~\ref{A1st} suggests that the corresponding condition on the propensity score estimator $\hat f_\ell$ would be
$\sqrt{n}\|\hat f_{\ell} - f_{T|X}\|_{F_{tX}} = o_p(1)$, which is not feasible.  
Therefore  we conjecture that the inference theory of the DML estimators
for the regular estimands in the semiparametric models of CCDDHNR cannot allow for misspecification of the nuisance functions.
In contrast, the DML estimator for the non-regular estimand here allows for doubly robust inference.
\cite{TanAS} develops doubly robust inference for the average binary treatment effect by carefully chosen loss functions to estimate the nuisance functions.
}\end{remark}

\begin{remark}[Common support] {\rm 
Assumption \ref{ACIACS}(b) implies that we need to observe sufficient individuals in the population who can find a match sharing the same value of the control variable~$X$ and receiving the counterfactual value~$t$.
An analogous assumption in the binary treatment case is that the propensity score is bounded away from zero, e.g.\ \cite{HIR03ETA}. 
Although the common support assumption is standard,  
we note that it should be made with care in practice and is strong especially with many control variables.
For the binary treatment case, \cite{KT} study extensively irregular identification and inverse weight estimation, when the propensity score can be close to zero as a small denominator.
For the continuous treatment case, the convergence rate of $\hat\beta_t$ might similarly be affected if the generalized propensity score can be close to zero.
We believe that this interesting extension is beyond the scope of the paper and is worthy of a separate research project.
See also \cite{SUZ} for a related discussion.
Another possible approach to relaxing Assumption~\ref{ACIACS}(b) is to define the object of interest by a common support via fixed trimming, e.g.\ \cite{Lee}.
}
\label{RCS}
\end{remark}

\begin{remark}[Kernel localization]{\rm
We discuss the construction of the doubly robust moment function (\ref{EDR})
by Gateaux derivative and a local Riesz representer; details are in Section~S2 in the online supplementary appendix.
From the literature on estimating regular parameters, the Gateaux derivative is fundamental to construct estimators with desired properties, such as bias reduction and double robustness in \cite{CLV} and \cite{IchimuraNewey22QE}. 
The partial mean $\beta_t$ is a marginal integration over 
the conditional distribution of $Y$ given $(T,X)$ and the marginal distribution of $X$, fixing the value of $T$ at $t$.
As a result, the Gateaux derivative and the Riesz representer depend on the choice of the distribution $f_{T}^h$ that belongs to a family of distributions approaching a point mass at $T$ as $h \rightarrow 0$.
We construct the locally robust estimator based on the influence function derived by the Gateaux derivative, so the asymptotic distribution of $\hat\beta_t$ depends on the choice of $f_{T}^h$ that is the kernel function $K_h(T-t)$.
Moreover, to construct the DML estimator of a linear functional of $\beta_t$ that preserves the good properties, the corresponding moment function is simply the linear functional of the moment function of $\beta_t$.
To our best knowledge, this is the first explicit calculation of Gateaux derivative for such a non-regular nonparametric parameter.
Alternatively \cite{Kennedy} construct a ``pseudo-outcome" 
that is motivated from the doubly robust and efficient influence function of the regular semiparametric parameter $\int \beta_t f_T(t)dt$.
 Then they locally regress the pseudo-outcome on $T$ at $t$ using a kernel to estimate $\beta_t$.
\cite{SC} illustrate in an example to estimate $\beta_t$ by the best linear projection of  an ``orthogonal signal of the outcome" which is the same ``pseudo-outcome" proposed by \cite{Kennedy}.

A second motivation of the moment function is 
adding to the influence function of the regression (or imputation) estimator $n^{-1}\sum_{i=1}^n \hat \gamma(t, X_i)$
the adjustment term from a kernel-based estimator $\hat \gamma$ under the low-dimensional case when the dimension of $X_i$ is fixed.
A series estimator $\hat \gamma$ yields a different adjustment.
These distinct features of continuous treatments are in contrast to the regular binary treatment case, where different nonparametric nuisance function estimators $\hat\gamma$ result in the same efficient influence function.
}
\label{RGD}
\end{remark}

\subsection{Conditional density estimation}
\label{SecGPS}
We propose an estimator for the {\it reciprocal of the generalized propensity score} (GPS), i.e.\  $1/f_{T|X}(t|x)$, when $d_T = 1$.
The estimator avoids plugging in a small estimate in the denominator, and the estimate is positive by construction.
When $d_T > 1$, we propose an estimator for the GPS.
We can use various nonparametric and ML methods designed for the conditional expectation.
We provide a root-mean-square convergence rate.
In Section~\ref{Sec1stEst}, we demonstrate these generic GPS estimators using the deep neural networks in \cite{Farrell18} and show how Assumption~\ref{A1st} can be verified. 

The theory of ML methods in estimating the conditional density is less developed compared with estimating the conditional expectation.
Alternative estimators for estimating the GPS can be the kernel density estimator, the artificial neural networks in \cite{ChenWhite}, the Lasso methods in \cite{SUZ} and \cite{BCK19JASA}, or the series cross-validated method in \cite{Zhang}.

It is known that for any CDF $F$, $\frac{d}{du} F^{-1}(u) = \frac{1}{F'(F^{-1}(u))}$ for $u\in(0,1)$.
So $\frac{1}{f_{T|X}(t|x)} = \frac{\partial}{\partial u} F^{-1}_{T|X}(u|x)\big|_{u=F_{T|X}(t|x)}$.
Inspired by the idea in \cite{Koenker94},
we estimate $\frac{1}{f_{T|X}(t|x)}$ by a numerical differentiation estimator, labelled as ReGPS,
\begin{align}
\widehat{\frac{1}{f_{T|X}(t|x)}} 
= 
\frac{\hat F_{T|X}^{-1}\big(\hat F_{T|X}(t|x) + \epsilon\big|x\big) - \hat F_{T|X}^{-1}\big(\hat F_{T|X}(t|x) - \epsilon\big|x\big)}{2\epsilon},
\tag{ReGPS}\label{ReGPS}
\end{align}
where
$\epsilon = \epsilon_{n}$ is a positive sequence vanishing as $n$ grows and
$\hat F_{T|X}(t|x) \pm \epsilon \in (0,1)$.
By a standard algebra, the conditional CDF $F_{T|X}(t|x) = \lim_{h_1 \rightarrow 0} \mathbb{E}\left[\Phi\left(\frac{t-T}{h_1}\right)\big|X=x \right]$, where
$\Phi$ is the CDF of a standard normal random variable and 
$h_1 = h_{1n}$ is a bandwidth sequence vanishing as $n$ grows.
Let $\hat\mu(W; x)$ be a generic estimator of the conditional expectation $\mathbb{E}\left[W|X= x\right]$ 
for an outcome variable~$W$.
Then we estimate $F_{T|X}(t|x)$ by $\hat F_{T|X}(t|x) = \hat \mu\left(\Phi\left(\frac{t-T}{h_1}\right);x\right)$ with a transformed outcome variable of $T$, $W = \Phi\left(\frac{t-T}{h_1}\right)$.
The conditional $u$-quantile function $F_{T|X}^{-1}(u|x)$ is estimated by the generalized inverse function $\hat F_{T|X}^{-1}(u|x) = \inf_{t\in\mathcal{T}}\{t: \hat F_{T|X}(t|x) \geq u\}$.
When $\hat F_{T|X}(t|x)$ is continuous in $t$, $\hat F^{-1}_{T|X}(u|x)$ is strictly increasing.
Then the resulting estimator $\widehat{1/f_{T|X}(t|x)} > 0$.

Denote the standard root-mean-square norm, or the $L_2(X)$ norm, of a random vector $X$ with distribution $F_X$ as $\left\|\hat\mu(W; X) - \mathbb{E}[W|X]\right\|_{F_X} \equiv \Big(\int_\mathcal{X} \big(\hat\mu(W; x) - \mathbb{E}[W|X=x]\big)^2 f_X(x)dx \Big)^{1/2}$ for a random variable $W$.

\begin{lemma}[ReGPS]
Consider an estimator of $\mathbb{E}\Big[\Phi\Big(\frac{t-T}{h_1}\Big)\Big|X=x\Big]$, denoted as $\hat \mu\Big(\Phi\Big(\frac{t-T}{h_1}\Big);x\Big)$, which is continuous in $t\in\mathcal{T}$ and satisfies $\sup_{t\in\mathcal{T}}
\Big\|\hat\mu\Big(\Phi\Big(\frac{t-T}{h_1}\Big); X\Big) - \mathbb{E}\Big[\Phi\Big(\frac{t-T}{h_1}\Big)\Big|X\Big]\Big\|_{F_X}
= O_p(R_{1n})$ for a sequence of constants $R_{1n}$.
Assume $F_{T|X}$ to be three-times differentiable with respect to $t$ with all three derivatives being bounded uniformly over $(t^\prime,x^\prime)^\prime\in\mathcal{T}\times\mathcal{X}$.
Then the ReGPS estimator $\widehat{1/f_{T|X}}$ satisfies 
$\sup_{t\in\mathcal{T}}\big\|\widehat{1/f_{T|X}(t|X)} - 1/f_{T|X}(t|X) \big\|_{F_X} = 
O_p(R_{1n}\epsilon^{-1} + h_1^2\epsilon^{-1} + \epsilon^2)$
and\ \
$\sup_{t\in\mathcal{T}}\big\| \widehat{1/f_{T|X}} - 1/f_{T|X} \big\|_{F_{tX}} = 
O_p(R_{1n}\epsilon^{-1} + h_1^2\epsilon^{-1} + \epsilon^2)$.
\label{LGPS}
\end{lemma}


Assumption~\ref{A1st} specifies the  root-mean-square convergence rate of our conditional density estimator. 
Thus as long as the root-mean-square convergence rate of a ML method  ($R_{1n}$) for the conditional expectation function is available, 
Assumption~\ref{A1st} can be satisfied with a suitable bandwidth $h_1$ and $\epsilon$.  
Then we are able to use such a ML method to estimate the conditional density, as illustrated in Section~\ref{Sec1stEst}.

Note that $\hat F^{-1}_{T|X}(u|x)$ estimates the conditional quantile function.
The ReGPS estimator inverses the CDF estimate and allows to apply various ML methods for conditional expectation functions. 
Alternatively we can estimate the conditional quantile function directly by a $\ell_1$-penalized quantile regression;
for example,  the conditional density function estimation in \cite{BCK19JASA}. 


When $d_T > 1$, we propose a direct estimator for the conditional density function $f_{T|X}(t|x)$ by 
\begin{align}
\hat f_{T|X}(t|x) 
= h_1^{-d_T}\hat\mu\left(h_1^{d_T}g_{h_1}(T-t); x\right), 
\tag{MultiGPS}
\label{MultiGPS}
\end{align}
labelled as MultiGPS,
where 
the bandwidth $h_1$ is a positive sequence vanishing as $n$ grows,
the product kernel $g_{h_1}(T_i-t) \equiv \Pi_{j=1}^{d_T} g((T_{ji} - t_j)/h_1)/h_1^{d_T}$.
We can choose $g()$ to be the Gaussian kernel.\footnote{A possible drawback of MultiGPS is that the estimate could be negative or small in finite samples. 
We may adopt the trimming/flooring approaches to addressing this concern in the literature.
For example, following \cite{HLL20ER}, we can use the estimate $\max\{\hat f_{T|X}(t|X_i), \delta_n\}$ for some positive sequence $\delta_n\rightarrow 0$.
 }


\begin{lemma}[MultiGPS]
Consider an estimator of $\mathbb{E}\big[h_1^{d_T}g_{h_1}(T-t)\big|X=x\big]$, denoted as \\
$\hat\mu(h_1^{d_T}g_{h_1}(T-t); x)$, which satisfies 
$\big\| \hat\mu\big(h_1^{d_T}g_{h_1}(T-t); X\big) - {\mathbb{E}}[h_1^{d_T}g_{h_1}(T-t)|X] \big\|_{F_X} = O_p(R_{1n})$
for a sequence of constants $R_{1n}$ and $t\in\mathcal{T}$.
Let $g()$ satisfy Assumption~\ref{Akernel} with $g()$ in place of $k()$ and with an unbounded support.
Assume $f_{T|X}(t|x)$ to be three-times differentiable with respect to $t$ with all three derivatives being bounded uniformly over $(t^\prime,x^\prime)^\prime\in\mathcal{T}\times\mathcal{X}$.
Then the MultiGPS estimator $\hat f_{T|X}$ satisfies 
$\big\| \hat f_{T|X}(t|X) - f_{T|X}(t|X)\big\|_{F_X} = O_p(R_{1n} h_1^{-d_T} + h_1^2)$  
and 
$\big\| \hat f_{T|X} - f_{T|X}\big\|_{F_{tX}} 
= O_p(R_{1n} h_1^{-d_T} + h_1^2)$ for $t\in\mathcal{T}$.  

Further assume $\sup_{t\in\mathcal{T}}\big\| \hat\mu\big(h_1^{d_T}g_{h_1}(T-t); X\big) - {\mathbb{E}}\big[h_1^{d_T}g_{h_1}(T-t)\big|X\big] \big\|_{F_X} = O_p(R_{1n})$ for a sequence of constants $R_{1n}$.
Then 
$\sup_{t\in\mathcal{T}} \big\| \hat f_{T|X}(t|X) - f_{T|X}(t|X)\big\|_{F_X} = O_p(R_{1n} h_1^{-d_T} + h_1^2)$  
and 
$\sup_{t\in\mathcal{T}} \big\| \hat f_{T|X} - f_{T|X}\big\|_{F_{tX}} 
= O_p(R_{1n} h_1^{-d_T} + h_1^2)$.  
\label{LGPSmulti}
\end{lemma}

\section{Asymptotic theory}
\label{SecAsy}
We present the asymptotically linear representation and asymptotic normality.
We provide low-level conditions for estimating the nuisance functions by the deep neural networks in \cite{Farrell18} in Section~\ref{Sec1stEst}. 
Conditions for kernel and series estimators are in Section S3.2 in the online supplementary appendix. 
Section~\ref{SecL2} provides sufficient rate conditions using the standard $L_2$ norm.

Let $\partial_t^\nu g(t, \cdot) \equiv \partial^\nu g(t, \cdot)/\partial t^\nu$ denote the $\nu$th partial derivative of a generic function $g$ with respect to $t$, and $\partial_t \equiv \partial_t^1$.
\begin{theorem}[Asymptotic normality] 
Let Assumptions~\ref{ACIACS}-\ref{A1st} hold.
Let $h \rightarrow 0$, $nh^{d_T} \rightarrow\infty$, and $nh^{d_T+4} \rightarrow C \in [0,\infty)$.
Then for any $t\in\mathcal{T}$, 
\begin{align}
\sqrt{nh^{d_T}} \left(\hat \beta_t - \beta_t\right) = \sqrt{\frac{h^{d_T}}{n}} \sum_{i=1}^n\bigg\{
&\frac{K_h(T_i-t)}{\bar f(t,X_i)} (Y_i - \bar\gamma(t, X_i)) 
+  \bar\gamma(t, X_i) - \beta_t  \bigg\}
+ o_p(1). \label{EIF}
\end{align}

Further let $\mathbb{E}\big[|Y-\bar\gamma(T,X)|^3\big|T=t, X\big]$ and its derivatives with respect to $t$ be bounded uniformly over $(t^\prime,x^\prime)^\prime\in \mathcal{T}\times\mathcal{X}$.
Let $\int_{-\infty}^\infty k(u)^3du < \infty$.
Then $\sqrt{nh^{d_T}} \left(\hat \beta_t - \beta_t - h^2 \mathsf{B}_t\right) \stackrel{d}{\longrightarrow} \mathcal{N}\left(0, \mathsf{V}_t\right)$,
where  
$\mathsf{V}_t \equiv \mathbb{E}\big[ \mathbb{E}[(Y-\bar\gamma(t,X))^2|T=t, X] f_{T|X}(t,X)/\bar f(t,X)^2\big] R_k^{d_T}$
and 
\\
$\mathsf{B}_t \equiv 
\sum_{j=1}^{d_T} \mathbb{E}\Big[
\Big(
2\partial_{t_j} \bar{\gamma}(t,X)\partial_{t_j} f_{T|X}(t|X)
+ \partial_{t_j}^2 \bar{\gamma}(t,X) f_{T|X}(t|X) + 
\big( \gamma(t,x) - \bar\gamma(t,x) \big)
\partial^2_{t_j} {\gamma}(t, X)
\Big)$\\$\Big/(2\bar f(t,X))
\Big]  \kappa$.
\label{TIF}
 \end{theorem}

 
Note that the second part in the influence function\footnote{
For our non-regular parameters, we borrow the terminology  ``influence function" in estimating a regular parameter that is $\sqrt{n}$-estimable. 
An influence function gives the first-order asymptotic effect of a single observation on the estimator.
The estimator is asymptotically equivalent to a sample average of the influence function.
See \cite{Hampel} and \cite{IchimuraNewey22QE}, for example.
} in (\ref{EIF}) $n^{-1}\sum_{i=1}^n \bar\gamma(t,X_i) - \beta_t = O_p(1/\sqrt{n}) = o_p(1/\sqrt{nh^{d_T}})$ and hence does not contribute to the first-order asymptotic variance~$\mathsf{V}_t$.
We keep these smaller-order terms to show that the nuisance function estimators have no first-order influence on the asymptotic distribution of $\hat \beta_t$.
This is in contrast to the binary treatment case where $K_h(T_i-t)$ is replaced by ${\bf 1}\{T_i - t\}$ in $\hat \beta_t$, 
so $\hat \beta_t$ converges at a root-$n$ rate.
Then the second part in (\ref{EIF}) is of first-order for a binary treatment, resulting in the well-studied efficient influence function in estimating the binary treatment effect in \cite{Hahn98ETA}.

Theorem~\ref{TIF} is fundamental for inference, such as constructing confidence intervals and the optimal bandwidth $h$ that minimizes the asymptotic mean squared error.
We propose an estimator for the leading bias  $\mathsf{B}_t$, inspired by the idea in \cite{PS96}.
Let the notation $\hat\beta_t = \hat\beta_{t,b}$ be explicit on the bandwidth $b$ and
\[
\hat{\mathsf{B}}_t \equiv \frac{\hat\beta_{t,b} - \hat\beta_{t,ab}}{b^2 (1-a^2)}
\] 
with a pre-specified fixed scaling parameter $a \in (0, 1)$.
Theorem~\ref{Tbw} below shows the consistency of $\hat{\mathsf{B}}_t$ under Assumption~\ref{AOptBW}(d).

We can estimate the asymptotic variance $\mathsf{V}_t$ by the sample variance of the estimated influence function $\hat{\mathsf{V}}_t \equiv h^{d_T} n^{-1} \sum_{\ell=1}^L\sum_{i \in I_\ell} \hat \psi_{i\ell}^2$, where 
$\hat \psi_{i\ell} \equiv K_h(T_i-t) (Y_i - \hat \gamma_\ell(t,X_i) )/\hat f_\ell(t|X_i) +  \hat \gamma_\ell(t,X_i) - \hat\beta_t$.
Then we propose a data-driven bandwidth $\hat h_t \equiv  \big(d_T \hat{\mathsf{V}}_t\big/\big(4\hat{\mathsf{B}}_t^2\big)\big)^{1/(d_T+4)} n^{-1/(d_T+4)}$
to consistently estimate the optimal bandwidth that minimizes the asymptotic mean squared error (AMSE) given in Theorem~\ref{Tbw}.

\begin{assumption}
For each $\ell=1,...,L$ and for any $t\in \mathcal{T}$, 
(a)~$\|(\hat\gamma_\ell - \bar \gamma)(\hat f_\ell - \bar f)\|_{F_{tX}}  = o_p(1)$, 
(b)~$\|(\hat\gamma_\ell - \bar \gamma)^2(\hat f_\ell - \bar f)^2\|_{F_{tX}} = O_p(1)$, 
$\|(\hat\gamma_\ell - \bar \gamma)^2\|_{F_{tX}} = O_p(1)$, and 
$\|(\hat f_\ell - \bar f)^2\|_{F_{tX}} = O_p(1)$;
(c)~$\mathbb{E}[(Y-\bar \gamma(t, x))^4|T=t, X=x]$ and its derivatives with respect to $t$ are bounded uniformly over $(t^\prime,x^\prime)^\prime \in \mathcal{T}\times\mathcal{X}$
and $\int_{-\infty}^\infty k(u)^4 du < \infty$;
(d)~$b\rightarrow 0$ and $nb^{d_T + 4} \rightarrow \infty$. $\int_{-\infty}^\infty k(u)k(u/a) du < \infty$ for $a\in (0,1)$.
\label{AOptBW}
\end{assumption}

\begin{theorem}[AMSE optimal bandwidth for $\hat\beta_t$]
Let the conditions in Theorem \ref{TIF} hold.
For $t\in\mathcal{T}$, if $\mathsf{B}_t$ is non-zero, then the bandwidth that minimizes the asymptotic mean squared error is 
$h^\ast_t =  \big(d_T \mathsf{V}_t\big/\big(4\mathsf{B}_t^2\big)\big)^{1/(d_T+4)} n^{-1/(d_T+4)}$.
Further let Assumption~\ref{AOptBW} hold.
Then $\hat{\mathsf{V}}_t - \mathsf{V}_t = o_p(1)$, $\hat{\mathsf{B}}_t - \mathsf{B}_t = o_p(1)$, and $\hat h_t/h_t^\ast -1 = o_p(1)$.
\label{Tbw}
\end{theorem}

Assumption~\ref{AOptBW}(a)-(c) are for the consistency of $\hat{\mathsf{V}}_t$. 
The condition~(a) strengthens Assumption~\ref{A1st}(a), and (b) is mild boundedness conditions that are implied when 
$\hat\gamma_\ell$ and $\hat f_\ell$ are bounded uniformly.  
In practice, we may use different Step~1 estimators in $\hat{\mathsf{V}}_t$ and $\hat\beta_t$ due to different high-level conditions.

%
%

A common approach is to choose an undersmoothing bandwidth $h$ smaller than $h^\ast_t$ such that the bias is first-order asymptotically negligible, i.e.\ $h^2\sqrt{nh^{d_T}} \rightarrow 0$.  
Then we can construct the usual $(1-\alpha)\times 100\%$ point-wise confidence interval $\Big[\hat \beta_t \pm \Phi^{-1}(1-\alpha/2) \sqrt{\hat{\mathsf{V}}_t/(nh^{d_T})} \Big]$, where $\Phi$ is the CDF of $\mathcal{N}(0,1)$.
Alternatively, we may consider a further bias correction to allow for a wider range of bandwidth choice so that we may implement $\hat h_t$ in practice. 
Specifically we may use the above bias estimator $\hat{\mathsf{B}}_t$ and account for its variation in the asymptotic theory of the bias-corrected estimator $\hat\beta_t - h^2\hat{\mathsf{B}}_t$.  
\cite{CCF} show that the AMSE optimal bandwidth of the original estimator is feasible in different contexts.
\cite{Westling} develop such robust bias-corrected inference for the local linear estimator in \cite{Kennedy}.


Next we present the asymptotic theory for $\hat \theta_t$.

\begin{theorem}[Partial effect]
Let the conditions in Theorem \ref{TIF} hold.
Assume that $f_{Z}(y,t,x)$ is four-times differentiable with respect to $t$ with all four derivatives being bounded uniformly over $(y,t^\prime,x^\prime)^\prime \in \mathcal{Z}$.
Let $\eta/h \rightarrow 0$, $nh^{d_T + 2} \rightarrow \infty$, $nh^{d_T + 2 }\eta^2 \rightarrow 0$, and $nh^{d_T+6}\rightarrow C\in [0,\infty)$.
Let $\int_{-\infty}^\infty k'''(u)^2 du < \infty$.
Assume that
for each $\ell= 1,...,L$ and for any $t \in \mathcal{T}$,
(a)~$\eta^{-1}h\big\|\hat \gamma_\ell - \bar \gamma \big\|_{F_{tX}} = o_p(1)$ and 
$\eta^{-1}h\big\|\hat f_\ell - \bar f \big\|_{F_{tX}} = o_p(1)$;
(b)~$\eta^{-1}h\sqrt{nh^{d_T}} \big\| \hat f_\ell - f_{T|X} \big\|_{F_{tX}}  \big\| \hat\gamma_\ell - \gamma\big\|_{F_{tX}}  \stackrel{p}{\rightarrow}~0$. 
Then for any $t \in \mathcal{T}$, 
\begin{align}
\sqrt{nh^{d_T + 2}}( \hat \theta_t - \theta_{t} ) = \sqrt{\frac{h^{d_T+2}}{n}}  \sum_{i=1}^n \frac{\partial}{\partial t_1} K_h(T_i - t) \frac{Y_i - \bar\gamma(t, X_i)}{\bar f(t,X_i)}
 + o_p(1)
 \label{EIFME}
 \end{align}
and 
$\sqrt{nh^{d_T + 2}}( \hat \theta_t - \theta_{t} - h^2 \mathsf{B}_t^\theta)   \stackrel{d}{\longrightarrow}  \mathcal{N}(0, \mathsf{V}_t^\theta)$,
where 
$\mathsf{V}_t^\theta \equiv \mathsf{V}_t \int k'(u)^2du R_k^{-1}$ with $\mathsf{V}_t$ given in Theorem~\ref{TIF}
and 
$\mathsf{B}^{\theta}_t \equiv\mathsf{B}_{\theta_t}  + \mathsf{B}_{\theta_t}^{mis}$ with
\begin{align*}
\mathsf{B}_{\theta_t} &\equiv \mathbb{E}\bigg[
\sum_{j=1}^{d_T}   
 \bigg\{
 \frac{1}{2}\partial_{t_j}^2 \partial_{t_1} \gamma(t, X) f_{T|X}(t|X)+ \partial_{t_j} \partial_{t_1} \gamma(t, x) \partial_{t_j}f_{T|X}(t|X)
\notag\\
&\ \ \  +  \partial_{t_j} \gamma(t, X) 
\bigg( \partial_{t_j} \partial_{t_1} f_{T|X}(t|X) - \partial_{t_j} f_{T|X}(t|X)
 \frac{\partial_{t_1}\bar f(t,X) }{\bar f(t,X)}\bigg)
\bigg\} \frac{\kappa}{\bar f(t,X)} \bigg], 
\end{align*}
\begin{align*}
\mathsf{B}_{\theta_t}^{mis} &\equiv\mathbb{E}\bigg[\sum_{j=1}^{d_T}   
 \bigg\{ \left(\partial_{t_1} \gamma(t,X) - \partial_{t_1}\bar \gamma(t, X)\right)  \partial_{t_j}^2 f_{T|X}(t|X)
\notag \\
& \ \ \ 
+  \left(\gamma(t,X) - \bar \gamma(t, X)\right)
\left( \partial_{t_j}^2 \partial_{t_1} f_{T|X}(t|X)
- \partial_{t_j}^2 f_{T|X}(t|X) \frac{\partial_{t_1}\bar f(t,X) }{\bar f(t,X)}\right)
\notag\\
&\ \ \ 
+ \partial_{t_j}^2 \gamma(t,X)\bigg(\partial_{t_1} f_{T|X}(t|X)-  \partial_{t_1}\bar f(t,X)\frac{f_{T|X}(t|X) }{\bar f(t,X)}\bigg)
\bigg\} \frac{\kappa}{2\bar f(t,X)} \bigg].
\end{align*}


\label{TME}
\end{theorem}

The conditions (a) and (b) in Theorem~\ref{TME} strengthen Assumption~\ref{A1st} for $\beta_t$ and 
imply that $\eta$ cannot be too small and depends on the precision of the nuisance function estimators.
The bias $\mathsf{B}_{\theta_t}^{mis}$ is due to misspecifying one of the nuisance functions 
and is zero when both nuisance functions are correctly specified or estimated by nonparametric methods.

We propose a data-driven bandwidth $\hat h_{\theta_t} \equiv  
\big((d_T+2) \hat{\mathsf{V}}_t^\theta\big/\big(4(\hat{\mathsf{B}}_t^\theta)^2\big)\big)^{1/(d_T+6)} n^{-1/(d_T+6)}$
to consistently estimate the optimal bandwidth that minimizes the AMSE given in Theorem~\ref{Tbwtheta}.
Following the same procedure for $\hat\beta_t$, 
let
$\hat{\mathsf{V}}_t^\theta \equiv 
\hat{\mathsf{V}}_t \int k'(u)^2du R_k^{-1}$
and 
$\hat{\mathsf{B}}^\theta_t  \equiv \big(\hat\theta_{t,b}
- \hat\theta_{t,ab}\big)/\big(b^2 (1-a^2)\big)$.
In practice, we may choose $\eta = h n^{-a}$, where $a > 1/(d_T + 6)$, such that the conditions in Theorem~\ref{TME} are satisfied.

\begin{theorem}[AMSE optimal bandwidth for $\hat\theta_t$]
Let the conditions in Theorem \ref{TME} hold.
For $t\in\mathcal{T}$, if $\mathsf{B}_t^\theta$ is non-zero, then the bandwidth that minimizes the asymptotic mean squared error is 
$h^\ast_{\theta_t} =  \big((d_T+2) \mathsf{V}_t^\theta\big/\big(4{\mathsf{B}_t^\theta}^2\big)\big)^{1/(d_T+6)} n^{-1/(d_T+6)}$.
Further let Assumption~\ref{AOptBW} hold.
Then $\hat{\mathsf{V}}_t^\theta - \mathsf{V}_t^\theta = o_p(1)$, $\hat{\mathsf{B}}^\theta_t - \mathsf{B}_t^\theta = o_p(1)$, and $\hat h_{\theta_t}/h_{\theta_t}^\ast -1 = o_p(1)$.
\label{Tbwtheta}
\end{theorem}


\subsection{Nuisance function estimators}
\label{Sec1stEst}
We show that the high-level conditions on the convergence rates in Assumption~\ref{A1st}
are attainable by the nonparametric and ML methods: kernel, series, 
and the deep neural networks in \cite{Farrell18}, where the dimension of the control variables $d_X$ is fixed.
Lasso methods have been extensively studied in \cite{SUZ}, \cite{SU_PRTE}, and \cite{SUZ_UQR}, where $d_X$ can grow with $n$.
These ML methods require different low-level conditions, such as dimensionality, smoothness, and tuning parameters.


Consider the conditional density estimator MultiGPS $\hat f_{T|X}$ given in Section~\ref{SecGPS} for example.
By Lemma~\ref{LGPSmulti}, Assumption~\ref{A1st}(b) requires 
\begin{align*}
\sqrt{nh^{d_T}}  \|\hat f - f_{T|X}\|_{F_{tX}} \|\hat\gamma-\gamma\|_{F_{tX}} =
\sqrt{nh^{d_T}} \left(  R_{1n}h_1^{-d_T} + h_1^2\right) \left\|\hat\gamma-\gamma\right\|_{F_{tX}} = o_p(1).
\end{align*}  
Therefore, we need to obtain the partial $L_2(tX)$ convergence rate $\|\hat\gamma-\gamma\|_{F_{tX}}$
and the standard $L_2(X)$ convergence rate $R_{1n}$.

We seek theoretical results, such as the above rate conditions, for insights on selection of the tuning parameters in practice
 that is challenging and under-developed in the ML literature.
We may use the optimal choices for the nuisance function estimators, as they do not affect the first-order asymptotics.
A common method is cross-validation.
We may choose the optimal rates for the bandwidths for $\hat\gamma$ and $\hat\mu(h_1^{d_T}g_{h_1}(T-t); x)$  that respectively minimize $\|\hat\gamma-\gamma\|_{F_{tX}}$ and $\big\| \hat\mu\big(h_1^{d_T}g_{h_1}(T-t); X\big) - {\mathbb{E}}[h_1^{d_T}g_{h_1}(T-t)|X] \big\|_{F_X}$, which might be available in the literature. 
Similarly we can derive the optimal $h_1^\ast \propto R_{1n}^{1/(2+d_T)}$.
%

Next we propose a deep MLP-ReLU network kernel estimator for $\gamma(t, x)$ (labelled as Kernel NN) and derive its $L_2(tX)$ convergence rate.
We illustrate the low-level conditions of conventional kernel and series estimators in Section~S3 in the online supplementary appendix, as the calculations are rather standard.
These results on the $L_2(tX)$ convergence rates for neural networks and series are new and non-trivial extensions of existing results in the literature.


We consider the deep neural networks in \cite{Farrell18} (FLM, hereafter) that use the 
fully connected feedforward neural networks (multilayer perceptron, or MLP) and the nonsmooth rectified linear units (ReLU) activation function.
We propose a deep MLP-ReLU network kernel estimator for $\gamma(t, x)$.
The proposed estimator serves the purpose to conveniently apply the $L_2(TX)$ convergence rate given in FLM to obtain the $L_2(tX)$ convergence rate.
So we can deliver valid asymptotic inference for $\beta_t$ and $\theta_t$ following deep learning.  
In this section, we closely follow the notations in FLM for easy reference, by slightly abusing our notations.

We consider a kernel-weighted loss function for any $t\in\mathcal{T}$,
\begin{align*}
\ell_{tb}(f, Z) \equiv \frac{1}{2}\left(Y- f(X)\right)^2 \mathsf{K}_b\left(T- t\right),
\end{align*} 
where a product kernel $\mathsf{K}_b(T-t) \equiv \Pi_{j=1}^{d_T} \mathsf{k}((T_j - t_j)/b)/b^{d_T}$ 
with a kernel function $\mathsf{k}$ and a positive sequence of bandwidth $b = b_n$ vanishing as $n$ grows. 
We define the {\it deep MLP-ReLU network kernel estimator} for any $t\in\mathcal{T}$ as
\begin{align}
\hat f_{tb} &\in \arg\min_{f_\theta \in \mathcal{F}_{MLP}, \|f_\theta\|_\infty \leq 2M} \sum_{i=1}^n \ell_{tb}(f_\theta, Z_i),
\label{EstDNN}
\end{align}
where 
$\mathcal{F}_{MLP}$ is the MLP class, $M$ is an absolute constant, and $\theta$ depending on $t$ collects the weights and constants over all nodes.
We refer the details of the MLP-ReLU network estimators to FLM.
Then we obtain $\hat \gamma(t, x) = \hat f_{tb}(x)$.

Denote the derivative of a function $f(x)$ as $\mathtt{D}^{\alpha}_x f(x) = \frac{\partial^{|\bm{\alpha}|} f(x)}{\partial x_1^{\alpha_1}\cdots \partial x_{d_X}^{\alpha_{d_X}}}$, where $\bm{\alpha} = (\alpha_1,...,\alpha_{d_X})$ and $|\bm{\alpha}| = \alpha_1 +...+\alpha_{d_X}$.

\begin{assumption}[DNN]
(a) For any $t\in\mathcal{T}$, the second derivatives of $f_{T|X}(t,x)$ and $\gamma(t,x)$ with respect to $t$ are bounded and continuous uniformly over $x\in\mathcal{X}$;
(b) $X$ are continuously distributed with support $\mathcal{X} = [-1,1]^{d_X}$ for fixed $d_X$,\footnote{To simplify exposition, $X$ is assumed to be continuous, but discrete variables with a fixed number of supporting points are allowed.
Theorem 1 in \cite{FLM2} find that the rate only depends on the dimension of the continuously distributed components.
}
and for an absolute constant $M > 0$, $\mathcal{Y} \subset [-M,M]$; 
(c) For $r\in\mathcal{N}_+$ and some finite positive constant $c$, $\max_{\bm{\alpha}, |\bm{\alpha}| \leq r}{\rm ess}\sup_{y \in \mathcal{Y}, t\in\mathcal{T}, x\in\mathcal{X},} |\mathtt{D}_{x}^\alpha f_{YT|X}(y,t|x)| \leq c$;
(d) $\|f_{T|X}(t|\cdot)\|_\infty \leq c$ for some finite positive constant $c$.
\label{ADNN}
\end{assumption}

%
%
%
%
%

Assumption~\ref{ADNN}(a) is due to the kernel weight $\mathsf{K}_b(T-t)$ in the loss function.
Assumption~\ref{ADNN}(b)-(d) collect assumptions for applying Theorem 1 in FLM.
Detailed discussion on these assumptions is referred to FLM.
As discussed in FLM, it is standard in nonparametric analysis to assume the true function to be estimated is bounded. 
The choice of $M$ may be arbitrarily large and is simply a formalization of the requirement that the optimizer is not allowed to diverge on the function level in the sup-norm sense. 
For practical implementation, we do not impose such bound.
But there is a practice for rescaling the output variable which is generally dependent on the activation function being used, e.g.\ the domain of the activation function.
A common practice for variable transformation is standardization (subtracting mean and dividing by standard deviation) or scaling to a specific range by an affine transformation (generally chosen between 0 and 1). 


\begin{theorem}[DNN]
Let $\hat f_{tb}$ be the deep MLP-ReLU network kernel estimator defined by (\ref{EstDNN}).
Let Assumption \ref{ADNN} hold.
Let $\mathsf{k}()$ satisfy Assumption~\ref{Akernel} with $\mathsf{k}()$ in place of $k()$ and with a bounded support.
Let width $H \asymp (nb^{2d_T})^{\frac{{d_X}}{2(r+{d_X})}}\log^2(nb^{2d_T})$ and depth $L\asymp \log (nb^{2d_T})$.
Then for any $t\in\mathcal{T}$,
$\|\hat f_{tb} - \gamma\|_{F_{tX}}^2 \leq C\cdot\Big\{ \big(nb^{2d_T}\big)^{-\frac{r}{r+d_X}}\log^8n  +\log\log n/\big(nb^{d_T}\big) + b^2\Big\}$ with probability approaching one as $n\rightarrow\infty$,
for a constant $C > 0$ independent of $n$, which may depend on $d_X$, $M$, and other fixed constants.
\label{TDNN}
\end{theorem}

We can apply deep neural networks to the GPS estimation proposed in Section~\ref{SecGPS}.
The ReGPS estimator can use $\hat\mu\left(\Phi((t-T)/h_1); x\right) = \hat f_{MLP-ReGPS}(x)$ the MLP estimator in FLM with the unweighted loss function:
\begin{align}
\hat f_{MLP-ReGPS} \in \arg\min_{f_\theta \in\mathcal{F}_{MLP}, \|f_\theta\|_\infty \leq 2M} \sum_{i=1}^n \big(\Phi((t-T_i)/h_1) - f_\theta(X_i)\big)^2.
\label{DNNReGPS}
\end{align}
The MultiGPS estimator can use $\hat\mu\left(h_1^{d_T} g_{h_1}(T-t); x\right) = \hat f_{MLP-MultiGPS}(x)$:
\begin{align}
\hat f_{MLP-MultiGPS} \in \arg\min_{f_\theta \in\mathcal{F}_{MLP}, \|f_\theta\|_\infty \leq 2M} \sum_{i=1}^n \big(h_1^{d_T} g_{h_1}(T_i-t) - f_\theta(X_i)\big)^2.
\label{DNNMultiGPS}
\end{align}


\begin{lemma}[Theorem 1 in FLM]
Let $X$ be continuously distributed with support $\mathcal{X} = [-1,1]^{d_X}$ for fixed $d_X$.
Let width $H \asymp n^{\frac{{d_X}}{2(r+{d_X})}}\log^2n$ and depth $L\asymp \log n$.
Let $R_{1n}^2 = n^{-\frac{r}{r+d_X}}\log^8n  +\log\log n/n$. 
\begin{enumerate}
\item[(a)]  
For $\hat f_{MLP-ReGPS}$ defined by (\ref{DNNReGPS}), assume 
(A)~$\max_{\bm{\alpha},|\bm{\alpha}|\leq r} {\rm ess}\sup_{t\in\mathcal{T}, x\in[-1,1]^{d_X}} $
$|\mathtt{D}_x^{\alpha} F_{T|X}(t|x) | \leq~c$ for some finite positive constant $c$.
Let $\mathcal{T}_0 = [\underline{T}, \overline{T}]\subset\mathcal{R}$.
Then 
$\sup_{t \in\mathcal{T}}\big\| \hat f_{MLP-ReGPS}(X) - \mathbb{E}[\Phi((t-T)/h_1)|X] \big\|_{F_X} = O_p(R_{1n})$.

\item[(b)]
For $\hat f_{MLP-MultiGPS}$ defined by (\ref{DNNMultiGPS}), 
assume 
(B)~$\max_{\bm{\alpha},|\bm{\alpha}|\leq r} {\rm ess}\sup_{t\in\mathcal{T}, x\in[-1,1]^{d_X}}$ $|\mathtt{D}_x^{\alpha} f_{T|X}(t|x) | \leq~c$ for some finite positive constant $c$.
Let $g()$ satisfy Assumption~\ref{Akernel} in place of $k()$ and with an unbounded support. 
Then $\sup_{t \in\mathcal{T}}\big\| \hat f_{MLP-MultiGPS}(X) - \mathbb{E}[ h_1^{d_T} g_{h_1}(t-T)|X] \big\|_{F_X} = O_p(R_{1n})$. 
 
 \end{enumerate}

\label{LDNNMultiGPS}
\end{lemma}
 
We are ready to show that Assumption~\ref{A1st} is attainable by the MLP-ReLU network estimators.
Take the MultiGPS estimator in (\ref{DNNMultiGPS}) for example, with $d_T=1$ for simplicity.
Assumption~\ref{A1st}(b) is
$nh
\Big( (nb^{2})^{-\frac{r}{r+d_X}}\log^8n  +\log\log n/(nb) + b^2 \Big)
\Big(
h_1^{-2}\big( n^{-\frac{r}{r+d_X}}$\\
$\log^8n  + \log\log n/n\big) + h_1^4 \Big)
\rightarrow 0$ 
by Theorem \ref{TDNN}, Lemmas~\ref{LGPSmulti} and \ref{LDNNMultiGPS}.
Assumption~\ref{A1st}(a) is implied by $nb^2\rightarrow\infty$ and $n^{r/(r+d_X)}h_1^2\rightarrow\infty$.
When $h = h_1 = b$, Assumption~\ref{A1st}(a) holds by letting $n^{r/(r+d_X)}h^2 \rightarrow\infty$, and (ii) holds by letting smoothness $r > {d_X}$.  
FLM discuss the same condition  $r > {d_X}$  for the average treatment effect of a binary treatment variable.
Similarly we note that this condition is not minimal but is sufficient to justify the practical use of the MLP-ReLU network estimators for valid inference on the average structural function and the partial effect of continuous treatments by our approach.

\subsection{Standard $L_2$ norm rate}
\label{SecL2}
We provide sufficient rate conditions using 
the standard $L_2(TX)$ norm in Assumption~\ref{A1stL2} to replace the partial $L_2(tX)$ norm in Assumption~\ref{A1st}.
Let the $L_2(TX)$ norm be 
$\|\hat\gamma_\ell-\gamma\|_{F_{TX}} \equiv \big(\int_\mathcal{T}\int_\mathcal{X}
\big(\hat\gamma_\ell(t,x) - \gamma(t,x) \big)^2  f_{TX}(t,x)dx dt\big)^{1/2}$.
We do not need to modify the rate condition on $\hat f_{\ell}$, as it equivalently uses the standard $L_2(X)$ norm for a given $t$. 
The cost of the more commonly used $L_2(TX)$ norm is losing the doubly robust inference and stricter regularity conditions.

\begin{assumption}
There exists a function $\bar f(t,x)$ that 
is three-times differentiable with respect to $t$ with all
three derivatives being bounded uniformly over $(t',x') \in\mathcal{T}_0\times\mathcal{X}$,
$\inf_{t\in\mathcal{T}}{\rm ess}\inf_{x\in\mathcal{X}}
 \bar f(t,x) \geq c$ for some positive constant $c$, and satisfies the following: 
for each $\ell= 1,...,L$ and for any $t \in \mathcal{T}$,
(a)~$h^{-d_T/2} \|\hat \gamma_\ell - \gamma\|_{F_{TX}}= o_p(1)$ and $\| \hat f_\ell - \bar f \|_{F_{tX}} = o_p(1)$;
(b)~$\sqrt{n}\|\hat f_\ell - \bar f\|_{F_{tX}} \|\hat\gamma_\ell -  \gamma\|_{F_{TX}} = o_p(1)$; 
(c)~$nh^{d_T + 2} \rightarrow 0$.
(d)~$\hat\gamma_\ell$ is twice differentiable with respective to $t$ with two derivatives being bounded uniformly over $\mathcal{T}_0\times\mathcal{X}$.
\label{A1stL2}
\end{assumption}

\begin{corollary}
Let the conditions in Theorem~\ref{TIF} hold.
Let Assumption~\ref{A1stL2} instead of Assumption~\ref{A1st} hold.
Then for any $t\in\mathcal{T}$, 
$\sqrt{nh^{d_T}} \left(\hat \beta_t - \beta_t \right) 
= \sqrt{h^{d_T}/n} \sum_{i=1}^n\Big\{
(Y_i - \gamma(t, X_i)) K_h(T_i-t)/\bar f(t,X_i)
+  \gamma(t, X_i) - \beta_t  \Big\}
+ o_p(1)
\stackrel{d}{\longrightarrow} \mathcal{N}\left(0, \mathsf{V}_t\right)$,
where $\mathsf{V}_t$ is given in Theorem~\ref{TIF} with $\bar\gamma=\gamma$.

Furthermore let the conditions in Theorem~\ref{TME} hold.
Instead of the conditions (a) and (b) in Theorem~\ref{TME}, assume 
(a$^\prime$)~$\eta^{-1}h^{1-d_T/2}\big\|\hat \gamma_\ell -  \gamma \big\|_{F_{TX}} = o_p(1)$ and 
$\eta^{-1}h\big\|\hat f_\ell - \bar f \big\|_{F_{tX}} = o_p(1)$;
(b$^\prime$)~$\eta^{-1}h\sqrt{n} \big\| \hat f_\ell - \bar f \big\|_{F_{tX}}  \big\| \hat\gamma_\ell - \gamma\big\|_{F_{TX}}  \stackrel{p}{\rightarrow}~0$. 
Then for any $t \in \mathcal{T}$, 
$\sqrt{nh^{d_T + 2}}( \hat \theta_t - \theta_{t} ) = \sqrt{h^{d_T+2}/n}  \sum_{i=1}^n  (Y_i - \gamma(t, X_i))
\partial_{t_1} K_h(T_i - t)
/\bar f(t,X_i)
 + o_p(1)   \stackrel{d}{\longrightarrow}  \mathcal{N}(0, \mathsf{V}_t^\theta)$,
where 
$\mathsf{V}_t^\theta$ is given in Theorem~\ref{TME} with $\bar\gamma = \gamma$.
\label{CL2}
\end{corollary}

Note that the rate condition Assumption~\ref{A1stL2}(b) 
 is the same rate condition
for the regular semiparametric models in CCDDHNR and \cite{CNS21ADML}, e.g.\ $\hat f_{T|X}(t|x)$ is replaced with the propensity score $P(T=t|X=x)$ for a discrete treatment.
We discuss the cost and implications of Assumption~\ref{A1stL2} based on the standard $L_2$ norm.  
First under Assumption~\ref{A1stL2}, the inference theory cannot allow $\gamma$ to be misspecified, while $f_{T|X}$ can be misspecified. 
Second, the convergence rates are faster than those required in Assumption~\ref{A1st}.
To learn intuition on the rate condition (a), we utilize the bounded kernel in the DML estimator, resulting in a penalty $h^{-d_T/2}$ in the loose bound $h^{-d_T/2} \|\hat \gamma_\ell - \gamma\|_{F_{TX}}$.
Third, the bandwidth $h$ choice is more restrictive. 
Assumption~\ref{A1stL2}(c) implies undersmoothing, i.e.\ the leading bias of $\hat\beta_t$ is first-order ignorable by $h^2\sqrt{nh^{d_T}}  \rightarrow 0$.

For a specific example, consider $\hat\gamma$ to be FLM's estimator for the conditional expectation function $\mathbb{E}[Y|T,X]$,
i.e.\ using the unweighted loss function $(Y-f(T,X))^2/2$.
FLM provide the corresponding $L_2(TX)$ convergence rate $\|\hat \gamma_\ell -\gamma\|_{F_{TX}}^2 = O_p\left(
n^{-\frac{r}{r+d_X+d_T}}\log^8n  +\log\log n/n\right)$ with the smoothness $r$ defined in Assumption~\ref{ADNN}(c).
We need  $r > d_T + d_X$ that is stronger than $r > d_X$, as discussed in Section~\ref{Sec1stEst}.
Moreover comparing the rate conditions in Assumption~\ref{A1st}(b) and Assumption~\ref{A1stL2}(b) for $d_T=1$, we can show that using the partial $L_2(tX)$ norm results in a tighter bound by $\sqrt{nh}\|\hat\gamma_\ell - \gamma\|_{F_{tX}} = o_p(\sqrt{n}\|\hat\gamma_\ell - \gamma\|_{F_{TX}})$ with $r = d_x + 1$, $b=h \propto n^{-a}$ and $a < 1/2$.

\section{Simulated DML estimator}
\label{SecSDML}
We introduce Simulated DML estimator $\check\beta_t$ that 
enables the high-level rate conditions based on the standard $L_2(TX)$ norm, rather than the partial $L_2(tX)$ norm, and also permits the doubly robust inference.
Following the estimation procedure given in Section~\ref{SecEst}, Step~1 computes the nuisance function estimators $\hat \gamma_\ell$ and $\hat f_\ell$.
In Step~2, let $U_i \equiv \mathsf{T}_i h_0 + t$ 
where $\{\mathsf{T}_i\}_{i\in I_\ell}$ are i.i.d. draws from $\{T_i\}_{i\in I_\ell}$ with replacement, 
for $i \in I_\ell$, $\ell\in\{1,..., L\}$,
and $h_0$ is a positive sequence converging to zero as $n\rightarrow\infty$.
The Simulated DML (SDML) estimator is defined as 
\begin{align*}
\check \beta_t \equiv \frac{1}{n}\sum_{\ell=1}^L \sum_{i \in I_\ell} \left\{ \hat \gamma_\ell (U_i, X_i) + \frac{K_h(T_i - t)}{ \hat f_\ell(t|X_i)} \left( Y_i - \hat \gamma_\ell(U_i, X_i) \right)\right\}.
\end{align*}
The corresponding partial effect estimator $\check\theta_t \equiv (\check\beta_{t^+} -\check\beta_{t^-})/\eta$ as in Step~3.

To get intuition, the SDML estimator $\check\beta_t$ uses $\hat\gamma_\ell(U_i, X_i)$ rather than $\hat\gamma_\ell(t, X_i)$ as in $\hat\beta_t$ that fixes the treatment value at the target $t$.
The simulated $U_i$ localizes the realized treatment values around $t$.
Introducing such local variation enables the standard $L_2$ rate of $\hat\gamma$.
Specifically $\mathsf{T}_i$ defined above follows the empirical distribution function of $\{T_i\}_{i\in I_\ell}$, denoted as $\hat F_{T\ell}$.
Therefore $U_i$ follows a CDF conditional on the sample $\{Z_i=(Y_i, X_i, T_i)\}_{i=1}^n$, $P(U_i \leq u|\{Z_i\}_{i=1}^n) = P(\mathsf{T}_i h_0 + t \leq u|\{Z_i\}_{i=1}^n) = \hat F_{T\ell}((u-t)/h_0)$.
%

Assumption~\ref{A1stU} gives the high-level conditions on the nuisance function estimators.

\begin{assumption}
There exist functions $\bar \gamma(t,x)$ and $\bar f(t,x)$ that 
are three-times differentiable with respect to $t$ with all
three derivatives being bounded uniformly over $(y,t',x') \in\mathcal{Z}$ such that 
for each $\ell = 1,..., L$ and for $t\in\mathcal{T}$,
(a) $h_0^{-d_T/2}\|\hat\gamma_\ell-\bar \gamma\|_{F_{TX}} = o_p(1)$
and $\|\hat f_\ell - \bar f\|_{F_{tX}}= o_p(1)$;
(b) $\sqrt{nh^{d_T}} h_0^{-d_T/2}\|\hat\gamma_\ell- \gamma\|_{F_{TX}}\|\hat f_\ell -  f_{T|X}\|_{F_{tX}} = o_p(1)$;
(c)
$\|(\hat\gamma_\ell - \bar \gamma)^2\|_{F_{TX}} = O_p(1)$;
(d) Either $\bar \gamma = \gamma$ or $\bar{f} = f_{T|X}$.
%
%
\label{A1stU}
\end{assumption}

%
%
%


\begin{theorem}[SDML-Asymptotic normality] 
Let Assumptions~\ref{ACIACS}, \ref{Akernel}, and \ref{A1stU} hold.
Let $h \rightarrow 0$, $nh^{d_T} \rightarrow\infty$, and $nh^{d_T+4} \rightarrow C \in [0,\infty)$.
Let $h_0\rightarrow 0$, $nh_0^{d_T} \rightarrow\infty$, and $\sqrt{nh^{d_T}}h^2h_0 \rightarrow 0$.
Then 
the results in Theorem~\ref{TIF} hold with $\check\beta_t$ in place of~$\hat\beta_t$.
\label{TIF-SDML}
 \end{theorem}

Assumption~\ref{A1stU}(a) strengthens the consistency condition in Assumption~\ref{A1st}(a) with a penalty $h_0^{-d_T/2}$. 
Interestingly when $h_0 = h$, the rate condition (b) $\sqrt{n}\|\hat\gamma_\ell-\gamma\|_{F_{TX}}
\|\hat f_\ell - f_{T|X}\|_{F_{tX}} = o_p(1)$, which is the same rate condition
for the regular semiparametric models in CCDDHNR and \cite{CNS21ADML}.
Assumption~\ref{A1stU}(c) is a boundedness condition that is implied if $\hat\gamma$ is uniformly bounded;
for example, deep neural networks in FLM for a uniformly bounded $Y$.
Specifically, 
FLM provide the corresponding $L_2(TX)$ convergence rate $\|\hat \gamma_\ell -\gamma\|_{F_{TX}}$
and illustrate its usefulness for semiparametric inference on the average treatment effect of a binary treatment.
We could use this rate to verify the high-level conditions in Assumption~\ref{A1stU} for the SDML estimator
or Assumption~\ref{A1stL2} for the DML estimator as discussed in Section~\ref{SecL2}.

We can estimate the AMSE optimal bandwidth $h_t^\ast$ given in Theorem~\ref{Tbw} by the SDML approach. 
We can estimate the leading bias $\mathsf{B}_t$ by
$\check{\mathsf{B}}_t \equiv \frac{\check\beta_{t,b} - \check\beta_{t,ub}}{b^2 (1-a^2)}$
 with a pre-specified positive scaling parameter $a\in(0,1)$.
We can estimate the asymptotic variance $\mathsf{V}_t$ by $\check{\mathsf{V}}_t \equiv h^{d_T} n^{-1} \sum_{\ell=1}^L\sum_{i \in I_\ell} \check \psi_{i\ell}^2$, where 
$\check \psi_{i\ell} \equiv K_h(T_i-t) (Y_i - \hat \gamma_\ell(U_i,X_i) )/\hat f_\ell(t|X_i) +  \hat \gamma_\ell(U_i,X_i) - \check\beta_t$.
Then a data-driven bandwidth $\check h_t \equiv  \big(d_T \check{\mathsf{V}}_t\big/\big(4\check{\mathsf{B}}_t^2\big)\big)^{1/(d_T+4)} n^{-1/(d_T+4)}$.
Theorem~\ref{Tbw-SDML} below shows the consistency of $\check{\mathsf{B}}_t$, $\check{\mathsf{V}}_t$, and $\check h_t$, under Assumption~\ref{AOptBW-SDML} that is modified from Assumption~\ref{AOptBW}.

\begin{assumption}
For each $\ell=1,...,L$ and for any $t\in \mathcal{T}$, 
(a) $h_0^{-d_T/2}\|(\hat\gamma_\ell - \bar \gamma)(\hat f_\ell(t|\cdot) - \bar f(t,\cdot))\|_{F_{TX}}  = o_p(1)$, 
(b)~$h_0^{-d_T/2}\|(\hat\gamma_\ell - \bar \gamma)^2(\hat f_\ell(t|\cdot) - \bar f(t,\cdot))^2\|_{F_{TX}} = O_p(1)$, 
$\|(\hat\gamma_\ell - \bar \gamma)^4\|_{F_{TX}} = O_p(1)$, and 
$\|(\hat f_\ell - \bar f)^2\|_{F_{tX}} = O_p(1)$;
(c)~$\mathbb{E}[(Y-\bar \gamma(t, x))^4|T=t, X=x] f_{T|X}(t|x)$ is bounded uniformly over $x \in \mathcal{X}$;
(d)~$b\rightarrow 0$ and $nb^{d_T + 4} \rightarrow \infty$.
\label{AOptBW-SDML}
\end{assumption}

\begin{theorem}[SDML-AMSE optimal bandwidth]
Let the conditions in Theorem~\ref{TIF-SDML} and Assumption~\ref{AOptBW-SDML}  hold.
Then for $t\in\mathcal{T}$, 
$\check{\mathsf{V}}_t - \mathsf{V}_t = o_p(1)$, $\check{\mathsf{B}}_t - \mathsf{B}_t = o_p(1)$, and $\check h_t/h_t^\ast -1 = o_p(1)$.
\label{Tbw-SDML}
\end{theorem}

We can similarly obatin the asymptotic theory for $\check\theta_t$.
\begin{theorem}[SDML-Partial effect]
Let the conditions in Theorems~\ref{TME} and \ref{TIF-SDML} hold.
Assume that 
for each $\ell= 1,...,L$ and for any $t \in \mathcal{T}$,
(a)~$\eta^{-1}h h_0^{-d_T/2}\big\|\hat \gamma_\ell - \bar \gamma \big\|_{F_{TX}} = o_p(1)$ and 
$\eta^{-1}h\big\|\hat f_\ell- \bar f \big\|_{F_{tX}} = o_p(1)$;
(b)~$\eta^{-1}h\sqrt{nh^{d_T}} h_0^{-d_T/2}\big\| \hat f_\ell- f_{T|X} \big\|_{F_{tX}}  \big\| \hat\gamma_\ell - \gamma\big\|_{F_{TX}}  \stackrel{p}{\rightarrow}~0$.
Then the results in Theorem~\ref{TME} hold
with $\check\theta_t$ in place of $\hat\theta_t$.
\label{TME-SDML}
\end{theorem}



\section{Numerical examples}
\label{SecNEx}
This section provides numerical examples of Monte Carlo simulations and an empirical illustration.
The estimation procedure of the proposed DML estimator is described in Section~\ref{SecEst}.
To estimate the first-step conditional expectation function $\gamma(t,x) = \mathbb{E}[Y|T=t, X=x]$ and the conditional density $f_{T|X}$ by MultiGPS as described in Section~\ref{SecGPS}, we employ three methods: Lasso,
the deep neural networks (NN) based on \cite{Farrell18},
and the Kernel NN proposed in Section~\ref{Sec1stEst}. 
We implement our DML estimator with these algorithms respectively.
Note that Lasso assumes certain sparsity specifications and NN is a nonparametric method.  Our doubly robust inference theory allows one of the first-step functions to be misspecified.
Software we develop is available at \url{https://github.com/KColangelo/Double-ML-Continuous-Treatment}.
Section~S1 in the online supplementary appendix provides the implementation details and additional results of ReGPS.

\subsection{Simulation study}
\label{SecMC}
We consider the data-generating process:
$\nu \sim \mathcal{N}(0,1)$,  $\varepsilon \sim \mathcal{N}(0,1)$, 
\begin{align*}
X &= (X_1,...,X_{100})' \sim \mathcal{N}(0,\Sigma), \ 
T = \Phi(3 X'\theta) + 0.75\nu - 0.5,   
\\ 
Y &= 1.2T + 1.2 X'\theta +  T^2 + TX_1 + \varepsilon*\sqrt{0.5+\Phi(X_1)}, 
\end{align*}
where 
$\theta_j = 1/j^2$,
$diag(\Sigma) = 1$, the $(i,j)$-entry $\Sigma_{ij} = 0.5$ for $|i-j|=1$ and $\Sigma_{ij} = 0$ for $|i-j|>1$  for $i,j=1,...,100$,
and $\Phi$ is the CDF of $\mathcal{N}(0,1)$.
Thus this is a nonseparable model and the potential outcome $Y(t) = 1.2t + 1.2 X'\theta +  t^2 + tX_1 + \varepsilon*\sqrt{0.5+\Phi(X_1)}$.
The parameters of interest are the average dose response function and the partial effect at $t = 0$, i.e.\ $\beta_0 = \mathbb{E}[Y(0)] = 0$ and $\theta_0 = \partial \mathbb{E}[Y(t)]/\partial t|_{t=0} = 1.2$.

We compare estimations with cross-fitting and without cross-fitting, and with a range of bandwidths to demonstrate robustness to bandwidth choice. 
We consider sample size $n \in \{1000,10000\}$ and the number of subsamples used for cross-fitting $L \in\{1, 5\}$. 
We use the second-order Epanechnikov kernel with bandwidth $h$. 
For the MultiGPS estimator described in Section~\ref{SecGPS}, we choose bandwidth $h_1=h$.
Let the bandwidth $h=c \sigma_T n^{-0.2}$ for a constant $c\in\{0.75, 1.0,1.25, 1.5\}$
and the standard deviation $\sigma_T$ of $T$.
We computed the AMSE-optimal bandwidth $h^\ast_0$ given in Theorem~\ref{Tbw} that has the corresponding $c^\ast = 1.45$.
Thus using some undersmoothing bandwidth with $c < c^\ast$, the 95\% confidence interval 
$\big[\hat \beta_t \pm 1.96 s.e.\big]$ is asymptotically valid, where
the standard error ($s.e.$) is computed using the sample analogue of the estimated influence function,
as described in Section~\ref{SecAsy}.

Table~\ref{TDGP-High} reports the results based on 1,000 Monte Carlo replications. 
Under no cross-fitting ($L=1$), the confidence intervals generally have lower coverage rates and the bias is larger than under cross-fitting. 
The estimators using NN and Kernel NN perform well in the case of fivefold cross-fitting, with coverage rates (Cov.) near the nominal 95\%, especially in a smaller sample size $n=1,000$. 
Intuitively Kernel NN estimates the conditional expectation function $\gamma(t,x)$ locally at $t$, resulting in a smaller effective sample size $nb^{d_T}$ than the full sample size $n$ used by NN.
Therefore NN may outperform Kernel NN in small samples.
The coverage rate and bias are improved the most for NN and Kernel NN with cross-fitting, but only marginally for Lasso. 
Cross-fitting should improve our estimation in the case that the machine learning algorithm is over-fitting. 
Given that cross-fitting does not improve Lasso, it might suggest that Lasso does not have a severe over-fitting problem for this data-generating process.



These methods seem  robust to bandwidth choice under cross-fitting.
Overall these results demonstrate consistency with the theoretical results of this paper, confirming the usefulness of cross-fitting for ML methods.


{\footnotesize
       \begin{table}
        \caption{\label{TDGP-High}Simulation Results.}
        \begin{center}
         \begin{tabular*}{0.9\textwidth}{@{}cc|ccc|ccc|ccc@{}}
                   \hline\hline
                   	\multicolumn{2}{c}{} 	      &\multicolumn{3}{c}{Lasso}  & \multicolumn{3}{c}{Neural Network} & \multicolumn{3}{c}{Kernel Neural Network}     \\[4pt]
	     L     & \multicolumn{1}{c}{c} & Bias  & RMSE   & \multicolumn{1}{c}{Cov.} 
	     & Bias  & RMSE   & \multicolumn{1}{c}{Cov.} 
	     & Bias  & RMSE  & Cov. \\
\hline
\hline
\\[-12pt]
		& \multicolumn{10}{c}{$\beta_0 = 0$} \\[2pt]
\hline
\\[-14pt]
    & \multicolumn{10}{c}{$n=1,000$} \\[2pt]  
           1       
          & 0.75  & 0.003 & 0.110 & 0.936 & 0.063 & 0.113 & 0.887 &	0.036	&	0.124	&	0.597\\
                 & 1.00  & 0.008 & 0.100 & 0.935 & 0.064 & 0.108 & 0.890&	0.037	&	0.111	&	0.689 \\
                 & 1.25  & 0.016 & 0.094 & 0.937 & 0.067 & 0.109 & 0.859 &	0.044	&	0.105	&	0.748\\
                 & 1.50  & 0.025 & 0.091 & 0.925 & 0.077 & 0.112 & 0.840 &	0.051	&	0.103	&	0.778\\
\hline
         5      & 0.75  & 0.001 & 0.111 & 0.936 & -0.025 & 0.143 & 0.946 &	0.003	&	0.146	&	0.959\\
                & 1.00  & 0.008 & 0.100 & 0.936 & -0.014 & 0.122 & 0.947 &	0.009	&	0.122	&	0.965\\
                 & 1.25  & 0.016 & 0.094 & 0.936 & -0.007 & 0.109 & 0.967 &	0.018	&	0.108	&	0.968\\
                 & 1.50  & 0.025 & 0.092 & 0.925 & 0.002 & 0.100 & 0.969 &	0.029	&	0.102	&	0.956\\
          \hline\\[-14pt]
& \multicolumn{10}{c}{$n=10,000$} \\[2pt]     
           1       & 0.75  & 0.003 & 0.040 & 0.955 & 0.002 & 0.035 & 0.961&	0.004	&	0.042	&	0.851 \\
                 & 1.00  & 0.005 & 0.036 & 0.956 & 0.005 & 0.032 & 0.957 &	0.006	&	0.037	&	0.885\\
                 & 1.25  & 0.008 & 0.034 & 0.954 & 0.009 & 0.032 & 0.937 &	0.010	&	0.035	&	0.895\\
                 & 1.50  & 0.012 & 0.033 & 0.944 & 0.014 & 0.032 & 0.918 &	0.014	&	0.034	&	0.899\\
\hline
          5       & 0.75  & 0.001 & 0.060 & 0.949 & 0.012 & 0.044 & 0.956 &	0.002	&	0.044	&	0.948\\
                 & 1.00  & 0.005 & 0.036 & 0.953 & 0.014 & 0.039 & 0.948 &	0.006	&	0.039	&	0.950\\
                & 1.25  & 0.008 & 0.034 & 0.952 & 0.018 & 0.039 & 0.916 &	0.009	&	0.036	&	0.948\\
                 & 1.50  & 0.012 & 0.033 & 0.944 & 0.023 & 0.039 & 0.899 &	0.014	&	0.035	&	0.945\\
\hline
\hline
\\[-12pt]
& \multicolumn{10}{c}{$\theta_0 = 1.2$} \\[2pt]
\hline\\[-14pt]
  &  \multicolumn{10}{c}{$n=1,000$} \\[2pt]    
          1     & 0.75  & 0.107 & 0.888 & 0.978 & 0.072 & 0.717 & 0.979 &	0.276	&	1.448	&	0.512\\
                 & 1.00  & 0.092 & 0.588 & 0.978 & 0.077 & 0.503 & 0.981&	0.206	&	0.943	&	0.620 \\
                 & 1.25  & 0.087 & 0.436 & 0.980 & 0.114 & 0.398 & 0.985 &	0.176	&	0.655	&	0.714\\
                 & 1.50  & 0.088 & 0.338 & 0.979 & 0.083 & 0.312 & 0.986 &	0.140	&	0.499	&	0.784\\
\hline
          5       & 0.75  & 0.098 & 0.892 & 0.977 & 0.054 & 0.869 & 0.997 &	0.327	&	1.792	&	0.908\\
                 & 1.00  & 0.084 & 0.588 & 0.980 & 0.084 & 0.568 & 0.997 &	0.242	&	1.131	&	0.902\\
                 & 1.25  & 0.077 & 0.434 & 0.979 & 0.088 & 0.426 & 0.993 &	0.206	&	0.836	&	0.897\\
                 & 1.50  & 0.077 & 0.335 & 0.982 & 0.061 & 0.336 & 0.996 &	0.177	&	0.616	&	0.920\\
              \hline\\[-14pt]
  &  \multicolumn{10}{c}{$n=10,000$} \\[2pt]    
          1     & 0.75  & -0.008 & 0.527 & 0.976 & 0.018 & 0.534 & 0.957&	0.099	&	0.664	&	0.822 \\
                 & 1.00  & 0.005 & 0.351 & 0.985 & 0.021 & 0.348 & 0.969 &	0.105	&	0.429	&	0.880\\
                 & 1.25  & 0.006 & 0.254 & 0.979 & 0.016 & 0.241 & 0.980&	0.086	&	0.295	&	0.917 \\
                 & 1.50  & 0.000 & 0.200 & 0.977 & 0.017 & 0.187 & 0.983 &	0.078	&	0.230	&	0.931\\\hline
          5     & 0.75  & -0.008 & 0.548 & 0.976 & 0.003 & 0.575 & 0.978&	0.099	&	0.757	&	0.931 \\
                 & 1.00  & 0.001 & 0.353 & 0.983 & 0.012 & 0.352 & 0.984 &	0.108	&	0.460	&	0.945\\
                 & 1.25  & 0.004 & 0.254 & 0.981 & 0.007 & 0.245 & 0.989 &	0.095	&	0.319	&	0.946\\
                 & 1.50  & -0.002 & 0.200 & 0.980 & 0.009 & 0.188 & 0.990 &	0.082	&	0.242	&	0.954\\
\hline
\hline 
   \end{tabular*}%
       \end{center}
       \footnotesize
       \renewcommand{\baselineskip}{11pt}
       \textbf{Note:} 
       The top panel shows the results for $\hat\beta_0$ and the bottom panel is for $\hat\theta_0$.
       $L=1$: no cross-fitting.  $L=5$: fivefold cross-fitting. 
The bandwidth is $h = c \sigma_T n^{-0.2}$, and $c = 1.45$ for the AMSE-optimal bandwidth for $\beta_t$.  
      The nominal coverage rate (Cov.) of the confidence interval is 0.95.
       \end{table}
}

\subsection{Empirical illustration}
We illustrate our method by re-analyzing the Job Corps program in the United States, which was conducted in the mid-1990s.
The Job Corps program is the largest publicly funded job training program, which targets disadvantaged youth.
The participants are exposed to different numbers of actual hours of academic and vocational training.
The participants' labor market outcomes may differ if they accumulate different amounts of human capital acquired through different lengths of exposure.
We estimate the average dose response functions to investigate the relationship between employment and the length of exposure to academic and vocational training.
As our analysis builds on \cite{FFGN12ReStat}, \cite{HHLP}, and \cite{Lee}, we refer the readers to the reference therein for further details of Job Corps.

We use the same dataset in \cite{HHLP}.
We consider the outcome variable ($Y$) to be the proportion of weeks employed in the second year following the program assignment.
The continuous treatment variable ($T$) is the total hours spent in academic and vocational training in the first year.
We follow the literature to assume the conditional independence Assumption~\ref{ACIACS}(a), meaning that selection into different levels of the treatment is random, conditional on a rich set of observed covariates, denoted by $X$. 
The identifying Assumption~\ref{ACIACS} is indirectly assessed in \cite{FFGN12ReStat}.
Our sample consists of 4,024 individuals who completed at least 40 hours (one week) of academic and vocational training.
There are 40 covariates measured at the baseline survey.
 In the online supplementary appendix, Figure~S1 shows the distribution of $T$ by a histogram, and Table~S2 provides brief descriptive statistics.
\\[5pt]
\textbf{Implementation details:}\
We estimate the average dose response function $\beta_t = \mathbb{E}[Y(t)]$ and partial effect $\theta_t = \partial \mathbb{E}[Y(t)]/\partial t$  by the proposed DML estimator with fivefold cross-fitting.
We implement three DML estimators: Lasso, the generalized random forests in \cite{ATW19AS}, the neural networks (NN) based on \cite{Farrell18},
and the Kernel NN proposed in Section~\ref{Sec1stEst}.  
The parameters for these methods are selected as described in Section~S1 in the online supplementary appendix.

We use the second-order Epanechnikov kernel with bandwidth $h$. 
For the MultiGPS estimator, we use the Gaussian kernel with bandwidth $h_1=h$.
We compute the optimal bandwidth that minimizes an asymptotic integrated MSE.
For practical implementation, consider a weight function $w(t) = {\bf 1}\{ t \in [\underline{t}, \bar t]\}/(\bar t - \underline{t})$ that is the density of $Uniform[\underline{t}, \bar t]$ on a subset of the support of $T$.
The bandwidth that minimizes the asymptotic integrated MSE $
\int_{\mathcal{T}} \big( \mathsf{V}_t/(nh^{d_T}) + h^4 \mathsf{B}_t^2\big) w(t) dt$
for an integrable weight function $w(t): \mathcal{T} \mapsto \mathcal{R}$
is $h^\ast_w =  \big(d_T\mathsf{V}_w\big/\big(4\mathsf{B}_w\big)\big)^{1/(d_T+4)} n^{-1/(d_T+4)}$,
where $\mathsf{V}_w \equiv \int_{\mathcal{T}} \mathsf{V}_t w(t) dt$ and 
$\mathsf{B}_w \equiv \int_{\mathcal{T}}   \mathsf{B}_t^2 w(t) dt$,  following Theorem~\ref{Tbw}.
Set $m$ equally spaced grid points over $[\underline{t}, \bar t]$: $\big\{\underline{t} = t_1, t_2,..., t_m = \bar t\big\}$.
Following the approach given in Section~\ref{SecAsy}, we estimate ${\mathsf{V}}_{t_j}$ with $h=3\hat\sigma_Tn^{-0.2} = 548.52$
and ${\mathsf{B}}_{t_j}$ with $b = 2h$ and $a=0.5$, for $j=1,...,m$.
A plug-in estimator $\hat h^\ast_w =  \big(\hat{\mathsf{V}}_w\big/\big(4\hat{\mathsf{B}}_w\big)\big)^{1/5} n^{-1/5}$, where
 $\hat{\mathsf{V}}_w = m^{-1}\sum_{j=1}^m \hat{\mathsf{V}}_{t_j}$
 and 
 $\hat{\mathsf{B}}_w = m^{-1}\sum_{j=1}^m \hat{\mathsf{B}}_{t_j}^2$.
We use $[\underline{t}, \bar t] = [160, 1840]$ and $t_j - t_{j-1} = 40$ in this empirical application.
We then obtain under-smoothing bandwidths  $0.8 \hat h^\ast_w$  that are 213.45 for Lasso,  224.36 for the generalized random forest, 223 for NN, and 225.11 for Kernel NN.

 \begin{figure}[!htp]
\centering
\caption{\small Estimated average dose response functions with 95\% confidence intervals}
\includegraphics[width=0.243\textwidth]{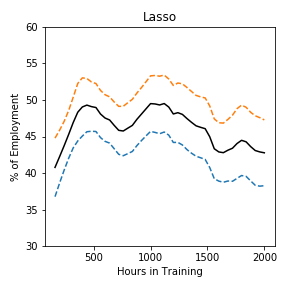}
\includegraphics[width=0.243\textwidth]{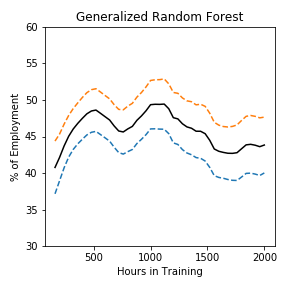}
\includegraphics[width=0.243\textwidth]{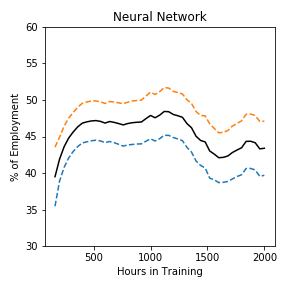}
\includegraphics[width=0.243\textwidth]{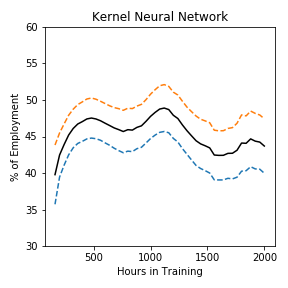}
\label{FMLbeta}
\end{figure}
 \begin{figure}[!htp]
\centering
\caption{\small Estimated partial effects  with 95\% confidence intervals}
\includegraphics[width=0.243\textwidth]{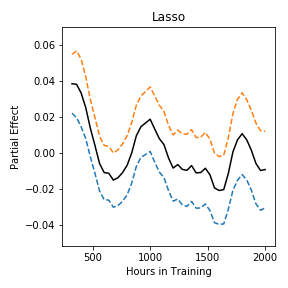}
\includegraphics[width=0.243\textwidth]{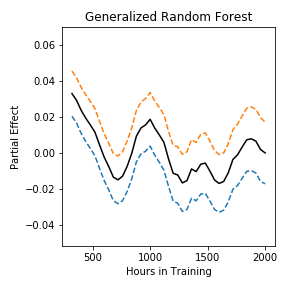}
\includegraphics[width=0.243\textwidth]{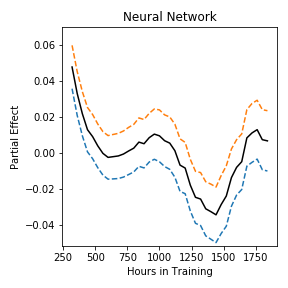}
\includegraphics[width=0.243\textwidth]{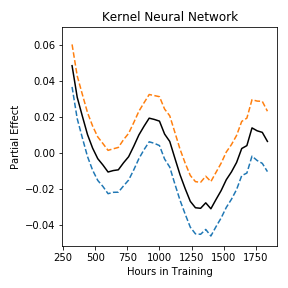}
\label{FMLtheta}
\end{figure}
 \textbf{Results:}\
Figure~\ref{FMLbeta} presents the estimated average dose response function $\beta_t$ along with 95\% point-wise confidence intervals.
The results for the three ML nuisance function  estimators have similar patterns.
The estimates suggest an inverted-U relationship between the employment and the length of participation.
DNN estimates appear to be the most erratic, possibly due to the smaller bandwidth compared with other estimators.

Figure~\ref{FMLtheta} reports the partial effect estimates $\hat\theta_t$ with step size $\eta =160$ (one month).
Across all procedures, we see positive partial effects when hours of training are less than around 500 (three months)
and negative partial effect around 1,500 hours (9 months).
Taking the estimates by NN for example, 
$\hat \beta_{400} = 46.83$ with standard error $s.e. = 1.38$ and $\hat \theta_{400} = 0.0217$ with $s.e. = 0.0123$ computed based on the result of Theorem~\ref{TME}.
This estimate implies that increasing the training from two months to three months increases the average proportion of weeks employed in the second year by $3.47\%$ (nearly two weeks) with $s.e. = 1.962\%$.

\cite{Lee09RES} finds that the program had a negative impact on employment propensities in the short term (104 weeks since random assignments) and a positive effect in the long term (104-208 weeks).
\cite{Lee09RES} considers a binary treatment variable of being in the program or not, with the outcome variable ${\bf 1}\{Y \geq 0\}$ in our notations.
We focus on the employment proportion in the second year following the program assignment (52-104 weeks)
and estimate the heterogenous effects of the total hours spent in academic and vocational training in the first year.

The empirical practice has focused on semiparametric estimation; see \cite{FFGN12ReStat}, \cite{HHLP}, \cite{Lee}, for example.
The semiparametric methods are subject to the risk of misspecification.
Our DML estimator provides a feasible approach to implementing a fully nonparametric inference in practice.

\section{Conclusion and outlook}

For a future extension, our DML estimator serves as the preliminary element for policy learning and optimization with a continuous decision, following \cite{Manski04}, \cite{HiranoPorter}, \cite{KT18ECMA}, \cite{Kallus}, \cite{DSLC}, \cite{AW_PL}, \cite{Farrell18}, among others. 

Another extension is robustness against multiway clustering, where the conventional cross-fitting does not ensure the independence between observations in $I_\ell$ from $I_\ell^c$.
We may adopt the $K^2$-fold multiway cross-fitting proposed by \cite{Chiang} that focus on regular DML estimators as in CCDDHNR. 
Since the form of estimators and the proofs of asymptotic theories for our continuous treatment case are similar to those studied in \cite{Chiang}, we expect that their proposed algorithm works for $\hat \beta_t$; a formal extension is out of the scope of this paper.

When unconfoundedness is violated, we can use the control function approach in triangular simultaneous equations models
by including in the covariates some estimated control variables using instrumental variables. 
For example, \cite{Lee09RES} studies the issue of sample selection for the wage effects of the Job Corps program.
To extend our empirical application to the wage effect of the length of exposure to the program, 
we may follow \cite{Lee09RES} to estimate bounds on the wage effect of the continuous treatment using the excess number of individuals who are induced to be selected.
A closer approach to our estimator is 
\cite{DNV03}, who
show that a nonparametric control function method accounts for both selection and endogeneity.
\cite{IN09ETA} show 
that the conditional independence assumption holds when the covariates $X$ include the additional control variable $V = F_{T|Z}(T|Z)$, the conditional distribution function of the endogenous variable given the instrumental variables $Z$. 
The influence function that accounts for estimating the control variables as generated regressors has derived in Corollary 2 in \cite{Lee15}.
\cite{Lee15} shows that the adjustment terms for the estimated control variables are of smaller order
in the influence function of the final estimator, but it may be important to include them to achieve local robustness.
This is a distinct feature of the average structural function of continuous treatments, as discussed in Section~\ref{SecAsy}. 
Using such an influence function to construct the corresponding DML estimator is left for future research.

%


\section*{Acknowledgements}
This work is not related to Kyle Colangolo's position at Amazon, and was completed outside of the regular duties of the position.
We are grateful to Max Farrell, Whitney Newey, Takuya Ura, Ted Westling, and Yichong Zhang for valuable discussion.
We thank Parush Arora for assistance.

{\small
\bibliographystyle{chicagoa}	
\bibliography{database}		
}

\newpage


    \section*{Appendix: Proofs of Results}
    \renewcommand{\theequation}{A.\arabic{equation}}
    \renewcommand{\thesection}{A}
    \setcounter{equation}{0}


%

%
\textbf{Proof of Lemma~\ref{LGPS}:}\  
We suppress subscripts for notational ease, e.g.\ $F(t|x) = F_{T|X}(t|x)$.
We first show that uniformly in $t\in\mathcal{T}$, 
$\|\mathbb{E}[\Phi((t-T)/h_1)|X=\cdot] - F_{T|X}(t|\cdot)\|_{F_X}^2
= 
  \int_\mathcal{X} \big(\int_{-\infty}^\infty\Phi((t-s)/h_1) f_{T|X}(s|x) ds  - F_{T|X}(t|x)\big)^2 f_X(x)dx
= 
 \int_\mathcal{X} (\int_{-\infty}^\infty h_1^{-1}\phi((t-s)/h_1) $
 \\
 $\times F_{T|X}(s|x) ds - F_{T|X}(t|x) )^2 f_X(x)dx
= 
\int_\mathcal{X} \big(\int_{-\infty}^\infty \phi(u) F_{T|X}(t+uh_1|x) du - F_{T|X}(t|x) \big)^2 $
\\
$f_X(x)dx
=  O(h_1^4)$,
using integration by parts, change of variables, a Taylor series expansion, and assuming $\partial^3 F_{T|X}(t|x)/\partial t^3$ to be uniformly bounded over $\mathcal{T}\times\mathcal{X}$.

By the triangle inequality,
\begin{align}
&\left\| \widehat{1/f_{T|X}(t|\cdot)}  - 1/f_{T|X}(t|\cdot)\right\|_{F_X} 
\notag\\
&\leq 
 \frac{1}{2\epsilon}
\bigg\{
\int_\mathcal{X}  \Big( \hat F^{-1}(\hat F(t|x) + \epsilon|x) - \hat F^{-1}(\hat F(t|x) - \epsilon|x)
\notag \\
&\ \ \ \ \ \ \ \ \ \ \ - \left(
 F^{-1}( F(t|x) + \epsilon|x) -  F^{-1}( F(t|x) - \epsilon|x) 
\right)\Big)^2 f_X(x) dx
\bigg\}^{1/2}\label{d1}\\
&\ \ \ + \left\{\int_\mathcal{X} \left(
\frac{ F^{-1}( F(t|x) + \epsilon|x) -  F^{-1}( F(t|x) - \epsilon|x) }{2\epsilon}
- \frac{1}{f(t|x)}\right)^2 f_X(x) dx
\right\}^{1/2}.
\label{d2}
\end{align}

For (\ref{d1}), we focus on $\big\| \hat F^{-1}(\hat F(t|\cdot) + \epsilon|\cdot) -  F^{-1}( F(t|\cdot) + \epsilon|\cdot) \big\|_{F_X}$.
Denote $s =\hat F^{-1}(\hat F(t|x) + \epsilon|x) \in \mathcal{T}$ by construction.
So $\hat F(s|x) = \hat F(t|x) + \epsilon$.
Denote the $p$th partial derivative of the conditional $u$-quantile function $F^{-1}(u|x)$ with respect to (w.r.t.) $u$ as $\partial^p F^{-1}(u|x) = \frac{\partial^p}{\partial v^p} F^{-1}(v|x)|_{v=u}$.
By the mean value theorem, $s = F^{-1}(F(s|x)|x) = F^{-1}(\hat F(s|x)|x) +  \big(F(s|x) - \hat F(s|x)\big) \partial^1F^{-1}(\bar F|x)$
with $\bar F$ between $F(s|x)$ and $\hat F(s|x)$.
So 
\begin{align*}
s &= \hat F^{-1}(\hat F(t|x) + \epsilon|x) 
\\
&= F^{-1}(\hat F(t|x) + \epsilon|x) + \big(F(s|x) - \hat F(s|x)\big) \partial^1F^{-1}(\bar F|x)
\\
&= F^{-1}(F(t|x) + \epsilon|x) 
+ \big(\hat F(t|x) - F(t|x)\big) \partial^1F^{-1}(\tilde F|x)
+ \big(F(s|x) - \hat F(s|x)\big) \partial^1F^{-1}(\bar F|x)
\end{align*}
with $\tilde F$ between $F(t|x) + \epsilon$ and $\hat F(t|x) + \epsilon$ by the mean value theorem.
As $\partial^1F^{-1}$ is assumed to be bounded above uniformly, 
we obtain that uniformly in $t\in\mathcal{T}$, $\big\| \hat F^{-1}(\hat F(t|\cdot) + \epsilon|\cdot) -  F^{-1}( F(t|\cdot) + \epsilon|\cdot) \big\|_{F_X}
= O_p\big(
\sup_{t\in\mathcal{T}}\big\| \hat F(t|\cdot) - F(t|\cdot) \big\|_{F_X}
\big) 
= O_p\big(
\sup_{t\in\mathcal{T}} \big\| \hat \mu(\Phi((t-T)/h_1);\cdot) - \mathbb{E}[\Phi((t-T)/h_1)|X=\cdot] \big\|_{F_X}
+\sup_{t\in\mathcal{T}} \|\mathbb{E}[\Phi((t-T)/h_1)|X=\cdot] - F(t|\cdot)\|_{F_X}
\big) = 
O_p(R_{1n} + h_1^2)$.
So the term in (\ref{d1}) is $O_p(R_{1n}/\epsilon + h_1^2/\epsilon)$ uniformly in $t\in\mathcal{T}$.

Next consider (\ref{d2}).
By a Taylor series expansion, 
\begin{align*}
F^{-1}(F(t|x)+\epsilon|x) &= F^{-1}(F(t|x) |x) + \epsilon \partial^1 F^{-1}(F(t|x)|x) + \epsilon^2 
 \partial^2 F^{-1}(F(t|x)|x)/2 \\
 &\ \ \ +  \epsilon^3 \partial^3  F^{-1}(\bar F(t|x)|x) /3!
 \\
F^{-1}(F(t|x) - \epsilon|x) &= F^{-1}(F(t|x) |x) - \epsilon \partial^1 F^{-1}(F(t|x)|x) + \epsilon^2 
 \partial^2 F^{-1}(F(t|x)|x)/2 \\
 &\ \ \ - \epsilon^3 \partial^3  F^{-1}(\tilde F(t|x)|x) /3!
 \end{align*}
with $\bar F(t|x)\in[F(t|x), F(t|x) + \epsilon]$
and $\tilde F(t|x)\in[F(t|x)-\epsilon, F(t|x)]$.
So
$\big(F^{-1}( F(t|x) + \epsilon|x) -  F^{-1}( F(t|x) - \epsilon|x)\big)\big(2\epsilon\big)
- 1/f(t|x) = 
\epsilon^2\big(\partial^3  F^{-1}(\tilde F(t|x)|x) + \partial^3  F^{-1}(\tilde F(t|x)|x) \big)/(2\cdot3!)$.
Since we assume $\partial^3  F^{-1}(u|x)$ to be uniformly bounded over $(u,x)\in (0,1)\times\mathcal{X}$, 
the term in (\ref{d2}) is $O(\epsilon^2)$ uniformly in $t\in\mathcal{T}$.

As we assume that there exists a positive constant $C$ such that $\sup_{(t, x) \in\mathcal{T}\times\mathcal{X}} f_{T|X}(t|x) \leq C$, 
we obtain that $\Big\{\int_\mathcal{X} \big( \widehat{1/f_{T|X}(t|x)}  - 1/f_{T|X}(t|x)
\big)^2 f_{TX}(t,x)dx \Big\}^{1/2} \leq
 \Big\| \widehat{1/f_{T|X}(t|\cdot)}  - 1/f_{T|X}(t|\cdot)\Big\|_{F_X} \sqrt{C}
= O_p(R_{1n}\epsilon^{-1} + h_1^2\epsilon^{-1} + \epsilon^2)$ uniformly in $t\in\mathcal{T}$.
\hfill $\square$
\\
\\
\textbf{Proof of Lemma~\ref{LGPSmulti}:}\ 
By the assumptions, there exists a finite positive constant $C$ such that $\sup_{(t, x) \in\mathcal{T}\times\mathcal{X}} f_{T|X}(t|x) \leq C$.
By the triangle inequality,
\begin{align}
&\left\{\int_\mathcal{X} \big( \hat f_{T|X}(t|x) - f_{T|X}(t|x) \big)^2 f_{TX}(t,x)dx \right\}^{1/2}\notag \\
&\leq \left\{ \int_\mathcal{X} \big( \hat f_{T|X}(t|x) - f_{T|X}(t|x) \big)^2 f_{X}(x)dx C\right\}^{1/2}
 \notag \\
&\leq \left\{ \frac{C}{h_1^{2d_T}}  \int_\mathcal{X}\left( \hat \mu\left( \Pi_{j=1}^{d_T}g\left(\frac{T_{j}-t_j}{h_1}\right);x \right) - \mathbb{E} \left[\Pi_{j=1}^{d_T}g\left(\frac{T_{j}-t_j}{h_1}\right)\Big| X=x \right] \right)^2 f_{X}(x)dx\right\}^{1/2}
\notag 
\\
&\ \ \ + \left\{ C\int_\mathcal{X}\left( \mathbb{E} \left[g_{h_1}\left(T-t\right)\Big| X=x \right]  - f_{T|X}(t|x) \right)^2 f_{X}(x)dx\right\}^{1/2}
\label{EGPS2}
 \\
&= O_p\left( h_1^{-d_T}R_{1n} + h_1^2\right). \notag
\end{align}
For the term in (\ref{EGPS2}) to be $O(h_1^2)$, we follow the standard algebra for kernel as the arguments in the proof of Lemma~\ref{LGPS}.
\hfill $\square$
\\
\\
%
%
\textbf{Asymptotically linear representation of $\hat\beta_t$:}\ 
We give an outline of deriving the asymptotically linear representation in Theorem~\ref{TIF}.
Let the moment function for identification $m(Z_i, \beta_t, \gamma) \equiv \gamma(t, X_i) - \beta_t$ by (\ref{IDreg}), i.e.\ $\mathbb{E}[m(Z_i, \beta_t, \gamma(t, X_i))] = 0$ uniquely defines $\beta_t$.
Let the adjustment term $\phi(Z_i, \beta_t, \gamma, \lambda) \equiv
K_h(T_i - t) \lambda(t, X_i) \big( Y_i - \gamma(t, X_i) \big)$, where $\lambda(t,x) \equiv 1/f_{T|X}(t|x)$.
The doubly robust moment function $\psi(Z_i, \beta_t, \gamma, \lambda) \equiv m(Z_i, \beta_t, \gamma(t, X_i)) + \phi(Z_i, \beta_t, \gamma(t, X_i), \lambda(t, X_i))$, as in equation (\ref{EDR}).

Let $Z_\ell^c$ denote the observations $z_i$ for $i \neq I_\ell$.
Let  $\hat \gamma_{i\ell} \equiv \hat \gamma_\ell(t, X_i)$ 
and $\hat\lambda_{i\ell} \equiv 1/\hat f_{\ell}(t|X_i)$ using $Z_\ell^c$ for $i \in I_\ell$.
Let $\bar \gamma_i \equiv \bar \gamma(t, X_i)$ and $\bar \lambda_i \equiv \bar \lambda(t, X_i)$.
We can write $\hat \beta_t = L^{-1}\sum_{\ell=1}^L \hat \beta_{t\ell}$, where $\hat \beta_{t\ell} = n_\ell^{-1}\sum_{i\in I_\ell}  \psi(Z_i, \beta_t, \hat \gamma_{i\ell}, \hat \lambda_{i\ell}) + \beta_t$ and $n_\ell = n/L$.
Then 
$\sqrt{nh^{d_T}}(\hat\beta_t - \beta_t) 
= \sqrt{nh^{d_T}} L^{-1}\sum_{\ell=1}^L (\hat \beta_{t\ell}- \beta_t)
=  L^{-1/2} \sum_{\ell=1}^L  \sqrt{n_\ell h^{d_T}}(\hat \beta_{t\ell}- \beta_t)$.
We show below $\sqrt{n_\ell h^{d_T}}(\hat \beta_{t\ell}- \beta_t) = 
\sqrt{h^{d_T}/n_\ell}\sum_{i \in I_\ell} \psi(Z_i, \beta_t, \bar \gamma_i, \bar \lambda_i) + o_p(1)
$ for each $\ell\in\{1,...,L\}$.
Since $L$ is fixed and $\{I_\ell\}_{\ell=1,...,L}$ are randomly partitioned distinct subgroups, the result follows from $\sqrt{nh^{d_T}}(\hat\beta_t - \beta_t) =
L^{-1/2} \sum_{\ell=1}^L  \sqrt{n_\ell h^{d_T}}(\hat \beta_{t\ell}- \beta_t)
= 
L^{-1/2} \sum_{\ell=1}^L  \sqrt{h^{d_T}/n_\ell}
$\\$\sum_{i \in I_\ell} 
\psi(Z_i, \beta_t, \bar \gamma_i, \bar  \lambda_i) + o_p(1)
= 
\sqrt{h^{d_T}/n}  \sum_{\ell=1}^L \sum_{i \in I_\ell} \psi(Z_i, \beta_t, \bar \gamma_i,\bar  \lambda_i) + o_p(1).$

We decompose the remainder term for each $\ell\in\{1,..., L\}$,
\begin{align}
&\sqrt{n_\ell h^{d_T}}\frac{1}{n_\ell}\sum_{i\in I_\ell}\left\{  \psi(Z_i, \beta_t, \hat \gamma_{i\ell}, \hat \lambda_{i\ell}) - \psi(Z_i, \beta_t, \bar \gamma_i, \bar \lambda_i) \right\} \notag \\
=&\ \sqrt{\frac{h^{d_T}}{n_\ell}} \sum_{i\in I_\ell}\bigg\{ 
\hat \gamma_{i\ell} - \bar \gamma_i - \mathbb{E}[\hat \gamma_{i\ell} - \bar \gamma_i|Z_\ell^c]   
\notag \\
& + K_h(T_i - t)\bar \lambda_i (\bar \gamma_i - \hat \gamma_{i\ell}) - \mathbb{E}\big[K_h(T_i - t)\bar \lambda_i (\bar \gamma_i -\hat \gamma_{i\ell})\big|Z_\ell^c\big] 
\tag{R1-1} \label{ER11} \\
& + 
K_h(T_i - t)(\hat \lambda_{i\ell} - \bar \lambda_i)(Y_i - \bar \gamma_i) - \mathbb{E}\big[K_h(T_i - t)(\hat \lambda_{i\ell} - \bar \lambda_i)(Y_i - \bar \gamma_i) \big|Z_\ell^c\big]
 \tag{R1-2} \label{ER12} \\
& + 
\mathbb{E}[ (\hat \gamma_{i\ell} - \bar \gamma_i)(1-K_h(T_i - t) \bar \lambda_i)|Z_\ell^c] 
+ \mathbb{E}[ (\hat \lambda_{i\ell} - \bar \lambda_i) K_h(T_i - t) (Y_i - \bar \gamma_i)|Z_\ell^c] 
\tag{R1-DR} \label{ER1DR} \\
& - 
K_h(T_i - t)\big(\hat \lambda_{i\ell} - \bar \lambda_i\big)\big(\hat \gamma_{i\ell} - \bar \gamma_i\big)\bigg\}. 
\tag{R2} \label{ER2}
\end{align}
The remainder terms (\ref{ER11}) and (\ref{ER12}) are stochastic equicontinuous terms that are controlled to be $o_p(1)$ by
the mean-square consistency conditions in Assumption~\ref{A1st}(a) and cross-fitting.
The second-order remainder term (\ref{ER2}) is controlled by Assumption~\ref{A1st}(b).
 
The remainder term (\ref{ER1DR}) is the key to doubly robust inference.
Note that in the binary treatment case when $K_h(T_i - t)$ is replaced by ${\bf 1}\{T_i = t\}$, the term (\ref{ER1DR}) is zero because $\psi$ is the Neyman-orthogonal influence function, under correct specification $\bar \gamma = \gamma$ and $\bar f = f_{T|X}$. 
In our continuous treatment case, the Neyman orthogonality holds as $h \rightarrow 0$.


We show that by a standard algebra of kernel and the assumed conditions, 
for any $t\in\mathcal{T}$, 
$\int_{\mathcal{T}_0} K_h(s - t) f_{TX}(s,x)ds = f_{TX}(t, x) + O(h^2)$
uniformly in $x\in \mathcal{X}$.
We will use the same arguments in the proofs.
We use change of variables $u = (u_1, u_2,..., u_{d_T}) = (T-t)/h$,
a Taylor expansion, the mean value theorem where $\bar t$ is between $t$ and $t+uh$,
$f_{TX}(t,x)$ being bounded away from zero, 
and the second derivatives of $f_{TX}(t,x)$ being bounded uniformly over $(t^\prime,x^\prime)^\prime \in \mathcal{T}\times\mathcal{X}$ to show that \begin{align}
&\int_{\mathcal{T}_0} K_h(s - t) f_{TX}(s,x)ds 
\notag\\
&= 
\int_{\mathcal{R}^{d_T}} \Pi_{j=1}^{d_T}k(u_j) f_{TX}(t+uh,x)du 
\notag \\
&= 
\int_{\mathcal{R}^{d_T}} \Pi_{j=1}^{d_T}k(u_j) \bigg( f_{TX}(t,x) + h \sum_{j=1}^{d_T}u_j\frac{\partial f_{TX}(t,x) }{\partial t_j} \notag\\ 
&\ \ \ + \frac{h^2}{2} \sum_{j=1}^{d_T}u_j^2\frac{\partial^2 f_{TX}(t,x)}{\partial{t_j^2}}  \Big|_{t=\bar t}
+ \frac{h^2}{2}\sum_{j=1}^{d_T}\sum_{l=1, l\neq j}^{d_T}u_j u_l \frac{\partial^2  f_{TX}(t,x)}{\partial t_j \partial t_l}  \Big|_{t=\bar t}
 \bigg) du_1\cdots du_{d_T} \notag \\
&= 
\int_{\mathcal{R}^{d_T}} \Pi_{j=1}^{d_T}k(u_j) 
f_{TX}(t,x) \bigg\{ 1 + h \sum_{j=1}^{d_T}u_j\frac{\partial f_{TX}(t,x) }{\partial t_j}
\frac{1}{f_{TX}(t,x)}  \notag\\ 
&\ \ \ + \frac{h^2}{2} \sum_{j=1}^{d_T}\bigg(u_j^2 \frac{\partial^2 f_{TX}(t,x)}{\partial{t_j^2}}  \Big|_{t=\bar t}
+ \sum_{l=1, l\neq j}^{d_T}u_j u_l \frac{\partial^2  f_{TX}(t,x)}{\partial t_j \partial t_l}  \Big|_{t=\bar t}\bigg)\frac{1}{f_{TX}(t,x)} 
 \bigg\} du_1\cdots du_{d_T} \notag \\
&= f_{TX}(t, x) (1 + h^2 C)
\label{algebra}
\\
&= f_{TX}(t, x) + O(h^2)
\notag
\end{align}
for some positive constant $C$, for any $t\in\mathcal{T}$, uniformly over $x\in\mathcal{X}$.

Our results are readily extended to include binary/multivalued treatments $D$ at the cost of notational complication, e.g.\ the low-dimensional setting in \cite{Cattaneo10JoE}.
Specifically, the frequency method replaces the kernel with an indicator function:
 $\hat \beta_{td} = n^{-1} \sum_{\ell=1}^L \sum_{i \in I_\ell} \big\{ \hat \gamma_l(t, d, X_i) + {\bf 1}\{D_i = d\} K_h(T_i - t)\big( Y_i - \hat \gamma_l(t, d, X_i)\big)/ \hat f_{{TD|X}_l}(t, d|X_i)\big\}$, where $\gamma(t, d, X_i) = \mathbb{E}[Y|T=t, D=d, X=X_i]$
and 
$f_{TD|X}(t, d|X_i) = f_{T|DX}(t|d, X_i)$\\$\times {\rm Pr}(D=d|X=X_i)$.\footnote{
There is a literature on the kernel smoothing of discrete (categorical) variables (\cite{AA}, \cite{OLR} and reference therein); such extension to smoothing discrete treatments is out of the scope of this paper.}
\\
\\
\textbf{Proof of Theorem~\ref{TIF}:}\
The statements in the following hold for $i \in I_\ell$, $\ell\in\{1,...., L\}$, and for all $t$.
For (\ref{ER1DR}), 
in the first part $
\mathbb{E}\big[ 1 -K_h(T_i - t) \bar{\lambda}_i \big|X_i\big] 
= 1- \mathbb{E}\big[K_h(T_i - t) \big|X_i\big] \bar{\lambda}_i
= 1 - \bar{\lambda}_i/\lambda_i + h^2 \partial_t^2 f_{T|X}(t|X_i) \bar{\lambda}_i \int_{-\infty}^\infty u^2k(u) du/2  + O_p(h^3)$.
So the first term in (\ref{ER1DR})
$\mathbb{E}[ (\hat \gamma_{i\ell} - \bar \gamma_i)(1-K_h(T_i - t) \bar \lambda_i)|Z_\ell^c] 
=\mathbb{E}[(\hat\gamma_{i\ell} - \bar{\gamma}_i)
(1 - \bar{\lambda}_i/\lambda_i + h^2 \partial_t^2 f_{T|X}(t|X_i) \bar{\lambda}_i \int_{-\infty}^\infty u^2k(u) du/2  + O_p(h^3))
|Z_\ell^c]$.
Under the conditions in Theorem~\ref{TIF}, $ \sqrt{nh^{4+d_T}}\big(\mathbb{E}[(\hat \gamma_\ell(t, X) - \bar \gamma(t, X) )|Z_\ell^c ]  + \mathbb{E}[ ( \hat \lambda_\ell(t,X) - \bar \lambda(t,X) )|Z_\ell^c]  \big)= o_p(1)$.
So when $\bar\lambda = \lambda$, this first term in (\ref{ER1DR}) is $o_p(1/\sqrt{n h^{d_T}})$.
When $\bar\lambda \neq \lambda$, 
this first term is $o_p(1/\sqrt{n h^{d_T}})$ by assuming $\|\hat\gamma_{\ell} - \bar{\gamma}\|_{F_{tX}} = \|\hat\gamma_{\ell} - {\gamma}\|_{F_{tX}} = o_p(1/\sqrt{n_\ell h^{d_T}})$
 in Assumption~\ref{A1st}(b)(c).

A similar argument yields the second term in (\ref{ER1DR}), 
$\mathbb{E}\Big[
(\hat\lambda_{i\ell} - \bar{\lambda}_i)
\Big((\gamma_i - \bar{\gamma}_i)/\lambda_i
+ h^2
\big((\gamma_i - \bar{\gamma}_i) \partial_t^2 f_{T|X}(t, X_i)/2 +  \partial_t f_{T|X}(t, X_i) \partial_t \gamma_i
+ f_{T|X}(t,X_i) \partial_t^2 \gamma_i/2
\big)\Big)\Big|Z_\ell^c\Big]$.
When $\bar{\gamma} = \gamma$, this second term of (\ref{ER1DR}) is $o_p(1/\sqrt{n h^{d_T}})$.
When $\bar{\gamma} \neq \gamma$, 
this second term of (\ref{ER1DR}) is $o_p(1/\sqrt{n_\ell h^{d_T}})$ by assuming $\|\hat\lambda_{\ell} - \bar{\lambda}\|_{F_{tX}} = \|\hat\lambda_{\ell} - {\lambda}\|_{F_{tX}} = o_p(1/\sqrt{n h^{d_T}})$ in Assumption~\ref{A1st}(b)(c).

Define $\Delta_{i\ell} = \hat \gamma_{i\ell} - \bar \gamma_i - \mathbb{E}\big[\hat \gamma_{i\ell} - \bar \gamma_i |Z_\ell^c\big]$.
By construction and independence of $Z_\ell^c$ and $Z_i$ for $i\in I_\ell$, $\mathbb{E}[\Delta_{i\ell}|Z_\ell^c] = 0$ and $\mathbb{E}[\Delta_{i\ell}\Delta_{j\ell}|Z_\ell^c] = 0$ for $i,j \in I_\ell$.
By Assumptions~\ref{ACIACS}(b) and \ref{A1st}(a),
$h^{d_T}\mathbb{E}[\Delta_{i\ell}^2|Z_\ell^c] = O_p( h^{d_T} \int_\mathcal{X} (\hat \gamma_{\ell}(t,x) - \bar \gamma(t,x))^2 f_X(x)dx) = o_p(1)$.
$\mathbb{E}\big[\big(\sqrt{h^{d_T}/n} \sum_{i\in I_\ell} \Delta_{i\ell}\big)^2 \big|Z_\ell^c\big]
= 
(h^{d_T}/n) \sum_{i\in I_\ell} \mathbb{E}\big[ \Delta_{i\ell}^2 \big|Z_\ell^c\big]
=
O_p\big(h^{d_T} \int_\mathcal{X} (\hat \gamma_{\ell}(t,x) - \bar \gamma(t,x))^2 f_X(x)dx\big)  
=
o_p(1)$.
The conditional Markov's inequality implies that $\sqrt{h^{d_T}/n} \sum_{i\in I_\ell}  \Delta_{i\ell} =o_p(1)$.

The analogous results hold for $\Delta_{i\ell} = K_h(T_i - t) \bar \lambda_i (\bar \gamma_i - \hat \gamma_{i\ell}) - \mathbb{E}\left[K_h(T_i - t) \bar \lambda_i (\bar \gamma_i - \hat \gamma_{i\ell})|Z_\ell^c\right]$  in (\ref{ER11}) and 
$\Delta_{i\ell} = K_h(T_i - t) (\hat \lambda_{i\ell} - \bar \lambda_i) (Y_i - \bar \gamma_i) - \mathbb{E}\big[K_h(T_i - t) (\hat \lambda_{i\ell} - \bar \lambda_i) (Y_i - \bar \gamma_i)\big|Z_\ell^c\big]$ in (\ref{ER12}).
In particular for (\ref{ER12}), a standard algebra using change of variables, a Taylor expansion, the mean value theorem, and Assumption~\ref{ACIACS} yields that 
{\small 
\begin{align}
&h^{d_T}\mathbb{E}[\Delta_{i\ell}^2|Z_\ell^c] 
\notag\\&\leq h^{d_T} \mathbb{E}\left[
K_h(T_i - t)^2 (\hat \lambda_{i\ell} - \bar{\lambda}_i)^2 (Y_i - \bar{\gamma}_i)^2\Big|Z_\ell^c 
\right]
\notag\\
&= h^{d_T} \int_\mathcal{X}
\int_{\mathcal{T}_0} K_h(s-t)^2 \left(\hat\lambda_\ell(t, x) - \bar{\lambda}(t,x)\right)^2
\mathbb{E}\left[\left(Y - \bar{\gamma}(t,x)\right)^2\big|T=s, X=x\right]
f_{TX}(s,x) dsdx
\notag\\
&=\int_\mathcal{X} \int_{\mathcal{R}^{d_T}} \Pi_{j=1}^{d_T}k(u_j)^{2} 
 \left(\hat\lambda_\ell(t, x) - \bar{\lambda}(t,x)\right)^2\mathbb{E}\left[\left(Y - \bar{\gamma}(t,x)\right)^2\big|T=t+uh, X=x\right]
\notag \\
&\ \ \ \times f_{TX}(t+uh,x) dudx
\notag\\
&=\int_\mathcal{X} \int_{\mathcal{R}^{d_T}} 
\bigg( \mathbb{E}\left[\left(Y - \bar{\gamma}(t,x)\right)^2\big|T=t, X=x\right]
+ \sum_{j=1}^{d_T}u_jh \frac{\partial}{\partial t_j} 
\mathbb{E}\left[\left(Y - \bar{\gamma}(t,x)\right)^2\big|T=t, X=x\right]
\big|_{t=\bar t}\bigg)
\notag\\ &\ \ \ \times 
\left( f_{TX}(t,x) +  \sum_{j=1}^{d_T}u_j h \frac{\partial}{\partial t_j}f_{TX}(t,x)\big|_{t=\grave t} \right)\Pi_{j=1}^{d_T}k(u_j)^{2} du 
\left(\hat\lambda_\ell(t, x) - \bar{\lambda}(t,x)\right)^2 dx
\notag\\
&= O_p\Big(\int_\mathcal{X} \big( \hat \lambda_{\ell}(t,x) - \bar{\lambda}(t,x) \big)^2 f_{TX}(t,x) dx\Big), 
\label{EThm1V}
\end{align}
}
where $\bar t$ and $\grave t$ are between $t$ and $t+uh$.
So $h^{d_T}\mathbb{E}[\Delta_{i\ell}^2|Z_\ell^c] = o_p(1)$ by Assumption~\ref{A1st}(a).
The conditional Markov's inequality implies that $\sqrt{h^{d_T}/n} \sum_{i\in I_\ell}  \Delta_{i\ell} =o_p(1)$.
Then (\ref{ER11})$=o_p(1)$ and (\ref{ER12})$=o_p(1)$.


For (\ref{ER2}),
\begin{align}
&\mathbb{E}\bigg[ \Big| \sqrt{h^{d_T}/n_\ell} \sum_{i \in I_\ell} K_h(T_i - t)(\hat \lambda_{i\ell} - \bar \lambda_i)(\bar \gamma_i - \hat \gamma_{i\ell})  \Big|  \bigg|Z_\ell^c\bigg] \notag\\
&\leq
\sqrt{n_\ell h^{d_T}} \int_\mathcal{X} \int_{\mathcal{T}_0} \Big| (\hat \lambda_{\ell}(t,x) - \bar \lambda(t,x))(\bar \gamma(t,x) - \hat \gamma_{\ell}(t,x)) \Big| K_h(s - t) f_{TX}(s,x)ds dx \notag\\
&\leq
\sqrt{n_\ell h^{d_T}} 
\Big(
\int_\mathcal{X} \int_{\mathcal{T}_0} (\hat \lambda_{\ell}(t,x) - \bar \lambda(t,x))^2 K_h(s - t) f_{TX}(s,x)ds dx \Big)^{1/2}
\notag \\
&\ \ \ \times \Big(
\int_\mathcal{X} \int_{\mathcal{T}_0} (\bar \gamma(t,x) - \hat \gamma_{\ell}(t,x))^2 K_h(s - t) f_{TX}(s,x)ds dx \Big)^{1/2}
\notag\\
&= \sqrt{n_\ell h^{d_T}}
\Big(\int_\mathcal{X} (\hat\lambda_{\ell}(t,x) - \bar \lambda(t,x))^2f_{TX}(t, x) dx \Big)^{1/2}
\Big(\int_\mathcal{X} (\hat\gamma_{\ell}(t,x) - \bar \gamma(t,x))^2f_{TX}(t, x) dx \Big)^{1/2} 
\notag\\
&\ \ \ + o_p(h^2) \notag
\\
&= o_p(1) 
\label{ER2pf}
\end{align}
by Cauchy-Schwartz inequality and Assumption~\ref{A1st}(b), and (\ref{algebra}).
So (\ref{ER2})$=o_p(1)$ follows by the conditional Markov's and triangle inequalities.

By the triangle inequality, we obtain the asymptotically linear representation\\ 
$\sqrt{nh^{d_T}} n^{-1}\sum_{i=1}^n \big( \hat \psi(Z_i, \beta_t, \hat \gamma, \hat \lambda) - \psi(Z_i, \beta_t, \bar\gamma, \bar\lambda) \big) = o_p(1)$.


For $\mathsf{B}_t$, first compute 
$ \mathbb{E}\Big[ 
\frac{K_h(T-t)}{\bar f(t,X)} \big(Y - \bar{\gamma}(t, X)\big)
\Big]
= 
\mathbb{E}\Big[ 
\frac{1}{\bar f(t,X)} \mathbb{E}\Big[K_h(T-t)\big({\gamma}(T,X) - \bar{\gamma}(t, X)\big)
|X\big]\Big]$.
A standard algebra for kernel as (\ref{algebra}) yields
\begin{align*}
&\mathbb{E}\left[K_h(T-t)\left({\gamma}(T,x) - \bar{\gamma}(t, x)\right)
|X=x\right]
\\
&=
\int_{\mathcal{T}_0} K_h(s-t)\left(\gamma(s,x) - \bar{\gamma}(t, x)\right) f_{T|X}(s|x) ds  \\
&= 
\int_{\mathcal{R}^{d_T}} k(u)\left(\gamma(t +uh,x) - \bar{\gamma}(t, x)\right) f_{T|X}(t+uh|x) du \\
&= 
\int_{\mathcal{R}^{d_T}}   \left( \gamma(t,x) - \bar\gamma(t,x) +\sum_{j=1}^{d_T} h u_j \partial_{t_j} {\gamma}(t, x) +  \frac{h^2}{2} u_j^2 \partial^2_{t_j} {\gamma}(t, x)
+  \frac{h^2}{2}\sum_{l=1, l\neq j}^{d_T} u_j u_l \partial_{t_j}\partial_{t_l} {\gamma}(t,x)
\right) 
\\
&\ \ \ \times
\left( f_{T|X}(t|x) +  \sum_{j=1}^{d_T} h u_j  \partial_{t_j} f_{T|X}(t|x) 
+ \frac{h^2}{2} u_j^2 \partial^2_{t_j} f_{T|X}(t|x)
+  \frac{h^2}{2}\sum_{l=1, l\neq j}^{d_T} u_j u_l \partial_{t_j}\partial_{t_l} f_{T|X}(t|x)
\right) 
\\
&\ \ \ \times
k(u_1)\cdots k(u_{d_T})du_1\cdots du_{d_T} + O(h^3) \\
&= 
\left(\gamma(t,x) - \bar\gamma(t,x)\right) f_{T|X}(t|x)
+ 
h^2  \sum_{j=1}^{d_T}\bigg(
\partial_{t_j} {\gamma}(t,X)\partial_{t_j} f_{T|X}(t|x)
+ \frac{1}{2}\partial_{t_j}^2 {\gamma}(t,x) f_{T|X}(t|x)
\\
&\ \ \ 
+
\left( \gamma(t,x) - \bar\gamma(t,x) \right)
\frac{1}{2}\partial^2_{t_j} {\gamma}(t, x)
\bigg)\int_{-\infty}^\infty u^2k(u) du
+ O(h^3)
\end{align*}
uniformly over $x \in \mathcal{X}$.
Thus 
\begin{align}
&\mathbb{E}\left[\psi(Z_i, \beta_t, \bar\gamma, \bar\lambda)\right]
\notag
\\
&= \mathbb{E}\left[ 
\frac{1}{\bar f(t,X)} \mathbb{E}\left[K_h(T-t)\left({\gamma}(T,X) - \bar{\gamma}(t, X)\right)
|X\right] + \bar\gamma(t, X) - \beta_t\right]
\notag \\
&= 
\mathbb{E}\left[
\left(\gamma(t,X) - \bar\gamma(t,X)\right) \frac{f_{T|X}(t|X)}{\bar f(t,X)}
+ \bar\gamma(t, X) - \beta_t
\right]
\label{AEDR}
\\
&\ \ \ + h^2\sum_{j=1}^{d_T} \mathbb{E}\bigg[
\partial_{t_j} \bar{\gamma}(t,X)\frac{\partial_{t_j} f_{T|X}(t|X)}{\bar f(t,X)}
+ \partial_{t_j}^2 \bar{\gamma}(t,X) \frac{f_{T|X}(t|X)}{2 \bar f(t,X)}
\notag
\\
&\ \ \ + 
\left( \gamma(t,x) - \bar\gamma(t,x) \right)
\frac{\partial^2_{t_j} {\gamma}(t, X)}{2\bar f(t,X)}
\bigg]  \int_{-\infty}^\infty u^2k(u) du  + O(h^3).
\notag
\end{align}
Due to the doubly robust property, (\ref{AEDR}) is zero.
Specifically, when $\bar f = f_{T|X}$ and $\bar \gamma \neq \gamma$, (\ref{AEDR}) becomes 
$\mathbb{E}\left[\left(\gamma(t,X) - \bar\gamma(t,X)\right) + \bar\gamma(t, X) - \beta_t\right] = 0$.
When $\bar f \neq f_{T|X}$ and $\bar \gamma = \gamma$, (\ref{AEDR}) becomes 
$\mathbb{E}\left[
\gamma(t, X) - \beta_t
\right] = 0$.
We then obtain $\mathsf{B}_t$ and $\lim_{h\rightarrow 0}\mathbb{E}[\psi(Z_i, \beta_t, \bar \gamma_i, \bar \lambda_i) ] = 0$.

The asymptotic variance is determined by $h^{d_T}\mathbb{E}\left[ \left( (Y-\bar{\gamma}(t, X))K_h(T_i - t)/\bar f(t,X) \right)^2\right]$.
A standard algebra for kernel as (\ref{EThm1V}) yields $\mathsf{V}_t$. 


Asymptotic normality follows from the Lyapunov central limit theorem with the third absolute moment.
Specifically,  
$\mathbb{E}\big[\big|\sqrt{nh^{d_T}}n^{-1}\psi(Z_i, \beta_t, \bar\gamma,\bar \lambda)\big|^3\big]
= O\big((n^{-1}h^{d_T})^{3/2}\mathbb{E}\big[
K_h(T-t)^3 |Y-\bar{\gamma}(t,X)|^3 \bar f(t,X)^{-3}
\big]
\big)
= O((n^3h^{d_T})^{-1/2})$, by the same arguments as (\ref{EThm1V}) 
under the condition that $\mathbb{E}[|Y-\bar{\gamma}(t,X)|^3|T=t, X]$ and its first derivative w.r.t. $t$ are bounded uniformly in $x\in\mathcal{X}$.
Let $s_n^2 \equiv \sum_{i=1}^n var\big( \sqrt{nh^{d_T}}n^{-1}\psi(Z_i, \beta_t, \bar\gamma, \bar\lambda)\big)
 = h^{d_T} var(\psi) = \mathsf{V}_t + o(1)$.
Thus the Lyapunov condition holds: $\sum_{i=1}^n\mathbb{E}\big[\big|\sqrt{nh^{d_T}}n^{-1}\psi(Z_i, \beta_t, \bar\gamma, \bar\lambda)\big|^3\big]/s_n^3 = O((nh^{d_T})^{-1/2}) = o(1)$.
\hfill $\square$
\\ \\
\textbf{Proof of Theorem~\ref{Tbw}:}\
By Theorem~\ref{TIF}, the asymptotic MSE is $h^4\mathsf{B}_t^2 + \mathsf{V}_t/(nh^{d_T})$.
Solving the first-order condition yields the optimal bandwidth $h_t^\ast$.  

Given the consistency of $\hat{\mathsf{V}}_t$ and $\hat{\mathsf{B}}_t$, the continuous mapping theorem implies 
$\hat h_t/h_t^\ast - 1 - o_p(1)$.
Below we show the consistency of $\hat{\mathsf{V}}_t$ and $\hat{\mathsf{B}}_t$.
\\[3pt] 
\textbf{Consistency of $\hat{\mathsf{V}}_t$:}\
Let $\hat{\mathsf{V}}_{t} = L^{-1}\sum_{\ell=1}^L\hat{\mathsf{V}}_{t\ell}$, where $\hat{\mathsf{V}}_{t\ell} \equiv h^{d_T} n_\ell^{-1} \sum_{i \in I_\ell} \hat \psi_{i\ell}^2$.
It suffices to show that $\hat{\mathsf{V}}_{t\ell}$ is consistent for $\mathsf{V}_{t}$ as $n_\ell\rightarrow\infty$, for $\ell=1,..., L$.
Toward that end, 
we show that 
(I)~$h^{d_T} n_\ell^{-1} \sum_{i \in I_\ell} \psi_{i}^2 - {\mathsf{V}}_{t} = o_p(1)$,
where $\psi_{i} \equiv K_h(T_i-t) (Y_i - \bar \gamma(t,X_i) )/\bar f(t,X_i) +  \bar \gamma(t,X_i) - \beta_t$,
(II)~$h^{d_T} n_\ell^{-1} \sum_{i \in I_\ell} \mathbb{E}[\hat \psi_{i\ell}^2 - \psi_{i}^2|Z_\ell^c] = o_p(1)$,
and (III)~$h^{d_T} n_\ell^{-1} \sum_{i \in I_\ell} \Delta_{i\ell} = o_p(1)$, 
where $\Delta_{i\ell} \equiv \hat \psi_{i\ell}^2 - \psi_{i}^2 - \mathbb{E}[\hat \psi_{i\ell}^2 - \psi_{i}^2|Z_\ell^c]$.


For (I), as computed in the proof of Theorem~\ref{TIF},
$h^{d_T}\mathbb{E}[\psi_{i}^2] =  \mathsf{V}_t + o(1)$.
By similar arguments in (\ref{EThm1V}) and Assumption~\ref{AOptBW}(c)(iv),
$\mathbb{E}[\psi_{i}^4] = h^{-3d_T} \mathbb{E}[\mathbb{E}[(Y -  \bar \gamma(t,X))^4|T=t, X]/ \bar f(t,X)^3] (\int_{-\infty}^{\infty} k(u)^4 du)^{d_T}+ o(h^{-3d_T}) =  O(h^{-3d_T})$.
By the Markov's inequality, for any $\epsilon > 0$, $P(|h^{d_T} n_\ell^{-1} \sum_{i \in I_\ell} \psi_{i}^2 - {\mathsf{V}}_{t} | > \epsilon) \leq \epsilon^{-2} 
h^{2d_T} n_\ell^{-1}var(\psi_{i}^2)
= O(h^{2d_T} n_\ell^{-1}\mathbb{E}[\psi_{i}^4]) = O(n_\ell^{-1} h^{-d_T}) = o(1)$.


For (II), first compute 
\begin{align*}
h^{d_T}\mathbb{E}[\hat\psi_{i\ell}^2|Z_\ell^c] = \mathbb{E}[\mathbb{E}[(Y - \hat \gamma_\ell(t,X) )^2|T=t, X, Z_\ell^c]f_{T|X}(t|X)/\hat f_\ell(t|X)^2|Z_\ell^c] R_k^{d_T}+ o_p(1).
\end{align*}
It suffices to show that
$\mathbb{E}[(\mathbb{E}[(Y - \hat \gamma_\ell(t,X))^2|T=t, X, Z_\ell^c] f_{T|X}(t|X)/\hat f_\ell(t|X)^2|Z_\ell^c]
-
\mathbb{E}[\mathbb{E}\big[ (Y - \bar\gamma(t,X))^2|T=t, X\big]f_{T|X}(t|X)/ \bar f(t,X)^2]$ 
\begin{align*}
&= \mathbb{E}\left[
\mathbb{E}\left[
\frac{(Y - \hat \gamma_\ell(t,X))^2}{\hat f_\ell(t|X)^2}
-
\frac{(Y - \bar\gamma(t,X))^2}{\bar f(t,X)^2}
\bigg|T=t, X, Z_\ell^c
\right]
f_{T|X}(t|X)\bigg|Z_\ell^c
\right]
= o_p(1).
\end{align*}

To simplify notations, let $\bar\lambda \equiv 1/\bar f(t,X)$, $\hat\lambda \equiv 1/\hat f_\ell(t,X)$, $\bar\gamma \equiv \bar \gamma(t,X)$, and $\hat\gamma\equiv\hat\gamma(t,X)$.
Decompose 
\begin{align}
\frac{Y - \hat \gamma_\ell(t,X)}{\hat f_\ell(t|X)} \equiv 
(Y-\hat \gamma)\hat\lambda = (Y-\bar \gamma)\bar \lambda 
 + (\bar \gamma-\hat\gamma)\bar \lambda + (Y-\bar \gamma)(\hat\lambda-\bar \lambda) 
+ (\bar \gamma-\hat\gamma)(\hat\lambda-\bar \lambda).
\label{OptBWde}
\end{align}
Therefore 
\begin{align*}
&\mathbb{E}\left[\mathbb{E}\left[
\frac{(Y - \hat \gamma_\ell(t,X))^2}{\hat f_\ell(t|X)^2}
- \frac{(Y -\bar \gamma(t,X))^2}{\bar f(t,X)^2}
\bigg|T=t, X, Z_\ell^c\right] f_{T|X}(t|X) \bigg|Z_\ell^c\right]\\
&= 
\mathbb{E}\bigg[\mathbb{E}\bigg[
\left( (Y-\bar \gamma)\bar \lambda 
 + (\bar \gamma-\hat\gamma)\bar \lambda + (Y-\bar \gamma)(\hat\lambda-\bar \lambda) 
+ (\bar \gamma-\hat\gamma)(\hat\lambda-\bar \lambda)
\right)^2
\\
&\ \ \ -  \left((Y-\bar \gamma)\bar \lambda\right)^2 
\bigg|T=t, X,  Z_\ell^c\bigg] f_{T|X}(t|X)\bigg|Z_\ell^c\bigg]
\end{align*}
is $ o_p(1)$ by Assumption~\ref{A1st} and Assumption~\ref{AOptBW}(a).


For (III),
$\mathbb{E}[\Delta_{i\ell}|Z_\ell^c] = 0$.
By (\ref{OptBWde}) and Assumption~\ref{AOptBW}(b)(c), the similar arguments as for $\mathbb{E}[\psi_{i}^4]$ in (I) yields
$\mathbb{E}[\Delta_{i\ell}^2|Z_\ell^c] = O_p\bigg(\mathbb{E}\bigg[K_h(T-t)^4\bigg(
\frac{(Y - \hat \gamma_\ell(t,X))^2}{\hat f_\ell(t|X)^2}
-
\frac{(Y - \bar \gamma(t,X))^2}{\bar f(t,X)^2}
\bigg)^2
\bigg|Z_\ell^c
\bigg]\bigg) 
= 
O_p(h^{-3d_T})$.
Then
$var(
h^{d_T} n_\ell^{-1} \sum_{i \in I_\ell} \Delta_{i\ell}|Z_\ell^c
) = O_p(h^{2d_T} n_\ell^{-1} h^{-3d_T}) = O_p(n_\ell^{-1} h^{-d_T}) = o_p(1)$.
The result follows by the conditional Markov's inequality.
\\[3pt]
\textbf{Consistency of $\hat{\mathsf{B}}_t$:}\
Theorem~\ref{TIF} provides $\mathbb{E}[\hat\beta_{t,ab}] = b^2 a^2 \mathsf{B}_t + o(b^2)$ and $\mathbb{E}[\hat{\mathsf{B}}_t] = \mathsf{B}_t + o(1)$.
Compute
\begin{align*}
&\mathbb{E}\big[\hat\beta_{t,b} \hat\beta_{t,ab}\big] \\
&= O\Big(n^{-1}
\mathbb{E}\left[
\int_{\mathcal{T}_0}K_b(s-t)K_{ab}(s-t) \mathbb{E}\Big[(Y - \hat\gamma_\ell(t, X))^2\big|T=s,X,Z_\ell^c\right] f_{T|X}(s|X) ds 
\\
&\ \ \ /\hat f_\ell(t|X)^2 \Big|Z_\ell^c\Big]
\Big)
\\
&= O\bigg(
n^{-1}b^{-d_T} a^{-d_T} \int_{\mathcal{R}^{d_T}} \Pi_{j=1}^{d_T} k(u_j) k\left(\frac{u_j}{a}\right) du \mathbb{E}\bigg[
 \mathbb{E}\big[(Y - \hat\gamma_\ell(t, X))^2\big|T=t,X,Z_\ell^c\big]
 \\
 &\ \ \ \times \frac{f_{T|X}(t|X)}{\hat f_\ell(t|X)^2}
\bigg|Z_\ell^c 
\bigg]
\bigg)
\\
&= O\big((nb^{d_T})^{-1}\big)
\end{align*}
by the same arguments as (\ref{algebra}).
So $cov(\hat\beta_{t,b}, \hat\beta_{t,ab})  = O(var(\hat\beta_{t,ab})) = O\big((nb^{d_T})^{-1}\big)$. 
It follows that $var(\hat{\mathsf{B}}_t) = b^{-4}(1-a^2)^{-2}\big( var(\hat\beta_{t,b}) + var(\hat\beta_{t,ab}) - 2cov(\hat\beta_{t,b}, \hat\beta_{t,ab}) \big) = O(b^{-4}(nb^{d_T})^{-1})$.

By the Markov's inequality and Assumption~\ref{AOptBW}(d), $P(|\hat{\mathsf{B}}_t - \mathbb{E}[\hat{\mathsf{B}}_t]| > \epsilon) \leq var(\hat{\mathsf{B}}_t)/\epsilon^2 = O((nb^{d_T+4})^{-1}) = o(1)$.
So $\hat{\mathsf{B}}_t - \mathsf{B}_t = o_p(1)$.
%
%
\hfill $\square$
\\ \\
\textbf{Proof of Theorem~\ref{TME}:}\
We decompose $\hat \theta_t - \theta_{t} = (\hat \theta_t - \theta_{t \eta}) + (\theta_{t \eta}  - \theta_{t})$,
where $\theta_{t \eta} \equiv ( \beta_{t^+} -  \beta_{t^-})/\eta$.
By a Taylor expansion of $\beta_{t^+}$ around $\beta_t$ and the mean value theorem, 
$\beta_{t^+} = \beta_t + \partial_{t_1}\beta_t \eta/2 + \partial_{t_1}^2 \beta_{\bar t} \eta^2/4$ for some $\bar t$ between $t$ and $t^+$.
Take the same expansion of  $\beta_{t^-}$ around $\beta_t$.
As we assume $\partial_{t_1}^2 \beta_{t}$ is uniformly bounded, the second part $\theta_{t \eta}  - \theta_{t} = O(\eta)$.
We assume that $\sqrt{nh^{d_T + 2}}(\theta_{t\eta} - \theta_{t}) = 
O(\sqrt{nh^{d_T + 2}} \eta) = o(1)$ under $\eta/h \rightarrow 0$.

Let $\hat \beta_t -\beta_t = n^{-1}\sum_{\ell=1}^L \sum_{i \in I_\ell} \hat \psi_{ti\ell}  = n^{-1}\sum_{\ell=1}^L \sum_{i \in I_\ell} \big(\psi_{ti} + R_{ti\ell}\big)$, where 
$\psi_{ti} \equiv \psi(Z_i, \beta_t, \bar\gamma_i, \bar \lambda_i)$, $\hat \psi_{ti\ell} \equiv \psi(Z_i, \beta_t, \hat \gamma_{i\ell}, \hat \lambda_{i\ell})$, and the remainder terms in $R_{ti\ell}$ are defined in (R1-1), (R1-2), (R1-DR), and (R2).  
Thus $\hat \theta_t - \theta_{t \eta} = \eta^{-1} n^{-1}\sum_{\ell=1}^L \sum_{i \in I_\ell} \big(\psi_{t^+i} - \psi_{t^-i} + R_{t^+i\ell} - R_{t^-i\ell}\big)$.
The goal is to show that the leading term of the variance of $\sqrt{nh^{d_T + 2}}\eta^{-1} n^{-1}\sum_{i=1}^n\big( \psi_{t^+i} - \psi_{t^-i}\big)$ is 
$\mathsf{V}_t^\theta$.

By Taylor expansions of  $\psi_{t^+i}$ and $\psi_{t^-i}$ around $\psi_{t}$ respectively and the mean value theorem, 
$\eta^{-1} n^{-1}\sum_{i=1}^n \big( \psi_{t^+i} - \psi_{t^-i}\big)
=
n^{-1}\sum_{i=1}^n \partial_{t_1}\psi_{ti} +  \eta^2 n^{-1}\sum_{i=1}^n  (\partial_{t_1}^3\psi_{\tilde ti}+ \partial_{t_1}^3\psi_{\check t i})/(2^33!)$,   
where $\tilde t \in (t, t+\eta/2)$ and $\check t \in (t-\eta/2, t)$.

%
We next show that $var\big(n^{-1}\sum_{i=1}^n \partial_{t_1}\psi_{ti}\big) 
= n^{-1}\mathbb{E}\big[(\partial_{t_1}\psi_{ti})^2\big] - n^{-1} \big( \mathbb{E}[\partial_{t_1}\psi_{ti}]\big)^2
= O\big(n^{-1}h^{-(d_T + 2)} + n^{-1}h^4 \big)
= O\big(n^{-1}h^{-(d_T + 2)} \big)$.
We first analyze 
$\mathbb{E}[\partial_{t_1}\psi_{ti}]  =
\mathbb{E}\Big[\int_{\mathcal{T}_0}  \Big\{ \partial_{t_1} K_h(s - t) \frac{\gamma(s,X)- \bar \gamma(t,X)}{\bar f(t,X)} + 
K_h(s - t)  \partial_{t_1}\Big( \frac{\gamma(s, X) -\bar  \gamma(t, X)}{\bar f(t,X)} \Big) \Big\} f_{T|X}(s|X) ds\Big]
+ \mathbb{E}[\partial_{t_1}\bar \gamma(t,X)] - \theta_t$.
We use $\partial K_h(T-t)/\partial t_1 = - \partial K_h(T-t)/\partial T_1$, integration by parts,
$K_h(T-t) = o(h^2)$ as $T$ approaching the boundary of $\mathcal{T}_0$, and the same arguments in (\ref{algebra}).
Denote $\bar f_{t|x} = \bar f(t,x)$ and $f_{t|x} = f_{T|X}(t|x)$.
Uniformly in $x\in\mathcal{X}$, 
{\small
\begin{align}
&\int_{\mathcal{T}_0}  \left\{ \partial_{t_1} K_h(s - t) \frac{\gamma(s, x)- \bar \gamma(t, x)}{\bar f_{t|x}} + 
K_h(s - t)  \partial_{t_1}\left( \frac{\gamma(s, x) -\bar  \gamma(t, x)}{\bar f_{t|x}} \right) \right\} f_{s|x} ds
\notag\\
=&\ 
\int_{\mathcal{T}_0}  \left\{ -\partial_{s_1} K_h(s - t) \frac{\gamma(s, x)- \bar \gamma(t, x)}{\bar f_{t|x}} 
-K_h(s - t)  \left( \frac{\partial_{t_1}\bar  \gamma(t, x)}{\bar f_{t|x}} 
+ \frac{\gamma(s,x) - \bar  \gamma(t, x)}{\bar f_{t|x}^2} \partial_{t_1} \bar f_{t|x} \right)
 \right\} f_{s|x} ds
\notag\\
=&\ 
\int_{\mathcal{T}_0}  K_h(s - t)\Bigg\{  
\frac{\partial_{s_1}\gamma(s, x)}{\bar f_{t|x}} f_{s|x} 
+
 \frac{\gamma(s, x) -\bar \gamma(t, x)}{\bar f_{t|x}} \partial_{s_1} f_{s|x}
\notag\\
&-  \frac{\partial_{t_1}\bar \gamma(t, x) }{\bar f_{t|x}}  f_{s|x}
-   \frac{\gamma(s, x) - \bar \gamma(t, x)}{\bar f^2_{t|x}} \partial_{t_1}f_{t|X} f_{s|x}\Bigg\} ds
+ o(h^2)
\notag\\
=&\ 
\int_{\mathcal{T}_0}  \frac{K_h(s - t)}{\bar f_{t|x}}\Bigg\{  
\left(\partial_{s_1} \gamma(s,x) - \partial_{t_1}\bar \gamma(t, x) \right) f_{s|x} 
+
\left(\gamma(s, x) -\bar \gamma(t, x)\right)\left( \partial_{s_1} f_{s|x} - f_{s|x} \frac{\partial_{t_1} \bar f_{t|x}}{\bar f_{t|x}}\right)
ds \notag \\
& + o(h^2)
\notag\\
=&\ 
 \int_{\mathcal{R}^{d_T}}  \bigg\{ 
 \bigg( f_{t|x} + \sum_{j=1}^{d_T} \bigg\{
 \partial_{t_j}f_{t|X} u_j h + \partial_{t_j}^2 f_{t|x} \frac{u_j^2h^2}{2}
  +\sum_{l=1, l\neq j}^{d_T}  \partial_{t_j}\partial_{t_l} f_{t|x} u_j u_l\frac{h^2}{2}\bigg\}
 \bigg)
 \notag\\
 &\times
 \bigg(\partial_{t_1} \gamma(t,x) - \partial_{t_1}\bar \gamma(t, x)
 \notag \\
&  + \sum_{j=1}^{d_T}\bigg\{ \partial_{t_j}\partial_{t_1} \gamma(t,x) u_j h + \partial_{t_j}^2\partial_{t_1} \gamma(t,x) \frac{u_j^2h^2}{2}
+ \sum_{l=1, l\neq j}^{d_T} \partial_{t_j}\partial_{t_l}\partial_{t_1} \gamma(t,x) u_j u_l\frac{h^2}{2} 
 \bigg\} \bigg)
\notag\\
&+ \bigg( \gamma(t, x) -\bar \gamma(t, x) + \sum_{j=1}^{d_T} \bigg\{ \partial_{t_j} \gamma(t, x)u_jh + \partial^2_{t_j} \gamma(t, X)\frac{u_j^2h^2}{2} 
+ \sum_{l=1, l\neq j}^{d_T}\partial_{t_j}\partial_{t_l} \gamma(t,x)u_j u_l \frac{h^2}{2} \bigg\}
\bigg)
\notag\\
&\times\bigg( \partial_{t_1} f_{t|x} + \sum_{j=1}^{d_T} \bigg\{ 
\partial_{t_j} \partial_{t_1} f_{t|x} u_j h 
+ \partial_{t_j}^2 \partial_{t_1} f_{t|x} \frac{u_j^2h^2}{2} 
+ \sum_{l=1, l\neq j}^{d_T}\partial_{t_j}\partial_{t_l} \partial_{t_1}f_{t|x} u_j u_l \frac{h^2}{2}\bigg\}
\notag\\
&-
\bigg( f_{t|x} + \sum_{j=1}^{d_T} \bigg\{
\partial_{t_j} f_{t|x} u_j h 
+ \partial_{t_j}^2 f_{t|x} \frac{u_j^2h^2}{2}
+ \sum_{l=1, l\neq j}^{d_T}\partial_{t_j}\partial_{t_l} f_{t|x} u_j u_l \frac{h^2}{2}
\bigg\}
\bigg)
 \frac{\partial_{t_1}\bar f_{t|x} }{\bar f_{t|x}}\bigg)\bigg\}\frac{ \Pi_{j=1}^{d_T}k(u_j)}{\bar f_{t|x}} 
 du\notag \\
& 
 + O(h^3) + o(h^2)
 \notag\\
 =&\ 
 h^2   
 \sum_{j=1}^{d_T}   
 \bigg\{
 \frac{1}{2}\partial_{t_j}^2 \partial_{t_1} \gamma(t, x) f_{t|x}+ \partial_{t_j} \partial_{t_1} \gamma(t, x) \partial_{t_j}f_{t|x}
 +  \partial_{t_j} \gamma(t, x) 
\bigg( \partial_{t_j} \partial_{t_1} f_{t|x} - \partial_{t_j} f_{t|x}
 \frac{\partial_{t_1}\bar f_{t|x} }{\bar f_{t|x}}\bigg)
\notag\\& 
+ \frac{1}{2} \left(\partial_{t_1} \gamma(t,x) - \partial_{t_1}\bar \gamma(t, x)\right)  \partial_{t_j}^2 f_{t|x}
+ \frac{1}{2}  \left(\gamma(t,x) - \bar \gamma(t, x)\right)\left( \partial_{t_j}^2 \partial_{t_1} f_{t|x} 
- \partial_{t_j}^2 f_{t|x}  \frac{\partial_{t_1}\bar f_{t|x} }{\bar f_{t|x}}\right)
\notag\\
&+
\frac{1}{2} \partial_{t_j}^2 \gamma(t,x)\bigg(\partial_{t_1} f_{t|x}-  \partial_{t_1}\bar f_{t|x}\frac{f_{t|x} }{\bar f_{t|x}}\bigg)
\bigg\} \frac{\int_{-\infty}^\infty u^2 k(u)du}{\bar f_{t|x}} 
\notag\\
 &+
\left\{ \left(\partial_{t_1} \gamma(t,x) - \partial_{t_1}\bar \gamma(t, x)\right)f_{t|x} 
+\left(\gamma(t,x) - \bar \gamma(t, x)\right)\left(\partial_{t_1}f_{t|x} - \partial_{t_1}\bar f_{t|x}\frac{f_{t|x}}{\bar f_{t|x}}\right)\right\}
 \frac{1}{\bar f_{t|x}} 
\label{DRbias1} \\
 & + o(h^2). \notag
\end{align}
}
Combining the term in (\ref{DRbias1}) with $\mathbb{E}[\partial_{t_1} \bar\gamma(t,X) - \theta_t]$, 
$\mathbb{E}\Big[
\Big\{ \Big(\partial_{t_1} \gamma(t,x) - \partial_{t_1}\bar \gamma(t, x)\Big)f_{t|x} 
+\big(\gamma(t,x) - \bar \gamma(t, x)\big)\Big(\partial_{t_1}f_{t|x} - \partial_{t_1}\bar f_{t|x}\frac{f_{t|x}}{\bar f_{t|x}}\Big)\Big\}
 \frac{1}{\bar f_{t|x}} + \partial_{t_1} \bar\gamma(t,X) - \theta_t\Big] = 0$ by Assumption~\ref{A1st}(c).
So we obtain $\mathbb{E}[\partial_{t_1}\psi_{ti}] = h^2 \mathsf{B}_t^\theta + o(h^2) = O(h^2)$.

In $\mathbb{E}[(\partial_{t_1} \psi_{ti})^2]$,
\begin{align*}
&\mathbb{E}\Big[
\big(
\partial_{t_1} K_h(T-t) (Y-\bar \gamma(t, X))/\bar f_{t|X}
\big)^2
\Big]
\\&= \mathbb{E}\left[
h^{-2d_T-2} k'((T_1-t_1)/h)^2 
\Pi_{j=2}^{d_T} k((T_j - t_j)/h)^2
\mathbb{E}\big[(Y-\bar \gamma(t, X))^2\big| T, X\big]/\bar f_{t|X}^2
\right]
\\&= \mathbb{E}\left[
h^{-d_T-2} \int_{\mathcal{R}^{d_T}} k'(u_1)^2 \Pi_{j=2}^{d_T}k(u_j)^2 \mathbb{E}\big[(Y-\bar \gamma(t, X))^2\big| T=t+uh, X\big]
\frac{f_{T|X}(t+uh|X)}{\bar f_{t|X}^2}du
\right]
\\&= O\left(h^{-d_T - 2} \mathsf{V}_t^\theta\right)
\end{align*}
by change of variables $u = (u_1, u_2,..., u_{d_T}) = (T-t)/h$ and 
the same arguments as (\ref{EThm1V}) using $\int_\mathcal{R} k'(u)^2 du < \infty$ implied by Assumption~\ref{Akernel}.
Thus the leading term of 
$var\big(n^{-1}\sum_{i=1}^n \partial_{t_1}\psi_{ti}\big)$  is $n^{-1}h^{-d_T-2} \mathsf{V}_t^\theta$.

By the same arguments and $\int_{-\infty}^{\infty} k'''(u)^2 du < \infty$, we can derive $var\big(\eta^2n^{-1}\sum_{i=1}^n \partial_{t_1}^3\psi_{ti}\big) 
= O\big(\eta^4n^{-1}\mathbb{E}\big[(\partial_{t_1}^3\psi_{ti})^2\big] \big)
= O\big(\eta^4 n^{-1}h^{-(d_T + 6)}\big)
= 
O\big((\eta/h)^4 \big) \times O\big(n^{-1}h^{-(d_T + 2)}\big)$.
Thus by $\eta/h\rightarrow 0$, the leading term of the variance of $\sqrt{nh^{d_T + 2}}\eta^{-1} n^{-1}\sum_{i=1}^n\big( \psi_{t^+i} - \psi_{t^-i}\big)$ is 
$\mathsf{V}_t^\theta$.

%


We follow the proof of Theorem~\ref{TIF} to control $\sqrt{nh^{d_T+2}} \eta^{-1} n^{-1} \sum_{\ell=1}^L\sum_{i\in I_\ell} \big(R_{t^+i\ell} - R_{t^-i\ell}\big) = o_p(1)$.
The conditions (a) and (b) give a crude bound such that $h\eta^{-1} \sqrt{h^{d_T}/n} \sum_{\ell=1}^L$\\
$ \sum_{i \in I_\ell} R_{ti\ell} = o_p(1)$ for $t = t^+, t^-$ respectively.


Asymptotic normality follows from the Lyapunov central limit theorem with the third absolute moment.
Specifically, a standard algebra as in the proof of Theorem~\ref{TIF} yields 
$\mathbb{E}\Big[\big|\sqrt{nh^{d_T+2}}n^{-1}
\partial_{t_1} K_h(T-t) (Y-\bar \gamma(t,X))/\bar f_{t|X}
\big|^3\Big]
= O((n^3h^{d_T})^{-1/2})$.
Let $s_n^2 \equiv \sum_{i=1}^n $\\$var\big( \sqrt{nh^{d_T+2}}n^{-1}
\partial_{t_1} K_h(T-t) (Y-\bar \gamma(t,X))/\bar f_{t|X}
\big)
 = h^{d_T+2} var(\partial_{t_1} K_h(T-t) (Y-\bar \gamma(t,X))$\\$/\bar f_{t|X}) = \mathsf{V}_t^\theta + o(1)$.
So the Lyapunov condition holds: $\sum_{i=1}^n
\mathbb{E}\big[\big|\sqrt{nh^{d_T+2}}n^{-1}
\partial_{t_1} K_h(T-t) (Y-\bar \gamma(t,X))/\bar f_{t|X}
\big|^3\big]/s_n^3 = O((nh^{d_T})^{-1/2}) = o(1)$.
We complete the proof.
\hfill $\square$
\\ \\
\textbf{Proof of Theorem~\ref{Tbwtheta}:}\
The proof follows the same arguments in Proof of Theorem~\ref{Tbw}, so we omit the repetition.\hfill $\square$
\\ \\
\textbf{Proof of Theorem~\ref{TDNN}:}\
Let $f_{\ast b} = f_{\ast t b} \equiv \arg\min_{f} \mathbb{E}[ \ell_{tb}(f, Z)]
= \arg\min_{f} \mathbb{E}\big[ \int_{\mathcal{T}_0} (\gamma(s,X) - f(X))^2$ $\mathsf{K}_b(s-t) f_{T|X}(s|X)ds\big]$
by the law of iterated expectations.
For any $x\in\mathcal{X}$, $t\in\mathcal{T}$, and $b > 0$, $\int_{\mathcal{T}_0} (\gamma(s,x) - f(x))^2 \mathsf{K}_b(s-t) f_{T|X}(s|x)ds$ is a quadratic function of $f(x)$.
Under the assumptions, the second-order condition $\int_{\mathcal{T}_0} \mathsf{K}_b(s-t) f_{T|X}(s|x) ds > 0$.
Thus $f_{\ast t b}(x) = \int_{\mathcal{T}_0} \gamma(s,x) \mathsf{K}_b(s-t) f_{T|X}(s|x)ds\big/\int_{\mathcal{T}_0} \mathsf{K}_b(s-t) f_{T|X}(s|x)ds$ is well-defined.  
We can show $f_{\ast tb} \in \mathcal{W}^{r,\infty}([-1,1]^{d_X}) \equiv \big\{f: \max_{\bm{\alpha}, |\bm{\alpha}| \leq r}{\rm ess}\sup_{x\in[-1,1]^{d_X}}$ $ |D^\alpha f(x)| \leq 1\big\}$, where $D^\alpha f$ is the weak derivative, as in Assumption 2 in FLM.  
Specifically, Assumption~\ref{ADNN} and the Leibniz integral rule imply that $|\mathtt{D}_x^\alpha f_{T|X}(t,x)|$ and $|\mathtt{D}_x^\alpha f_{\ast tb}(x)|$ are bounded uniformly over $x\in\mathcal{X}$, $t\in\mathcal{T}$, and $b > 0$.
Note that $f_{\ast tb}\notin\mathcal{F}_{DNN}$.

The smoothness condition Assumption~\ref{ADNN}, with an unknown constant $c$,  
implies that $\gamma(t,\cdot)$ lies in the H\"{o}lder ball $\mathcal{W}^{r,\infty}([-1,1]^{d_X}) \equiv \Big\{f: \max_{\bm{\alpha}, |\bm{\alpha}| \leq r}{\rm ess}\sup_{x\in[-1,1]^{d_X}} |D^\alpha f(x)|$\\$ \leq 1\Big\}$
for any $t\in\mathcal{T}$, 
where $D^\alpha f$ is the weak derivative \citep{Yarotsky}, as in Assumption 2 in FLM.

Let $f_n = f_{ntb} \equiv  \arg\min_{f\in\mathcal{F}_{DNN}, \|f\|_\infty \leq 2M} \|f-f_{\ast b}\|_\infty$,
and $\epsilon_{n}  = \epsilon_{ntb} \equiv  \epsilon_{DNN} \equiv  \|f_n - f_{\ast b}\|_\infty$.
Since we show the pointwise results for $t\in\mathcal{T}$,  
to follow the notations in FLM for easy reference, we suppress the subscript $t$ (and $b$) in the proof when there is no confusion.

Let the bounded kernel function $\mathsf{k}() < \bar{\mathsf{k}}$ for some constant $\bar{\mathsf{k}} > 0$.

We modify equation (2.1) in FLM to the following:
\begin{align}
&\left| \ell_{tb}(f, Z) - \ell_{tb}(g, Z) \right| \leq 
\mathsf{K}_b\left(T-t\right) M \left| f(X) - g(X) \right|
\leq \frac{\bar{\mathsf{k}}^{d_T}}{b^{d_T}} M \left| f(X) - g(X) \right|, 
\label{FLM2.1-1} \tag{2.1-1}\\
&2\left(\mathbb{E}[\ell_{tb}(f, Z)] - \mathbb{E}[\ell_{tb}(f_{\ast b}, Z)] \right) = \|f-\gamma\|_{F_{tX}}^2 + O(b^2).
\label{FLM2.1-2}
\tag{2.1-2}
\end{align}
Lemma 8 in FLM and the bounded kernel $\mathsf{k}()$ imply the Lipschitz condition (\ref{FLM2.1-1}).
The key of our modification is the condition (\ref{FLM2.1-2}) that replaces $c_1\mathbb{E}[(f-f_\ast)^2] \leq \mathbb{E}[\ell(f,Z)] - \mathbb{E}[\ell(f_\ast, Z)] \leq  c_2 \|\hat f-f_\ast\|^2_{L_2(X)}$ in FLM's (2.1).
We prove (\ref{FLM2.1-2}) holds uniformly in uniformly bounded $f$ at the end of this proof.
In the proof of Theorem 1 in FLM, 
the main decomposition in equation~(A.1) starts with the inequality in their equation~(2.1):
$c_1 \|\hat f-f_\ast\|^2_{L_2(X)}  \leq \mathbb{E}[\ell(\hat f,Z)] - \mathbb{E}[\ell(f_\ast, Z)]$.
This is the only place where this inequality is used.  
We modify it to (\ref{FLM2.1-2}) that implies 
$\|\hat f_b-\gamma\|_{F_{tX}}^2 = O_p\left(\mathbb{E}[\ell_{tb}(\hat f_b, Z)] - \mathbb{E}[\ell_{tb}(f_{\ast b}, Z)] \right) + O_p(b^2)$. 
Thus we can bound $\|\hat f_b-\gamma\|_{F_{tX}}^2$ 
using the bound of $\mathbb{E}[\ell_{tb}(\hat f_b, Z)] - \mathbb{E}[\ell_{tb}(f_{\ast b}, Z)]$.
We modify (A.1) in FLM and bound
\begin{align}
&\mathbb{E}[\ell_{tb}(\hat f_b, Z)] - \mathbb{E}[\ell_{tb}(f_{\ast b}, Z)]  \notag
\\
&\leq 
(\mathbb{E} - \mathbb{E}_n)[\ell_{tb}(\hat f_b, Z)] - \mathbb{E}[\ell_{tb}(f_{\ast b}, Z)]
+ 
\mathbb{E}_n[\ell_{tb}(f_n, Z) - \ell_{tb}(f_{\ast b}, Z)].
\label{FLMA1} 
\end{align}

To bound the second bias term in (\ref{FLMA1}), FLM only use the inequality in (2.1) $\mathbb{E}[\ell(f,Z)] - \mathbb{E}[\ell(f_\ast, Z)] \leq  c_2 \|\hat f-f_\ast\|^2_{L_2(X)}$ in the second inequality in their (A.2). 
We use (\ref{FLM2.1-2}) that implies $\mathbb{E}[\ell_{tb}(f_n, Z) - \ell_{tb}(f_{\ast b}, Z)] = O_p(\|f_n-\gamma\|_{F_{tX}}^2) + O(b^2) = O_p(\epsilon_n^2 + b^2)$.

Next we modify the proof of Theorem~1 in FLM by replacing all $(\ell, \hat f, f_\ast)$ with $(\ell_{tb}, \hat f_b, f_{\ast b})$ for any $t\in \mathcal{T}$ and $b$.
We only point out the key modifications of their proof in the following.

By (\ref{FLM2.1-1}), $var[\ell_{tb}(f_n, z) - \ell_{tb}(f_{\ast b}, z)] = O_p\left( M^2\|f_n -f_{\ast b}\|_\infty^2 /b^{d_T}\right)$.
Thus (A.2) in FLM from Bernstein's inequality is modified to $c_2 \epsilon_n^2 + \epsilon_n \sqrt{\frac{2C_\ell^2 \tilde \gamma}{nb^{d_T}}} + \frac{7 C_\ell M\tilde \gamma}{nb^{d_T}}$.

Similarly the last equation on page 201 in FLM is modified to
$\mathbb{V}[g] = \mathbb{E}\Big[\big|\ell_{tb}(f, z) - \ell_{tb}(f_{\ast b}, z)\big|^2 \Big] \leq C_\ell^2\mathbb{E}[(f-f_{\ast b})^2\mathsf{K}_b(T-t)^2] = O\left(C_\ell^2 M r_0^2/b^{d_T}\right)$.
Thus the statement for (A.7) in FLM is modified to the following:
we find that $(\mathbb{E} - \mathbb{E}_n)[\ell_{tb}(\hat f,z) - \ell_{tb}(f_{\ast b}, z)] = O_p\Big( 6 \mathbb{E}_\eta R_n \mathcal{G} + \sqrt{\frac{2C_\ell^2 r_0^2 \tilde \gamma}{nb^{d_T}}} + \frac{23\cdot 3MC_\ell}{3} \frac{\tilde \gamma}{nb^{d_T}}\Big)$.


On page 202 of FLM, the bound of $\mathbb{E}_n R_n\mathcal{G}$ is multiplied by $b^{-d_T}$.
This is because in Lemma 2 and Lemma 3 in FLM, the Lipschitz condition is modified by
$|\phi(f_1) - \phi(f_2)| \leq L|f_1-f_2| (\bar{\mathsf{k}}/b)^{d_T}$.
It follows that 
(A.9) in FLM is modified to $r_0\cdot\Big(
\frac{K\sqrt{C}}{b^{d_T}} \sqrt{\frac{WL\log W}{n}\log n} + \sqrt{\frac{2C_\ell \tilde\gamma}{nb^{d_T}}}\Big)
+ c_2 \epsilon_n^2 + \epsilon_n\sqrt{\frac{2C_\ell^2 \tilde\gamma}{nb^{d_T}}} + 30MC_\ell\frac{\tilde\gamma}{nb^{d_T}}$.
And the bound in (A.10) in FLM is multiplied by $b^{-d_T}$.
Thus (A.14) in FLM is modified to 
\begin{align*}
\bar r =&\ \frac{8}{c_1}\bigg(\frac{K\sqrt{C}}{b^{d_T}}
\sqrt{\frac{WL\log W}{n} \log n} + \sqrt{\frac{2 C_\ell^2 \tilde \gamma}{nb^{d_T}}}
\bigg)
+  \bigg(
\sqrt{\frac{2 (c_2 \vee1)}{c_1}} \epsilon_n + \sqrt{\frac{120MC_\ell}{c_1} \frac{\tilde \gamma}{nb^{d_T}}}  \bigg)
\\
&+ \frac{r_\ast}{b^{d_T}}.
\end{align*}
Therefore (A.17) in FLM is modified to 
\begin{align*}
C'\left(
\sqrt{\frac{WL\log W}{nb^{2d_T}} \log n} 
+ 
\sqrt{\frac{\log\log n + \gamma}{nb^{d_T}}}
+ \epsilon_n 
\right)
\end{align*}
with some constant $C' >0$ that does not depend on $n$.
Then we can optimize the upper bound on page 206 of FLM,
\begin{align*}
\bar r \leq C'\left(
\sqrt{\frac{\epsilon_n^{-\frac{2d_X}{r}}(\log(1/\epsilon_n) + 1)^7}{nb^{2d_T}} \log n} 
+ 
\sqrt{\frac{\log\log n + \gamma}{nb^{d_T}}}
+ \epsilon_n
\right)
\end{align*}
by choosing $\epsilon_n = (nb^{2d_T})^{-\frac{r}{2(r+d_X)}}$, $H \asymp \cdot (nb^{2d_T})^{\frac{d_X}{2(r+d_X)}} \log^2(nb^{2d_T}), L\asymp \cdot \log(nb^{2d_T})$.
Hence, w.p.a.1, we can bound (\ref{FLMA1})
\[
\mathbb{E}[\ell_{tb}(\hat f_b, Z)] - \mathbb{E}[\ell_{tb}(f_{\ast b}, Z)]  
\leq \bar r^2 \leq C \left( (nb^{2d_T})^{-\frac{r}{r+d_X}} \log^8 n  + \frac{\log\log n + \gamma}{nb^{d_T}}\right).
\]


The remaining proof is to show (\ref{FLM2.1-2}).
We add and subtract $\gamma(T,X)$ to $(Y-f(X))$ in the loss function, and by the law of iterated expectations, we obtain 
$2 \mathbb{E}[\ell_{tb}(f, Z)] = \mathbb{E}[var(Y|T,X) \mathsf{K}_b(T-t)] + \mathbb{E}[(\gamma(T,X) - f(X))^2 \mathsf{K}_b(T-t)]$.
Since the first term does not depend on $f$, we focus on the second term $Q_b(f) \equiv \mathbb{E}[(\gamma(T,X) - f(X))^2 \mathsf{K}_b(T-t)]$.

Let $Q(f) \equiv \|f-\gamma\|^2_{F_{tX}}$. 
We show that 
$Q_b(f) = Q(f) + b^2\mathsf{B}(f) + o(b^2)$ 
uniformly in uniformly bounded $f$, 
where $\mathsf{B}(f) \equiv \mathbb{E}\big[
(\partial_t    \gamma)^2 f_{t|X} + (\gamma - f)^2 \partial_t^2  f_{t|X}/2 + 2(\gamma -f) \partial_t \gamma \partial_t   f_{t|X}
+ (\gamma-f) \partial_t^2 \gamma f_{t|X} \big] \int u^2\mathsf{k}(u)du$.  
By change of variables, a Taylor expansion, and the mean value theorem,
$Q_b(f) = \mathbb{E}\big[\int_{\mathcal{T}}(\gamma(T,X) - f(X))^2 \mathsf{K}_b(T-t) f_{T|X}(T|X) dT\big]
= \mathbb{E}\big[\int_{\mathcal{T}}\big\{ \gamma(t,X) - f(X)
+ ub \partial_t \gamma(t, X)
+ u^2b^2 \partial_t \gamma(\bar t, X)/2
\big\}^2 \mathsf{k}(u)
\big\{ f_{T|X}(t|X) + ub \partial_t f_{T|X}(t|X) + u^2b^2 \partial_t^2 f_{T|X}(\tilde t|X)$\\$/2\big\} du\big]$, where $\bar t$ and $\tilde t$ are between $t$ and $t + ub$. 
Under Assumption~\ref{ADNN}(a),  apply the dominated convergence theorem to the terms associated with  $\partial_t \gamma(\bar t, X)$ and $\partial_t^2 f_{T|X}(\tilde t|X)$ that converge to $\partial_t^2 \gamma(t, X)$ and $\partial_t^2 f_{T|X}(t|X)$, respectively, as $b\rightarrow 0$.  

Thus uniformly in uniformly bounded $f$, 
\begin{align*}
2\left(\mathbb{E}[\ell_{tb}(f, Z)] - \mathbb{E}[\ell_{tb}(f_{\ast b}, Z)] \right) 
&= Q_b(f) - Q_b(f_{\ast b})\notag \\
&= \|f-\gamma\|_{F_{tX}}^2 - \|f_{\ast b} - \gamma\|_{F_{tX}}^2 + b^2(\mathsf{B}(f) - \mathsf{B}(f_{\ast b})) + o(b^2).
\end{align*}

To show (\ref{FLM2.1-2}), it suffices to show that $- \|f_{\ast b}- \gamma\|_{F_{tX}}^2 + b^2(\mathsf{B}(f) - \mathsf{B}(f_{\ast b})) = O(b^2)$.
Note that $\gamma(t, x) = \arg\min_f \lim_{b\rightarrow 0}\mathbb{E}[ \ell_{tb}(f, Z)] = \arg\min_f Q(f)$.
By the definition of the minimizers: $Q(f_{\ast b}) \geq Q(\gamma) = 0$ and $Q_b(\gamma) \geq Q_b(f_{\ast b})$.
So $- \|f_{\ast b}- \gamma\|_{F_{tX}}^2 + b^2(\mathsf{B}(f) - \mathsf{B}(f_{\ast b}))
= -Q_b(f_{\ast b}) + b^2\mathsf{B}(f) + o(b^2)
\geq -Q_b(\gamma) + b^2\mathsf{B}(f) + o(b^2)
= -Q(\gamma) - b^2 \mathsf{B}(\gamma) + b^2\mathsf{B}(f) + o(b^2) 
= b^2(\mathsf{B}(f) - \mathsf{B}(\gamma)) + o(b^2).
$ 
And 
$- \|f_{\ast b} - \gamma\|_{F_{tX}}^2 + b^2(\mathsf{B}(f) - \mathsf{B}(f_{\ast b}))  \leq b^2(\mathsf{B}(f) - \mathsf{B}(f_{\ast b}))$.
%
We obtain (\ref{FLM2.1-2}).
\hfill $\square$
\\
\\
\textbf{Proof of Lemma~\ref{LDNNMultiGPS}:}\
We use Theorem 1 in FLM by verifying their assumptions.  
\\[5pt]
\textbf{(a)}\
 In FLM's notation, $f_\ast(x) = \mathbb{E}[\Phi((t-T)/h_1)|X=x]$ for $t\in\mathcal{T}$ and $h_1 > 0$.
Assumption 1 in FLM holds, because $\sup_{t \in\mathcal{T}, h_1 > 0, x\in[-1,1]^{d_X}}| \mathbb{E}[\Phi((t-T)/h_1)|X=x] | \leq 1$ and $\sup_{t \in\mathcal{T}, h_1 > 0,} \Phi((t-T)/h_1) \leq 1$.

By integration by parts, change of variables, the Leibniz integral rule, and the condition (A), 
$| \mathtt{D}_x^{\alpha} f_\ast(x) | 
=\big| \mathtt{D}_x^{\alpha} \int_{\underline{T}}^{\overline{T}} \Phi((t-s)/h_1) f_{T|X}(s|x) ds \big|
= \big| \mathtt{D}_x^{\alpha} \big\{
\Phi((t-\overline T )/h_1) + \int_{\underline{T}}^{\overline{T}}  h_1^{-1}\phi((t-s)/h_1)F_{T|X}(s|x) ds
\big\} \big|
= 
\int_{-\infty}^\infty  \phi(u) \big|\mathtt{D}_x^{\alpha}F_{T|X}(t+uh_1|x) \big| du  \leq 1$,
where $\phi$ is the standard normal probability density function.
Thus for any $t\in\mathcal{T}$ and $h_1 > 0$, 
$f_\ast \in \mathcal{W}^{r,\infty}([-1,1]^{d_X})$ and Assumption~2 in FLM holds.

All the conditions in Theorem 1 in FLM hold for any $t\in\mathcal{T}$ and $h_1 > 0$.
By Theorem~1 in FLM,  
for any $t\in\mathcal{T}$ and $h_1 > 0$, with probability at least $1-\exp(-n^{\frac{d_X}{r+d_X}} \log^8n)$, for $n$ large enough,   
$\big\|\hat f_{MLP-ReGPS}(X)-\mathbb{E}[\Phi((T-t)/h_1)|X]\big\|_{F_{X}} \leq C \cdot R_{1n}$, 
 for a constant $C > 0$ independent of $n$, which may depend on $d_X, M$, and other fixed constants.
 Since $t$ and $h_1$ do not appear as fixed constants in the conditions of Theorem 1 in FLM,
 $C$ does not depend on $t$ and $h_1$.
 Therefore we obtain the first result.
\\[5pt]
\textbf{(b)}\
In FLM's notation, $f_\ast(x) = E[h_1^{d_T} g_{h_1}(T-t)|X=x]$ for $t\in\mathcal{T}$ and $h_1 > 0$.
Assumption~\ref{Akernel} implies that for an absolute constant $M > 0$, $\sup_{x\in\mathcal{X}}| E[h_1^{d_T} g_{h_1}(T-t)|X=x] | \leq M$ and $h_1^{d_T} g_{h_1}(T-t) < M$, so Assumption 1 in FLM holds. 

By change of variables, the Leibniz integral rule, and the condition (B), we show
$\big| \mathtt{D}_x^{\alpha} f_\ast(x) | 
=\big| \mathtt{D}_x^{\alpha} \int_{\mathcal{T}} h_1^{d_T} g_{h_1}(s-t) f_{T|X}(s|x) ds \big|
=  \int_{\mathcal{T}} g(u) \big|\mathtt{D}_x^{\alpha}  f_{T|X}(t+uh_1|x) \big| du h_1^{d_T}
\leq h_1^{d_T}$.
Thus for any $t\in\mathcal{T}$ and $h_1 \in (0,1]$, $f_\ast \in \mathcal{W}^{r,\infty}([-1,1]^{d_X})$, and Assumption 2 in FLM holds.
All the conditions in Theorem~1 in FLM hold for $t\in\mathcal{T}$ and $h_1 \in (0,1]$.
The same arguments in (a) yield the second result.
\hfill $\square$ 
\\ \\
\textbf{Proof of Corollary~\ref{CL2}:}\
We have shown in (\ref{ER2pf}) that 
$\int_\mathcal{X}\int_{\mathcal{T}_0} (\hat \gamma(t,x) - \gamma(t,x))^2 K_h(s-t) f_{TX}(s,x) ds dx = O_p(\|\hat\gamma - \gamma\|_{F_{tX}}^2)$.
Next we show that we can replace
$\|\hat\gamma - \gamma\|_{F_{tX}}$ with $h^{-d_T/2}\|\hat \gamma - \gamma\|_{F_{TX}} + h$ to obtain the rate conditions of Assumption~\ref{A1stL2}.
By the triangular inequality,
\begin{align*}
&\int_\mathcal{X}\int_{\mathcal{T}_0} (\hat \gamma(t,x) - \gamma(t,x))^2 K_h(s-t) f_{TX}(s,x) ds dx \\
\leq &
\int_\mathcal{X}\int_{\mathcal{T}_0} (\hat \gamma(t,x) - \hat \gamma(s,x))^2 K_h(s-t) f_{TX}(s,x) ds dx
\\
&+ \int_\mathcal{X}\int_{\mathcal{T}_0} (\hat \gamma(s,x) - \gamma(s,x))^2 K_h(s-t) f_{TX}(s,x) ds dx
\\
&+ \int_\mathcal{X}\int_{\mathcal{T}_0} (\gamma(s,x) - \gamma(t,x))^2 K_h(s-t) f_{TX}(s,x) ds dx
\\
\leq &
\int_\mathcal{X}\int_{\mathcal{R}^{d_T}} (\hat \gamma(t,x) - \hat \gamma(t+uh,x))^2 k(u) f_{TX}(t+uh,x) du dx 
\\
&+ h^{-d_T} \bar k \int_\mathcal{X}\int_{\mathcal{T}_0} (\hat \gamma(s,x) - \gamma(s,x))^2 f_{TX}(s,x) ds dx 
\\
&\int_\mathcal{X}\int_{\mathcal{R}^{d_T}} \big(uh \partial_t \gamma(t,x) + u^2 h^2 /2 \partial_t^2 \gamma(\bar t,x)\big)^2 k(u) \big(f_{TX}(t,x)  + uh \partial_t f_{TX}(\grave t, x)\big) du dx 
\\
= &\ 
O_p(h^{-d_T}\|\hat \gamma - \gamma\|_{F_{TX}}^2) + O_p(h^2)
\end{align*}
for a bounded kernel $\sup_u |k(u)| \leq \bar k$ for some positive constant $\bar k$,
for some $\bar t$ and $\grave t$ between $t$ and $t + uh$ using the mean value theorem.
Therefore we can replace Assumption~\ref{A1st}(a)(b) with Assumption~\ref{A1stL2}(a)(b).

Consider (\ref{ER1DR}).
When $\bar\lambda \neq \lambda$ and $\bar \gamma = \gamma$, 
for this first term of (\ref{ER1DR}) $\mathbb{E}[(\hat\gamma_{i\ell} - \bar{\gamma}_i)
(1 - \bar{\lambda}_i/\lambda_i + h^2 \partial_t^2 f_{T|X}(t|X_i) \bar{\lambda}_i \int_{-\infty}^\infty u^2k(u) du/2  + O_p(h^3))
|Z_\ell^c]$ to be $o_p(1/\sqrt{n h^{d_T}})$,
we need $\sqrt{nh^{d_T}}\|\hat \gamma - \gamma\|_{F_{tX}} = o_p(1)$.
That requires $\sqrt{n}\|\hat \gamma - \gamma\|_{F_{TX}} = o_p(1)$, which is not feasible.

When $\bar{\gamma} \neq \gamma$ and $\bar{\lambda} = \lambda$, 
the second term of (\ref{ER1DR}) $\mathbb{E}\Big[
(\hat\lambda_{i\ell} - \bar{\lambda}_i)
\Big((\gamma_i - \bar{\gamma}_i)/\lambda_i
+ h^2
\big((\gamma_i - \bar{\gamma}_i) \partial_t^2 f_{T|X}(t, X_i)/2 +  \partial_t f_{T|X}(t, X_i) \partial_t \gamma_i
+ f_{T|X}(t,X_i) \partial_t^2 \gamma_i/2
\big)\Big)\Big|Z_\ell^c\Big] = o_p(1/\sqrt{n_\ell h^{d_T}})$ by assuming $\|\hat\lambda_{\ell} - \bar{\lambda}\|_{F_{tX}} = o_p(1/\sqrt{n h^{d_T}})$.
\hfill $\square$
\\
\\
{\textbf{Proof of Theorem~\ref{TIF-SDML}:}\
Define $\check \psi_{i\ell} \equiv \hat \gamma_\ell (U_i, X_i) + \frac{K_h(T_i - t)}{ \hat f_\ell(t|X_i)} \left( Y_i - \hat \gamma_\ell(U_i, X_i) \right) - \beta_t$, $\tilde \psi_i \equiv  \bar\gamma (U_i, X_i) + \frac{K_h(T_i - t)}{ \bar f(t,X_i)} \left( Y_i -  \bar \gamma(U_i, X_i) \right)- \beta_t$, and 
$\psi_i \equiv  \bar \gamma(t, X_i) + \frac{K_h(T_i - t)}{  \bar f(t,X_i)} \left( Y_i - \bar \gamma(t, X_i) \right)
- \beta_t$.
To show the asymptotically linear representation of $\sqrt{nh^{d_T}}(\check \beta_t-\beta_t)
= \sqrt{h^{d_T}/n} \sum_{\ell=1}^L$\\$\sum_{i\in I_\ell} \check\psi_{i\ell} = \sqrt{h^{d_T}/n} \sum_{i=1}^n \psi_i  + o_p(1)$,
we decompose 
$\check\psi_{i\ell}  = (\check\psi_{i\ell} - \tilde\psi_i) + (\tilde\psi_i - \psi_i) + \psi_i $.


We first show that $\sqrt{nh^{d_T}} n^{-1}\sum_{\ell=1}^L\sum_{i\in I_\ell} (\check \psi_{i\ell} - \tilde\psi_i) = o_p(1)$. 
The proof closely follows the proof of Theorem~\ref{TIF}, by replacing $\bar\gamma_i \equiv \bar \gamma(t, X_i)$ with $\bar\gamma_i \equiv \bar \gamma(U_i, X_i)$, replacing $\hat\gamma_{i\ell}\equiv\hat\gamma_{\ell}(t, X_i)$ with $\hat\gamma_{i\ell}\equiv\hat\gamma_\ell(U_i, X_i)$, and using the rate conditions in Assumption~\ref{A1stU}.
Specifically, for $\ell\in\{1,..., L\}$, conditional on $Z_\ell^c$, we need a rate for bounding
\begin{align}
&\int_\mathcal{X}\int_{-\infty}^\infty \left( \hat\gamma_\ell(u,x)-\bar \gamma(u,x)\right)^2 f_{UX}(u,x) dudx 
\notag\\
&= 
 \int_\mathcal{X}\int_{\mathcal{T}_0} \left( \hat\gamma_\ell(\mathsf{T}_i h_0 + t,x)-\bar \gamma(\mathsf{T}_i h_0 + t,x)\right)^2  d\hat F_{T\ell}(\mathsf{T}_i)f_{X}(x) dx 
\notag
\\
&= 
 \frac{1}{n_\ell}\sum_{i\in I_\ell} \int_\mathcal{X} \left( \hat\gamma_\ell({T}_i h_0 + t,x)-\bar \gamma({T}_i h_0 + t,x)\right)^2 f_{X}(x) dx
\notag \\
&\equiv \frac{1}{n_\ell}\sum_{i\in I_\ell} W_{i\ell}, 
\label{1ineq}
\end{align}
where $W_{i\ell} \equiv \int_\mathcal{X} \left( \hat\gamma_\ell({T}_i h_0 + t,x)-\bar \gamma({T}_i h_0 + t,x)\right)^2 f_{X}(x) dx $.
We show that conditional on $Z_\ell^c$, $n_{\ell}^{-1}\sum_{i\in I_\ell} W_{i\ell} = \mathbb{E}\big[W_{i\ell}\big|Z_\ell^c\big]  + o_p(1)$ below by the conditional Markov's inequality.
First we compute 
\begin{align}
\mathbb{E}\left[W_{i\ell}\big|Z_\ell^c\right]
&= \int_{\mathcal{T}_0}\int_\mathcal{X} \left( \hat\gamma_\ell(s h_0 + t,x)-\bar \gamma(s h_0 + t,x)\right)^2 f_{X}(x) dx  f_T(s) ds 
\notag
\\
&=
\int_{\mathcal{R}^{d_T}} \int_\mathcal{X} \left( \hat\gamma_\ell(u,x)-\bar \gamma(u,x)\right)^2 f_{X}(x) dx  f_T\left(\frac{u-t}{h_0}\right) du h_0^{-d_T} 
\notag
\\
&\leq
\int_{\mathcal{T}_0}\int_\mathcal{X} \left( \hat\gamma_\ell(u,x)-\bar \gamma(u,x)\right)^2 f_{TX}(u, x) dx  du h_0^{-d_T} \bar f_T/\underline{f_{T|X}}
\notag
\\
& = O_p\left(h_0^{-d_T}\|\hat\gamma_\ell-\bar\gamma\|_{F_{TX}}^2 \right).
\notag
\end{align}
The inequality is by Assumption~\ref{ACIACS}(b)(iii) so that $\bar f_T/\underline{f_{T|X}} < \infty$, 
where $\bar f_T \equiv \sup_{t\in\mathcal{T}_0} f_T(t)$ and $\underline{f_{T|X}} \equiv \inf_{(t,x)\in\mathcal{T}\times\mathcal{X}}f_{T|X}(t|x)$.
Similarly
\begin{align*}
\mathbb{E}\big[ W_{i\ell}^2 \big| Z_\ell^c\big] 
&= \int_{\mathcal{T}_0} \left(  \int_\mathcal{X} \left(\hat\gamma_\ell(sh_0+t, x) - \bar \gamma(sh_0+t, x)\right)^2 f_X(x) dx\right)^2 f_T(s) ds
\\
&\leq 
\int_{\mathcal{T}_0} \int_\mathcal{X} \left(\hat\gamma_\ell(sh_0+t, x) - \bar \gamma(sh_0+t, x)\right)^4 f_X(x) dx f_T(s) ds
\\
&\leq \int_{\mathcal{R}^{d_T}} \int_\mathcal{X} \left(\hat\gamma_\ell(u, x) - \bar \gamma(u, x)\right)^4 f_X(x) dx f_T\left(\frac{u-t}{h_0}\right) du h_0^{-d_T}
\\
&\leq  \int_{\mathcal{T}_0} \int_\mathcal{X} \left(\hat\gamma_\ell(u, x) - \bar \gamma(u, x)\right)^4 f_{TX}(u, x) dx  du h_0^{-d_T}\bar f_T/\underline{f_{T|X}} 
\\
&= O_p(h_0^{-d_T})
\end{align*}
by Assumption~\ref{A1stU}(c) and Jensen's inequality.  

Using the same arguments for (R1-1) in the proof of Theorem~\ref{TIF}, let $\Delta_{i\ell} \equiv W_{i\ell} - \mathbb{E}\big[W_{i\ell}\big|Z_\ell^c\big]$.
Then $\mathbb{E}\big[\big( n_\ell^{-1}\sum_{i\in I_\ell}\Delta_{i\ell}\big)^2\big|Z_\ell^c\big] =
n_\ell^{-1}\big(
\mathbb{E}\big[ W_{i\ell}^2 \big| Z_\ell^c\big] - \mathbb{E}\big[W_{i\ell}\big|Z_\ell^c\big]^2
\big) = o_p(1)$ by assuming $nh_0^{d_T}\rightarrow\infty$.
By the conditional Markov's inequality, $n_\ell^{-1}\sum_{i\in I_\ell}\Delta_{i\ell} = o_p(1)$.

We obtain a rate for (\ref{1ineq}),  
$\int_\mathcal{X}\int_{\mathcal{U}} \left( \hat\gamma_\ell(u,x)-\bar \gamma(u,x)\right)^2 f_{UX}(u,x) dudx
= n_{\ell}^{-1}\sum_{i\in I_\ell} W_{i\ell} = \mathbb{E}\big[W_{i\ell}\big|Z_\ell^c\big]  + o_p(1)
= O_p\left(h_0^{-d_T}\|\hat\gamma_\ell-\gamma\|_{F_{TX}}^2 \right).$
This gives the modified conditions in Assumption~\ref{A1stU}(a)(ii).

The same arguments in the proof of Theorem~\ref{TIF} yield $\sqrt{nh^{d_T}} n^{-1}\sum_{\ell=1}^L\sum_{i\in I_\ell} (\check \psi_{i\ell} - \tilde\psi_i) = o_p(1)$. 


Next we show
$\sqrt{nh^{d_T}} n^{-1}\sum_{i=1}^n ( \tilde\psi_i - \psi_i)
\equiv \sqrt{nh^{d_T}} n^{-1}\sum_{i=1}^n W_i = \sqrt{nh^{d_T}} \mathbb{E}[W_i] + o_p(1) = o_p(1)$, where
$W_i \equiv (\bar \gamma(U_i, X_i) - \bar \gamma(t, X_i))(1-K_h(T_i -t)/\bar f(t,X_i))$.

We first calculate $\mathbb{E}[K_h(T-t)|X=x] = \int_{-\infty}^\infty k(u) f_{T|X}(t+uh|x) du = f_{T|X}(t|x) + O_p(h^2)$ uniformly over $x \in \mathcal{X}$ by Assumption~\ref{ACIACS}(c).
\begin{align*}
\mathbb{E}[W_i] &= 
\mathbb{E}\left[ \mathbb{E}\left[\bar \gamma(U_i, X_i)
- \bar \gamma(t, X_i)
|X_i\right] 
\mathbb{E}\left[ 1-K_h(T_i -t)/
\bar f(t,X_i)|X_i\right]\right] 
\\&=
\mathbb{E}\left[ 
\int_{\mathcal{T}_0}\left(\bar \gamma(sh_0 + t, X_i)  - \bar \gamma(t, X_i)\right) f_T(s) ds
\left( 1- \mathbb{E}[K_h(T_i -t)|X_i] /\bar f(t,X_i)
\right)
\right] 
\\&
= O\left( h^2 \int_\mathcal{X}
\int_{\mathcal{T}_0} sh_0 \partial_t\bar \gamma(\bar t, x)  f_T(s) 
f_X(x) 
dsdx\right)
\\&= O(h^2h_0)
\end{align*}
for some $\bar t$ between $t$ and $t+sh_0$.
Let $\sqrt{nh^{d_T}}h^2h_0 = o(1)$.

Following the same arguments, let $\Delta_{i\ell} \equiv W_i - \mathbb{E}[W_i]$.
We show $\sqrt{nh^{d_T}} n^{-1}\sum_{i=1}^n \Delta_{i\ell} = o_p(1)$ by the conditional Markov's inequality.

\begin{align}
&\mathbb{E}\left[\left(
\sqrt{nh^{d_T}} n^{-1}\sum_{i=1}^n \Delta_{i\ell} 
\right)^2\right]
= h^{d_T}\mathbb{E}[\Delta_{i\ell}^2] 
\leq h^{d_T} \mathbb{E}[W_i^2 ]
\notag \\
&=
\mathbb{E}\left[
\mathbb{E}\left[
(\bar \gamma(U_i, X_i) - \bar \gamma(t, X_i))^2|X_i
\right]
h^{d_T}\mathbb{E}\left[
(1-K_h(T_i -t)/\bar f(t,X_i))^2
|X_i\right]
\right]
\notag\\
&=
O_p\left( \int_{\mathcal{X}}  n_\ell^{-1} \sum_{i\in I_\ell} 
(\bar \gamma(T_ih_0+t, x) - \bar \gamma(t, x))^2
f_X(x) dx\right) 
\label{2kernel}\\
&=
O_p\left( \int_{\mathcal{T}_0}\int_{\mathcal{X}}
(\bar \gamma(sh_0+t, x) - \bar \gamma(t, x))^2
f_X(x) dx f_T(s) ds\right) 
\label{Markov}\\
&=
O_p\left( \int_{\mathcal{T}_0}\int_{\mathcal{X}}
s^2 h_0^2 \partial_t \bar \gamma(\bar t, x)^2
f_X(x) dx f_T(s) ds\right) 
\notag\\
&= O(h_0^2)
\notag
\end{align}
for some $\bar t$ between $t$ and $t+sh_0$.
The equality for (\ref{2kernel}) is because 
$h^{d_T}\mathbb{E}[(1-K_h(T-t))^2|X=x]
= h^{d_T}\int_{\mathcal{R}^{d_T}} (1-h^{-d_T}\Pi_{j=1}^{d_T}k(u_j))^2 f_{T|X}(t+uh|x) du h^{d_T}
= O(1)$.
The equality for (\ref{Markov}) comes from $n_\ell^{-1} \sum_{i\in I_\ell} A_i - \mathbb{E}[A_i] = o_p(1)$, where $A_i \equiv  \int_{\mathcal{X}} (\bar \gamma(T_ih_0+t, x) - \bar \gamma(t, x))^2 f_X(x) dx$, 
by Markov's inequality:
$\mathbb{E}[(n_\ell^{-1} \sum_{i\in I_\ell} A_i - \mathbb{E}[A_i] )^2] 
= var(n_\ell^{-1} \sum_{i\in I_\ell}A_i)
$\\$\leq  n_\ell^{-1}\mathbb{E}[A_i^2] = n_\ell^{-1} \mathbb{E}\big[\big(
\int_{\mathcal{X}} T_i^2 h_0^2 \partial_t\bar \gamma(\bar t, x)^2 f_X(x)$ $dx
\big)^2\big] = O(n^{-1} h_0^2) = o(1)$.
\hfill$\square$
\\
\\
\textbf{Proof of Theorem~\ref{Tbw-SDML}:}\
The proof of the consistency of $\check{\mathsf{V}}_t$ follows the proof of Theorem~\ref{Tbw} by modifying Assumption~\ref{AOptBW}(a)(ii) to Assumption~\ref{AOptBW-SDML}(a)(ii).
Assumption~\ref{AOptBW-SDML}(a)(ii) are obtained by the same arguments for (\ref{1ineq}).
The consistency of $\check{\mathsf{B}}_t$ and $\check h_t$ are directly implied by the proof of Theorem~\ref{Tbw}. 
\hfill$\square$
\\
\\
\textbf{Proof of Theorem~\ref{TME-SDML}:}
The proof follows the proof of Theorem~\ref{TME} directly by modifying the conditions (a) and (b) in Theorem~\ref{TME}.
\hfill$\square$

\newpage

\renewcommand{\theequation}{S.\arabic{equation}}
\renewcommand{\thesection}{S\arabic{section}}
\renewcommand{\thepage}{S\arabic{page}}

\renewcommand{\thetheorem}{S\arabic{theorem}}
\renewcommand{\theassumption}{S\arabic{assumption}}
\renewcommand{\theremark}{S\arabic{remark}}
\renewcommand{\thetable}{S\arabic{table}}
\renewcommand{\thefigure}{S\arabic{figure}}

\setcounter{equation}{0}
\setcounter{page}{1}
\setcounter{section}{0}


%



\begin{center}
{\Large 
Online supplement for  ``Double Debiased Machine Learning Nonparametric Inference with Continuous Treatments"
\\[10pt]
\large Kyle Colangelo 
\hspace{0.6cm} 
Ying-Ying Lee\footnote{Ying-Ying Lee, Department of economics, University of California Irvine.
E-mail: \href{yingying.lee@uci.edu}{yingying.lee@uci.edu}.
Kyle Colangelo, Amazon.
}
\\[5pt]
\normalsize September 2023
} 
\maketitle

\end{center}

\begin{center}
{\bf \small Abstract}
\end{center}


Section~\ref{ASecMC} presents supplemental material for Section~5 in the main text.
Section~\ref{SecM} discusses the kernel localization to justify our kernel-based DML estimators.
Section~\ref{ASec1stEst} presents the kernel and series nuisance function  estimators.
Section~\ref{SecUnif} presents uniform inference theory.


\section{Numerical examples}
\label{ASecMC}

We implement  our DML estimator in Python, using the packages scikit-learn, pytorch, numpy, pandas, rpy2 and scipy. We use the R package ``grf'' for the generalized random forest implementation, implementing it in Python via the rpy2 package.

We describe the nuisance function estimators in more detail. \\[3pt]
\textbf{Lasso:}
The penalization parameter is chosen via grid search utilizing tenfold cross validation for $\hat\gamma$ and $\hat f_{T|X}$ separately.
The basis functions contain third-order polynomials of $X$ and $T$, and interactions among $X$ and $T$.\\[3pt]
\textbf{Neural Network:} 
We use rectified linear units (ReLU) activation functions and the tuning parameters in Table~\ref{TNN}.
Learning rate, momentum and weight decay are tuned via cross validation over repeated simulations in the simulation study and the empirical analysis. 
The weights are fit using stochastic gradient descent with a weight decay and a learning rate.\footnote{
Weight decay is a form of regularization to prevent overfitting.
Weight decay is a penalty where after each iteration the weights in the network are multiplied by $(1-\alpha\lambda)$ before adding the adjustment in the direction of the gradient, where $\alpha$ is the learning rate (step size) and $\lambda$ is the weight decay.  }
We use only one hidden layer as more complex deep neural networks require more sample size.
\begin{table}[h]
        \caption{\label{TNN} Neural Network}
        \begin{center}
         \begin{tabular*}{\textwidth}{@{}cccccc@{}}
                   \hline\hline
&model &neurons &learning rate &momentum  &weight decay  \\
Simulation study &&&&\\
n=1,000 &$\gamma$ &10 &0.01&0.9 &0.05\\
&$f_{T|X}$ &10 &0.01&0.9 &0.3
\\[3pt]
n=10,000 &$\gamma$ &100 &0.05&0.95 &0.05\\
&$f_{T|X}$ &100 &0.4 &0 &0.075\\[3pt]
\hline
\\[-5pt]
Empirical analysis &$\gamma$ &25 &0.15&0.9 &0.05\\
&$f_{T|X}$  &25 &0.05&0.3&0.15\\
\hline
\hline 
   \end{tabular*}%
       \end{center}
       \end{table}
\\[3pt]
\textbf{Generalized Random forest (GRF):}
We use the generalized random forests in \cite{ATW19AS}, with 2,000 trees and all other parameters chosen via cross validation in every Monte Carlo replication. The parameters tuned via cross validation are: The fraction of data used for each tree, the number of variables tried for each split, the minimum number of observations per leaf, whether or not to use ``honesty splitting,'' whether or not to prune trees such that no leaves are empty, the maximum imbalance of a split, and the amount of penalty for an imbalanced split. Unlike Lasso, we do not add any additional basis functions as inputs into GRF.


       \begin{table}
        \caption{\label{TDGP-regps}Simulation Results with ReGPS for Lasso.}
        \begin{center}
         \begin{tabular*}{0.78\textwidth}{@{}rc|ccc|ccc@{}}
          \hline\hline
		&\multicolumn{1}{c}{} &\multicolumn{3}{c}{$\beta_0=0$}     
		&\multicolumn{3}{c}{$\theta_0=1.2$}     \\
		\hline
		 L     &\multicolumn{1}{c}{c} & Bias  & RMSE  & \multicolumn{1}{c}{Coverage} 
		 & Bias  & RMSE  & \multicolumn{1}{c}{Coverage} \\
\hline
\hline
	   1    & 0.75  & 0.008 & 0.106 & 0.957   & 0.051 & 1.197 & 0.985 \\
                 & 1.00  & 0.013 & 0.096 & 0.954 & 0.020 & 1.191 & 0.981\\
                 & 1.25  & 0.020 & 0.091 & 0.955 & 0.008 & 1.183 & 0.977\\
                 & 1.50  & 0.029 & 0.088 & 0.944  & -0.002 & 1.174 & 0.979\\
    \hline
           5    & 0.75  & 0.006 & 0.107 & 0.957 & 0.042 & 1.199 & 0.983\\
                 & 1.00  & 0.011 & 0.097 & 0.956 & 0.012 & 1.193 & 0.982\\
                 & 1.25  & 0.019 & 0.091 & 0.951 & 0.001 & 1.184 & 0.976\\
                 & 1.50  & 0.029 & 0.089 & 0.945 & -0.010 & 1.174 & 0.978\\
         \hline\hline
         \end{tabular*}
       \end{center}
       \footnotesize
       \renewcommand{\baselineskip}{11pt}
       \textbf{Note:} $n=1,000$.  $L=1$: no cross-fitting.  $L=5$: fivefold cross-fitting. 
		The bandwidth is $h = c \sigma_T n^{-0.2}$, and $c = 1.43$ for the AMSE-optimal bandwidth.  
		The nominal coverage rate of the confidence interval is 0.95.
       \end{table}

Table \ref{TDGP-regps} presents results for Lasso using the ReGPS estimator. The performance
is comparable with the MultiGPS estimator and 
is improved in terms of coverage rate for $\beta_0$.

       \begin{table}
        \caption{\label{TDGP-SDML}Simulation Results with SDML for Lasso.}
        \begin{center}
         \begin{tabular*}{0.78\textwidth}{@{}rc|ccc|ccc@{}}
          \hline\hline
		&\multicolumn{1}{c}{} &\multicolumn{3}{c}{$\beta_0=0$}     
		&\multicolumn{3}{c}{$\theta_0=1.2$}     \\
		\hline
		 L     &\multicolumn{1}{c}{c} & Bias  & RMSE  & \multicolumn{1}{c}{Coverage} 
		 & Bias  & RMSE  & \multicolumn{1}{c}{Coverage} \\
\hline
\hline

         \hline\hline
         \end{tabular*}
       \end{center}
       \footnotesize
       \renewcommand{\baselineskip}{11pt}
       \textbf{Note:} $n=1,000$.  $L=1$: no cross-fitting.  $L=5$: fivefold cross-fitting. 
		The bandwidth is $h = c \sigma_T n^{-0.2}$, and $c = 1.43$ for the AMSE-optimal bandwidth.  
		The nominal coverage rate of the confidence interval is 0.95.
       \end{table}

\begin{figure}[!htp]
\centering
\caption{Histogram of Hours of Training}
\includegraphics[width=0.6\textwidth]{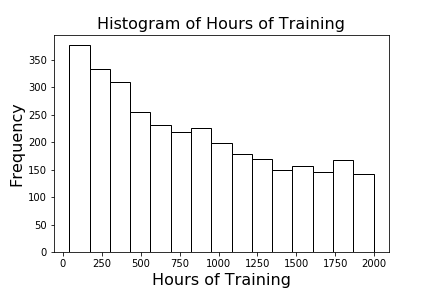}
\label{HistT}
\end{figure}

       \begin{table}
        \caption{\label{Tss}Descriptive statistics.}
        \begin{center}
         \begin{tabular*}{\textwidth}{@{}lccccc@{}}
                   \hline\hline
Variable  & \multicolumn{1}{c}{Mean} & \multicolumn{1}{c}{Median} & \multicolumn{1}{c}{StdDev} & \multicolumn{1}{c}{Min} & \multicolumn{1}{c}{Max} 
\\ 
\hline \\[-1.8ex] 
  share of weeks employed in 2nd year ($Y$)                     &44.00 &40.38  &37.88 &0 &100  
  \\
   total hours spent in 1st-year training ($T$)                    &1219.80 &992.86 &961.62  &40  &6188.57 
   \\
\hline
         \end{tabular*}
       \end{center}
       \footnotesize
       \renewcommand{\baselineskip}{11pt}
       \textbf{Note:} Summary statistics for 4,024 individuals who completed at least 40 hours of academic and vocational training.
       \end{table}

\section{Kernel localization}
\label{SecM}
We discuss the construction of the doubly robust moment function
by Gateaux derivative and a local Riesz representer.
Importantly the expression in (\ref{EGD}) is the building block to construct estimators for $\beta_t$ and the linear functionals of $\beta_t$.
Section \ref{SecKernel}
discusses the adjustment for the first-step kernel estimators in the moment functions of the regression estimator and inverse probability weighting estimator that do not use the doubly robust moment function and cross-fitting. 
We illustrate how the DML estimator assumes weaker conditions.
A series estimator $\hat \gamma$ yields a different adjustment.
These distinct features of continuous treatments are in contrast to the regular binary treatment case, where different nonparametric nuisance function estimators $\hat\gamma$ result in the same efficient influence function.

One way to obtain the influence function
 is to calculate the limit of the Gateaux derivative with respect to a smooth deviation from the true distribution, as the deviation approaches a point mass, following \cite{CLV} and \cite{IchimuraNewey22QE}.
For any $t\in\mathcal{T}$, let $\beta_t(\cdot): \mathcal{F} \rightarrow \mathcal{R}$, where $\mathcal{F}$ is a set of CDFs of $Z$ that is unrestricted except for regularity conditions. 
The estimator converges to $\beta_t(F)$ for some $F \in \mathcal{F}$, which describes how the limit of the estimator varies as the distribution of a data observation varies. 
Let $F^0$ be the true CDF of $Z$.
Let the CDF $F_Z^h$  approach a point mass at $Z$ as $h \rightarrow 0$. 
Consider $F^{\tau h} = (1-\tau) F^0 + \tau F_Z^h$ for $\tau \in [0,1]$ such that for all small enough $\tau$, $F^{\tau h} \in \mathcal{F}$
and the corresponding pdf $f^{\tau h} = f^0 + \tau(f_Z^h - f^0)$.
We calculate the Gateaux derivative of the functional $\beta_t(F^{\tau h})$ with respect to a deviation $F_Z^h - F^0$ from the true distribution $F^0$.

Below we show that the Gateaux derivative for the direction $f^h_Z - f^0$ is
\begin{align}
\lim_{h\rightarrow 0}
\frac{d}{d \tau}\beta_t(F^{\tau h})\Big|_{\tau = 0} 
=&\ \gamma(t,X) - \beta_t + \lim_{h\rightarrow 0}\int_{\mathcal{X}}\int_{\mathcal{Y}} 
 \frac{y - \gamma(t,x)}{f_{T|X}(t|x)} f_{Z}^h(y,t,x) dy dx \label{EGD} \\
=&\ \gamma(t,X) - \beta_t + 
\frac{Y - \gamma(t,X)}{f_{T|X}(t|X)} \lim_{h\rightarrow 0} f_T^h(t). \notag
\end{align}
Note that the last term in (\ref{EGD}) is a partial mean that is a marginal integration over $\mathcal{Y} \times \mathcal{X}$, fixing the value of $T$ at $t$.
Thus the Gateaux derivative depends on the choice of $f_{T}^h$.
We then choose $f^h_Z(z) = K_h(Z-z) {\bf 1}\{ f^0(z) > h\}$, following \cite{IchimuraNewey22QE},
so $\lim_{h \rightarrow 0}  f_T^h(t) = \lim_{h \rightarrow 0}  K_h(T-t)$.


\cite{IchimuraNewey22QE} show that if a semiparametric estimator is asymptotically linear and locally regular, then the influence function is $\lim_{h\rightarrow 0} d\beta_t(F^{\tau h})/d\tau|_{\tau=0}$.
Here, we use the Gateaux derivative limit calculation to motivate our moment function that depends on $F_{T}^h$.
Then we show that our estimator is asymptotically equivalent to a sample average of the moment function.

%


\begin{remark}[Linear functional of $\beta_t$]
{\rm
Consider a non-regular function-valued linear functional of $\beta_t$, denoted by $\alpha_t = \mathbb{A}[\beta_t]$ for a linear operator $\mathbb{A}$.  
So $\alpha_t$ is also a nonparametric function of $t$.
 To construct the DML estimator of $\alpha_t$, the Gateaux derivative of $\alpha_t$ is simply the linear functional of the Gateaux derivative of $\beta_t$ in (\ref{EGD}), i.e.,
 \begin{align*} 
\lim_{h\rightarrow 0}
\frac{d}{d \tau}\alpha_t(F^{\tau h})\Big|_{\tau = 0}  &= \lim_{h\rightarrow 0}\frac{d}{d \tau}\mathbb{A}\left[\beta_t(F^{\tau h})\right]\Big|_{\tau = 0} 
\\&=  \mathbb{A}[\gamma(t,X)] -  \mathbb{A}[\beta_t] + \lim_{h\rightarrow 0} \mathbb{A}\left[\frac{Y - \gamma(t,X)}{f_{T|X}(t|X)} f_T^h(t)\right].
 \end{align*}
We may work out the close-form expression of this Gateaux derivative of $\alpha_t$ and use its estimated sample analogue to construct the DML estimator of $\alpha_t$. 
An alternative DML estimator is simply $\hat\alpha_t = \mathbb{A}\big[\hat\beta_t\big]$.
Note that the partial effect can be expressed as a linear functional of $\beta_t$: $\theta_t = \mathbb{A}[\beta_t] = \partial \beta_t/\partial {t_1}$.
}
\label{Ralpha}
\end{remark}


\begin{remark}[Local Riesz representer]
{\rm 
The above discussion on the Gateaux derivative suggests that the Riesz representer for the non-regular $\beta_t$ is not unique and depends on the kernel or other methods for localization at $t$.
We define the ``{\it local Riesz representer}" to be $\alpha_{th}(T,X) =  f_T^h(t)/f_{T|X}(T|X) = K_h(T-t)/f_{T|X}(T|X)$ indexed by the evaluation value~$t$ and the bandwidth of the kernel $h$.
Our local Riesz representer $\alpha_{th}(T,X)$ satisfies  $ \beta_t 
= \int_{\mathcal{X}} \gamma(t, X) dF_X(X)
= \lim_{h \rightarrow 0} \int_{\mathcal{X}} \int_{\mathcal{T}}  \alpha_{th}(T,X) \gamma(T,X) dF_{TX}(T, X)$ for all $\gamma$ with finite second moment,
following the insight of the local Riesz representation theorem for a regular parameter \citep{Newey94ETA}.
Then we can obtain the influence function by adding an adjustment term $\alpha_{th}(T,X)(Y-\gamma(T,X))$, which is the product of the local Riesz representer and the regression residual.
For a series localization, \cite{CLS14JoE} define the {\it sieve Riesz representer} for plug-in sieve M estimators of irregular functionals;
see also \cite{ChenPouzo15} for general semi/nonparametric conditional moment models. 
}
\end{remark}
\textbf{Gateaux derivative:}\
Let the Dirac delta function $\delta_t(T) = \infty$ for $T=t$, $\delta_t(T) = 0$ for $T\neq t$,
and $\int g(s)\delta_t(s) ds = g(t)$, for any continuous compactly supported function $g$.\footnote{Note that a nascent delta function to approximate the Dirac delta function is  $K_h(T-t) = k((T-t)/h)/h$ such that $\delta_t(T) = \lim_{h \rightarrow 0} K_h(T-t)$.} 
For any $F\in \mathcal{F}$,
\begin{align*}
\beta_t(F) &= 
\int_{\mathcal{X}}  \mathbb{E}[Y|T=t, X=x] f_X(x) dx\\
&= \int_{\mathcal{X}}\int_{\mathcal{T}_0}  \mathbb{E}[Y|T=s, X=x] \delta_t(s)ds f_X(x) dx \\
&=
\int_{\mathcal{X}}\int_{\mathcal{T}_0} \int_{\mathcal{Y}} y\delta_t(s)  \frac{f_{Z}(y,s,x) f_{X}(x)}{f_{TX}(s,x)} dy dsdx.
\end{align*}

\begin{align}
\frac{d}{d \tau}\beta_t(F^{\tau h})\Big|_{\tau = 0} =&\ 
\int_{\mathcal{X}}\int_{\mathcal{T}_0} \int_{\mathcal{Y}} y\delta_t(s) \frac{d}{d \tau} \left( \frac{f_{Z}(y,s,x) f_{X}(x)}{f_{TX}(s,x)}\right) dy dsdx\Big|_{\tau = 0}\notag\\
=&\ 
\int_{\mathcal{X}}\int_{\mathcal{T}_0} \int_{\mathcal{Y}} \frac{y\delta_t(s)}{f_{TX}(s,x)}\big( \left( - f^0_{Z}(y,s,x) + f_{Z}^h(y,s,x) \right) f_X(x) \notag\\
&+ f_{Z}(y,s,x) \left( - f_{X}^0(x) + f_X^h(x)\right) \big)dy dsdx\notag\\
&- \int_{\mathcal{X}}\int_{\mathcal{T}_0} \int_{\mathcal{Y}} y\delta_t(s)  \frac{f_{Z}(y,s,x) f_{X}(x)}{f_{TX}(s,x)^2}\left( -f_{TX}^0(s,x) + f_{TX}^h(s,x)\right) dy dsdx.
\label{EGD3}
\end{align}
The influence function can be calculated as 
\begin{align}
\lim_{h \rightarrow 0}
\frac{d}{d \tau}\beta_t(F^{\tau h})\Big|_{\tau = 0} 
=&\ 
\gamma(t,X) - \beta_t +\lim_{h \rightarrow 0} \int_{\mathcal{X}}\int_{\mathcal{Y}} 
 \frac{y - \gamma(t,x)}{f_{T|X}(t|x)} f_{Z}^h(y,t,x) dy dx \notag \\
=&\ 
\gamma(t,X) - \beta_t +
 \frac{Y - \gamma(t,X)}{f_{T|X}(t|X)} \lim_{h \rightarrow 0} f_{T}^h(t).
 \label{EGDinf}
\end{align}
Specifically the term associated with $f_{TX}^h$ in (\ref{EGD3}) contributes $\gamma(t,X)/f_{T|X}(t|X) \lim_{h\rightarrow 0} f_T^h(t)$ in (\ref{EGDinf}) by calculating 
\begin{align*}
&\lim_{h\rightarrow 0} \int_{\mathcal{X}}\int_{\mathcal{T}_0} \int_{\mathcal{Y}} y\delta_t(s)  \frac{f_{Z}(y,s,x) f_{X}(x)}{f_{TX}(s,x)^2} f_{TX}^h(s,x) dy dsdx
\\
&= \lim_{h\rightarrow 0}  \int_{\mathcal{X}} \int_{\mathcal{Y}} y  \frac{f_{Z}(y,t,x) f_{X}(x)}{f_{TX}(t,x)^2} f_{TX}^h(t,x) dy dx
\\
&=  \lim_{h\rightarrow 0} \int_{\mathcal{X}}   \frac{\gamma(t,x)}{f_{T|X}(t|x)} f_{TX}^h(t,x) dx
\\
&= \frac{\gamma(t,X)}{f_{T|X}(t|X)} \lim_{h\rightarrow 0}  f_{T}^h(t).
\end{align*}
Other terms in (\ref{EGDinf}) are calculated by the same arguments.

In particular, we specify $F_Z^h$ following \cite{IchimuraNewey22QE}. 
Let $K_h(Z) = \Pi_{j=1}^{d_z} k(Z_j/h)/h$, where $Z = (Z_1, ..., Z_{d_z})'$ and $k$ satisfies Assumption~2.2
and is continuously differentiable of all orders with bounded derivatives.
Let $F^{\tau h} = (1-\tau)F^0 + \tau F_Z^h$
with pdf 
with respect to a product measure 
given by $f^{\tau h}(z) = (1-\tau) f^0(z) + \tau f^0(z) \delta_Z^h(z)$,
where $\delta_Z^h(z) = K_h(Z-z)/f^0(z)$, a ratio of a sharply peaked pdf to the true density.
Thus $f_{Z}^h(y,t,x) = K_h(Y-y)K_h(T-t)K_h(X-x)$.
It follows that 
$\lim_{h \rightarrow 0} f_{T}^h(t) = \lim_{h \rightarrow 0}  K_h(T-t)$ and
\begin{align*}
\lim_{h \rightarrow 0} \int_{\mathcal{X}}\int_{\mathcal{Y}} 
 \frac{y - \gamma(t,x)}{f_{T|X}(t|x)} f_{Z}^h(y,t,x) dy dx 
 =\frac{Y - \gamma(t,X)}{f_{T|X}(t|X)} \lim_{h \rightarrow 0}  K_h(T-t).
\end{align*}

$\mathbb{E}\big[
\frac{d}{d \tau}\beta_t(F^{\tau h})\big|_{\tau = 0} \big]
= 
\mathbb{E}\big[ \gamma(t,X) - \beta_t 
+ 
\frac{Y - \gamma(t,X)}{f_{T|X}(t|X)} K_h(T-t)
\big] = O(h^2)$.
So Neyman orthogonality holds a $h \rightarrow 0$.


%
%

\subsection{Adjustment for first-step kernel estimation}
\label{SecKernel}
We discuss another motivation of our moment function.
We consider two alternative estimators for $\beta_t$: the regression estimator 
$\hat \beta^{REG}_t = n^{-1}\sum_{i=1}^n \hat \gamma(t, X_i)$
that is based on the identification in equation (2.2) in the main text,
and the inverse probability weighting (IPW) estimator 
$\hat \beta^{IPW}_t = n^{-1}\sum_{i=1}^n K_h(T_i-t) Y_i/\hat f_{T|X}(t|X_i)$
that is based on the identification in (2.3) in the main text.
Adding the influence function that accounts for the first-step estimation partials out the first-order effect of the first-step estimation on the final estimator, as discussed in \cite{CEINR} and \cite{BEV20AS} for the semiparametric empirical likelihood inference in a low dimensional nonparametric setting.

For $\hat \beta^{REG}_t$, consider $\hat \gamma(t,x)$ to be a local constant or local polynomial estimator with bandwidth $h$ for low-dimensional $X$.
\cite{Newey94ET} and \cite{Lee} have derived the asymptotically linear representation of $\hat \beta^{REG}_t$ that is first-order equivalent to that of our DML estimator given in Theorem~3.1.
Specifically we can obtain the adjustment term by the influence function of the partial mean 
$\int_{\mathcal{X}} \hat \gamma(t,x) f(x) dx = n^{-1}\sum_{i=1}^n K_h(T_i - t)(Y_i - \gamma(t, X_i))/f_{T|X}(t|X_i) + o_p((nh^{d_T})^{-1/2})$ with a suitably chosen $h$ and regularity conditions.
Thus the moment function can be constructed by adding the influence function adjustment for estimating 
the nuisance function $\gamma(t, X)$ to the original moment function $\gamma(t, X)$.

Similarly for $\hat \beta^{IPW}_t$, when $\hat f_{T|X}$ is a standard kernel density estimator with bandwidth $h$, \cite{HHLP} derive the asymptotically linear representation of $\hat \beta^{IPW}_t$ that is first-order equivalent to our DML estimator.
We can show that the partial mean $\int_\mathcal{Z} K_h(T-t) Y/\hat f_{T|X}(t|X) dF_{Z} = n^{-1}\sum_{i=1}^n \gamma(t, X_i)\left(1 - K_h(T_i - t)/f_{T|X}(t|X_i) \right) + o_p((nh^{d_T})^{-1/2})$ with a suitably chosen $h$ and regularity conditions.
Thus the moment function
can be constructed by adding the influence function adjustment for estimating 
the nuisance function $f_{T|X}$ to the original moment function $K_h(T-t) Y/f_{T|X}(t|X)$.

\begin{remark}[First-step bias reduction]{\rm
In general, nonparametric estimation of an infinite-dimensional nuisance parameter contributes a finite-sample bias to the final estimator.
It is noteworthy that 
$\hat \beta_t$ requires a weaker bandwidth condition for controlling the bias of the first-step estimator than the regression estimator $\hat \beta^{REG}_t$ and the IPW estimator $\hat \beta^{IPW}_t$.
Our DML estimator for continuous treatments inherits this advantageous property from the DML estimator for a binary treatment.
Therefore the DML estimator can be less sensitive to variation in tuning parameters of the first-step estimators.
To illustrate with an example of $\hat \beta^{REG}_t$, consider the first-step $\hat \gamma$ to be a local constant estimator with bandwidth $h_1$ and a kernel of order $r$. 
To control the bias of $\hat \gamma$ to be asymptotically negligible for $\hat \beta^{REG}_t$, we assume $h_1^r \sqrt{nh_1^{d_T}} \rightarrow 0$.
In contrast, when $\hat \gamma$ and $\hat f_{T|X}$ in the DML estimator $\hat \beta_t$ 
are local constant estimators with bandwidth $h_1$ and a kernel of order $r$, Assumption~2.3(b) requires $h_1^{2r}\sqrt{nh^{d_T}}\rightarrow 0$.
It follows that the DML estimator need not undersmooth the nuisance function estimators, while the regression estimator $\hat \beta^{REG}_t$ requires an undersmoothing $\hat \gamma$.
Moreover we observe that the condition is weaker than $h_1^r\sqrt{n} \rightarrow 0$ for the binary treatment that has a regular root-$n$ convergence rate.
}
\end{remark}

\begin{remark}[First-step series estimation]
{\rm 
When $\hat \gamma(t,x)$ is a series estimator in $\hat \beta_t^{REG}$, computing the partial mean $\int_\mathcal{X} \hat \gamma(t,x) f(x) dx$ for the influence function results in a different adjustment term than the kernel estimation discussed above.
For example, \cite{LeeLi} derive the asymptotic theory of a partial mean of a series estimator, in estimating the average structural function with a special regressor.
Heuristically, let us consider a basis function $b(T,X)$, including raw variables $(T,X)$ as well as interactions and other transformations of these variables.
Computing $\int_{\mathcal{X}} \hat \gamma(t, x) f(x) dx$ implies the adjustment term of the form
 $ \mathbb{E}[b(t,X)]\big( n^{-1}\sum_{i=1}^nb(T_i,X_i)b(T_i,X_i)' \big)^{-1}$
$\times n^{-1}\sum_{i=1}^n b(T_i, X_i)' \big(y_i - \gamma(T_i, X_i)\big) = n^{-1}\sum_{i=1}^n \lambda_{ti} \big(y_i - \gamma(T_i, X_i)\big)$, resulting in a form of an average weighted residuals in estimation or projection of the residual on the space generated by the basis functions. 
Notice that the conditional density $f_{T|X}(t|X)$ is not explicit in the weight $\lambda_{ti}$. 
Such adjustment term may motivate different estimators of $\beta_t$; for example, the approximate residual balancing estimator in \cite{AIW}, \cite{CEINR}, and \cite{DSLC}.

}\end{remark}


\section{Supplement for Section~3.1 Nuisance function estimators}
\label{ASec1stEst}

\subsection{Kernel}
\label{SecKernelA}
The asymptotic distribution of a kernel-based estimator is well-studied (see, e.g.\ Chapter 19 in \cite{HansenBook}).
Consider the Nadaraya-Watson regression estimator $\hat\gamma(t,x) = \sum_{i=1}^n Y_i K_{h_\gamma}(T_i - t) K_{h_\gamma}(X_i - x)\big/\sum_{i=1}^n K_{h_\gamma}(T_i - t) K_{h_\gamma}(X_i - x)$ with a bandwidth $h_\gamma$ and a kernel of order $r_\gamma$. 
Estimate the GPS $f_{T|X}$ by the standard kernel estimator 
$\hat f_{T|X}(t|x) =  \sum_{i=1}^n K_{h_f}(T_i - t) K_{h_f}(X_i - x)\big/\sum_{i=1}^n K_{h_f}(X_i - x)$ with a bandwidth $h_f$ and a kernel of order $r_f$.
Let $d = d_T + d_X$.

\begin{assumption}[First-step kernel] 
Over $(t,x) \in \mathcal{T}\times\mathcal{X}$ that is bounded,
(a) $var(Y|T=t,X=x)$ is continuous and uniformly bounded,
(b) the $r_\gamma$th derivatives of $\gamma(t,x)$ exist and are continuous, uniformly bounded,
(c) the $r_f$th derivatives of $f_{T|X}(t|x)$ exist and are continuous, uniformly bounded;
(d) $h_\gamma \rightarrow 0$, $nh_\gamma^d \rightarrow\infty$, $nh_\gamma^{d+2r_\gamma} = O(1)$;
(e) $h_f \rightarrow 0$, $nh_f^d \rightarrow\infty$, $nh_f^{d+2r_f} = O(1)$.
\label{Akernel1}
\end{assumption}

\begin{theorem}
Under Assumption~\ref{Akernel1} and $r_f \geq r_\gamma-1$, 
$\|\hat\gamma - \gamma\|_{F_{tX}} = O_p\big((nh_\gamma^d)^{-1/2} + h_\gamma^{r_\gamma}\big)$
and 
$\|\hat f_{T|X} - f_{T|X}\|_{F_{tX}} = O_p\big((nh_f^d)^{-1/2} + h_f^{r_f} \big)$,
for $t\in\mathcal{T}$.
\label{Tkernel}
\end{theorem}

Assumption~2.3(b) requires $\sqrt{nh^{d_T}} 
\big((nh_f^{d} )^{-1/2}  + h_f^{r_f}\big) 
\big((nh_\gamma^d)^{-1/2} + h_\gamma^{r_\gamma}\big)
\rightarrow 0$.
It further implies that we can choose the AMSE optimal bandwidths for $\hat\gamma$ and $\hat f_{T|X}$ respectively.
Assumption~\ref{Tkernel} implies the smoothness of the nuisance functions increases with $d$
that is a common feature of nonparametric estimation. 
To see the curse of dimension, consider an example for $d_T = 1$ and $r_\gamma = r_f = r$.
We choose the optimal rates for the bandwidths 
$h_\gamma \propto n^{-1/(2r + d)}$, $h_f \propto n^{-1/(2r + d)}$, and $h \propto n^{-1/5}$.
Assumption~2.3(b) implies that $r > d/3$.
That is, a high-order kernel is required for a large $d$ and is known not to perform well in practice.


\subsection{Series}
\label{SecSeriesA}
We illustrate how a series estimator satisfies Assumption~2.3 using the results in \cite{Newey97JoE} summarized in Chapter 20 in \cite{HansenBook}.
For a detailed review of series methods, see \cite{Chen07}.

Let $Z_K \equiv Z_K(T,X)$ be a $K\times 1$ vector of regressors obtained by making transformations of $(T,X)$, such as polynomial. 
The series approximation to $\gamma(t,x)$ is $\gamma_K(t,x) \equiv Z_K(t,x)'\beta_K$, where 
$\beta_K \equiv \mathbb{E}[Z_KZ_K']^{-1}\mathbb{E}[Z_KY]$.
Consider a least squares estimator $\hat \beta_K \equiv \left(\sum_{i=1}^n Z_{Ki}Z_{Ki}'\right)^{-1}\sum_{i=1}^n Z_{Ki}Y_i$ and  
$\hat \gamma_K(t,x) \equiv Z_K(t,x)'\hat\beta_K$.

To analyze the asymptotic properties, define  
${\bf Q}_{K} \equiv \int_{\mathcal{T}\times \mathcal{X}} Z_K(t,x) Z_K(t,x)' dF_{TX}(t,x)$, 
$\zeta_K \equiv \sup_{(t,x)}\big(Z_K(t,x)'{\bf Q}_K^{-1}Z_K(t,x)\big)^{1/2}$, and the projection approximation error $r_K(t,x) \equiv \gamma(t,x) - Z_K(t,x)'\beta_K$.

\begin{assumption}[Series]
(a) The smallest eigenvalue of ${\bf Q}_K$ is bounded away from zero;
(b) $\zeta_K^2\log(K)/n\rightarrow 0$ as $n, K\rightarrow\infty$;
(c)~There are $\alpha$ and $\beta_K$ such that $\sup_{(t,x)\in\mathcal{T}\times\mathcal{X}}$
\\
$| r_K(t,x) | = O(K^{-\alpha})$ as $K\rightarrow\infty$;
(d)~$var(Y|T=t, X=x)$ and $f_{T|X}(t, x)$ is bounded above uniformly over $\mathcal{T}\times\mathcal{X}$. 
%
%
%
%
\label{Aseries}
\end{assumption}

\begin{theorem}[Series]
Under Assumption~\ref{Aseries}, $\|\hat\gamma_K - \gamma\|_{F_{tX}} = O_p(\sqrt{K/n} + K^{-\alpha})$ for $t\in\mathcal{T}$.
\label{Tseries}
\end{theorem}

Theorem 20.7 in \cite{HansenBook} provides the convergence rate of an integrated squared error which is defined as $\|\hat\gamma_K - \gamma\|_{F_{TX}}^2$, i.e.\ under Assumption~\ref{Aseries}, the $L_2(TX)$ convergence rate $\|\hat\gamma_K - \gamma\|_{F_{TX}} = O_p(\sqrt{K/n} + K^{-\alpha})$.
Theorem~\ref{Tseries} contributes the convergence rate for the partial $L_2(tX)$ norm for a series estimator
that is the same as the standard $L_2(TX)$ convergence rate.

We can apply series methods to the GPS estimation proposed in Section~2.1.
The ReGPS estimator can use a logistic series estimator for the conditional mean of $\Phi((t-T)/h_1)$ given $X$, with the standard $L_2(X)$ convergence rate $R_{1n}$ implied in \cite{HIR03ETA}.
When $d_T > 1$, the MultiGPS estimator can be a series regression of $h_1^{d_T}g_{h_1}(T_i - t)$ on $X$, with the standard $L_2(X)$ convergence rate $R_{1n}$ given in Theorem 20.7 in \cite{HansenBook}.

To verify Assumption~\ref{Aseries}(b), 
$\zeta_K \leq O(K)$ for power series and $\zeta_K \leq O(K^{1/2})$ for splines 
under the assumption that $f_{TX}(t,x)$ is strictly positive on $\mathcal{T}\times\mathcal{X}$ 
(Theorem~20.3 in \cite{HansenBook}).
Assumption~\ref{Aseries}(c) is from Assumption~3 in \cite{Newey97JoE} and is satisfied for splines and power series by $\alpha = s/d$, where $s$ is the number of continuous derivatives of $\gamma(t,x)$.
It implies that $\|r_K\|_{F_{TX}} = O(K^{-\alpha})$
and $\|r_K\|_{F_{tX}} = O(K^{-\alpha})$.
Note that Assumption~\ref{Aseries}(c) is sufficient to obtain $\|r_K\|_{F_{tX}}$ for the partial $L_2(tX)$ norm, 
but it may be stronger than necessary; see Theorem 20.2 in \cite{HansenBook} for the $L_2(TX)$ norm $\|r_K\|_{F_{TX}}$.

\subsection{Proofs}
\textbf{Proof of Theorem~\ref{Tkernel}:}\
The result is obtained by Markov's inequality with a bound on $\mathbb{E}\big[\|\hat\gamma(t,x) - \gamma(t,x)\|_{F_{tX}}^2\big]$.
Note that $\mathbb{E}\big[\|\hat\gamma(t,x) - \gamma(t,x)\|_{F_{tX}}^2\big]$ equals the integrated mean squared error 
\\
$\int_{\mathcal{X}} \mathbb{E}\big[(\hat\gamma(t,x) - \gamma(t,x))^2\big] f_{TX}(t, x) dx
=
\int_{\mathcal{X}} 
\Big(
h_\gamma^{2r_\gamma} B(t,x)^2 + (nh_\gamma^d)^{-1}\int_{-\infty}^\infty K(u)^2 du\cdot var(Y|T=t, X=x)/f_{TX}(t,x)
\Big)f_{TX}(t, x) dx + o\big(h_\gamma^{2r_\gamma}  + (nh_\gamma^d)^{-1}\big)
= O\big(h_\gamma^{2r_\gamma}  + (nh_\gamma^d)^{-1}\big)$, 
where the leading bias $B(t,x)$ is analyzed below.

The explicit expression of $B(t,x)$ for $r_\gamma=2$ is given in Theorem 19.1 in \cite{HansenBook}.
It is straightforward to derive $B(t,x)$ for $r_\gamma > 2$ at the cost of notational complication.
 The derivation is based on the proof of Theorem 19.1 in \cite{HansenBook}, and we only note the modification to conserve space.
To simplify notation, let $r = r_\gamma$, $z = (t,x)$, $f(z) = f_{TX}(t,x)$, and $\gamma^{(r)}(z)$ be the $r$th partial derivative of $\gamma(z)$.
By two Taylor series expansions, equation (19.31) in \cite{HansenBook} becomes 
$\int_{-\infty}^\infty K(u)(\gamma(z + hu) - \gamma(z)) f(z + hu) du
= 
\int_{-\infty}^\infty K(u)(\gamma^{(1)}(z) hu + \gamma^{(2)}(z) h^2 u^2/2!+...+ \gamma^{(r)}(z) h^ru^r/r! + o(h^r) \big)\big( f(z) + f^{(1)}(z) hu + ...+ f^{(r-1)}(z)h^{r-1}u^{r-1}/(r-1)! + o(h^{r-1}) \big) du 
= h^r B(t,x) + o(h^r)$.
Under Assumption~\ref{Akernel1} and $r_f \geq r_\gamma-1$, $B(t,x)$ is uniformly bounded.  
Thus the above asymptotic integrated mean squared error is finite.

The same arguments apply to the result for $\hat f_{T|X}$.
\hfill$\square$
\\ \\
\textbf{Proof of Theorem~\ref{Tseries}:}\
The proofs of Theorem 20.6 and Theorem 20.7 in \cite{HansenBook} analyze $\|\hat\gamma_K - \gamma\|_{F_{TX}}^2$.
We follow the same argument to analyze the $L_2(tX)$ norm. 

We can write $Y = Z_K'\beta_K + e_K$, where $e_K$ is the projection error.
Define 
$\widetilde Z_{Ki} \equiv {\bf Q}_K^{-1/2} Z_{Ki}$, 
$\widetilde {\bf Q}_K \equiv n^{-1}\sum_{i=1}^n \widetilde Z_{Ki}\widetilde Z_{Ki}'$,
and ${\bf Q}_{Kt} \equiv \int_\mathcal{X} Z_K(t,x) Z_K(t,x)' f_{TX}(t,x)dx$.

\begin{align}
\|\hat\gamma_K - \gamma\|_{F_{tX}}^2
=&\ \int_\mathcal{X}\big( Z_K(t,x)' (\hat\beta_K - \beta_K) - r_K(t,x)\big)^2 f_{TX}(t,x)dx \notag\\
=&\ (\hat\beta_K - \beta_K)' \left( \int_\mathcal{X} Z_K(t,x) Z_K(t,x)' f_{TX}(t,x)dx\right) (\hat\beta_K - \beta_K) \label{Eseries1} \\
&-2(\hat\beta_K - \beta_K)'\left( \int_\mathcal{X} Z_K(t,x) r_K(t,x) f_{TX}(t,x)dx\right) \label{Eseries2}\\
&+ \int_\mathcal{X} r_K(t,x)^2 f_{TX}(t,x)dx \notag \\
=&\ O_p\left( (\hat\beta_K - \beta_K)' {\bf Q}_{Kt} (\hat\beta_K - \beta_K) + \|r_K\|_{F_{tX}}^2 \right). \notag
\end{align}

Assumption~\ref{Aseries}(c) implies $\|r_K\|_{F_{tX}} = O(K^{-\alpha})$.
Consider the term in (\ref{Eseries2}).
For the $L_2(TX)$ norm, $\int_{\mathcal{T}\times\mathcal{X}} Z_K(t,x) r_K(t,x) dF_{TX}(t,x)=0$ due to the regression and projection errors. 
But this is not the case for the $L_2(tX)$ norm.
\begin{align*}
\left| (\hat\beta_K - \beta_K)'\left( \int_\mathcal{X} Z_K(t,x) r_K(t,x) f_{TX}(t,x)dx\right) \right| 
&\leq
\int_\mathcal{X} \left| \left(\hat\gamma_K(t,x) - \gamma_K(t,x)\right) r_K(t,x) \right|  f_{TX}(t,x)dx  
\\&\leq 
\|  \left(\hat\gamma_K - \gamma_K\right) r_K  \|_{F_{tX}}
\\&\leq 
\|\hat\gamma_K - \gamma_K\|_{F_{tX}} \|r_K\|_{F_{tX}},
\end{align*}
where
$\|\hat\gamma_K - \gamma_K\|_{F_{tX}}^2 = \int_\mathcal{X}\big(Z_k(t,x)'(\hat\beta_K-\beta_K)\big)^2 f_{TX}(t,x)dx 
= (\hat\beta_K - \beta_K)' {\bf Q}_{Kt} (\hat\beta_K - \beta_K)$.

Write $\hat \beta_K = \left({\bf Z}_K'{\bf Z}_K\right)^{-1}{\bf Z}_K' {\bf Y}$, where ${\bf Z}_K \equiv (Z_1,..., Z_n)'$ and ${\bf Y} \equiv (Y_1,...,Y_n)'$.
We show the term in (\ref{Eseries1})
\begin{align*}
(\hat\beta_K - \beta_K)'  {\bf Q}_{Kt}  (\hat\beta_K - \beta_K) &=
({\bf e}_K' {\bf Z}_K)({\bf Z}_K'{\bf Z}_K)^{-1} {\bf Q}_{Kt} ({\bf Z}_K'{\bf Z}_K)^{-1}({\bf Z}_K'{\bf e}_K)\\
&= n^{-2} ({\bf e}_K' \widetilde{\bf Z}_K)\widetilde{\bf Q}_{K}^{-1} {\bf Q}_{K}^{-1/2}{\bf Q}_{Kt} {\bf Q}_{K}^{-1/2}\widetilde{\bf Q}_{K}^{-1}(\widetilde{\bf Z}_K'{\bf e}_K)\\
&\leq \left(\lambda_{\max}\left( 
\widetilde{\bf Q}_{K}^{-1} {\bf Q}_{K}^{-1/2}{\bf Q}_{Kt} {\bf Q}_{K}^{-1/2}\widetilde{\bf Q}_{K}^{-1} 
 \right)\right)^2 \left( n^{-2} {\bf e}_K'\widetilde{\bf Z}_K \widetilde{\bf Z}_K'{\bf e}_K \right) \\
& = O_p(1) \left( n^{-2} {\bf e}_K'{\bf Z}_K{\bf Q}_{K}^{-1} {\bf Z}_K'{\bf e}_K \right)\\
& = O_p(K/n).
\end{align*}
The last equality follows the proof of Theorem 20.7 in \cite{HansenBook}.
The above inequality is by the Quadratic Inequality, where $\lambda_{\max}({\bf Q})$ denotes the largest eigenvalue of a matrix ${\bf Q}$.
By the Schwarz Matrix Inequality,
$\lambda_{\max}\big(  \widetilde{\bf Q}_{K}^{-1} {\bf Q}_{K}^{-1/2}{\bf Q}_{Kt} {\bf Q}_{K}^{-1/2}\widetilde{\bf Q}_{K}^{-1} 
 \big) \leq \lambda_{\max}\big( \widetilde{\bf Q}_{K}^{-1} \big)\lambda_{\max}\big({\bf Q}_{K}^{-1/2}\big)\lambda_{\max}\big({\bf Q}_{Kt} \big)$
 \\
 $\times\lambda_{\max}\big({\bf Q}_{K}^{-1/2}\big)\lambda_{\max}\big(\widetilde{\bf Q}_{K}^{-1} 
 \big)$, which is $O_p(1)$ by Assumption~\ref{Aseries}(a) and Theorem 20.5 in \cite{HansenBook}.
We notice that ${\bf Q}_{Kt}$ does not affect the bound.

Putting together, we obtain $\|\hat\gamma-\gamma\|_{F_{tX}}^2 = O_p(K/n + K^{-2\alpha})$.
\hfill $\square$

\section{Uniform asymptotic theory}
\label{SecUnif}
We extend the asymptotic theory to uniformity over 
$t \in \mathcal{T}$ which is a compact subset of the support of $T$.
The uniform asymptotic representation in Theorem~\ref{TIFunif} is the basis for a uniform inference procedure for $\beta_t$ and $\theta_t$.
Assumption~\ref{Aunif} strengthens Assumption~2.3 for the nuisance estimators. 

\begin{assumption}
There exist functions $\bar \gamma(t,x)$ and $\bar f(t,x)$ that 
are three-times differentiable with respect to $t$ with all
three derivatives being bounded uniformly over $(y,t',x') \in\mathcal{Z}$,
$\inf_{t\in\mathcal{T}}{\rm ess}\inf_{x\in\mathcal{X}}
$\\$ \bar f(t,x) \geq c$ for some positive constant $c$, and satisfy the following: 
For each $\ell= 1,...,L$,\\
(a)~$\sup_{t\in\mathcal{T}} \sqrt{nh^{d_t}} 
 \big\| \hat f_\ell - \bar f \big\|_{F_{tX}}
\big\| \hat\gamma_\ell - \bar\gamma\big\|_{F_{tX}}
\stackrel{p}{\rightarrow}0$;
(b)~There exist positive sequences $A_{1n}\rightarrow 0$ and $A_{2n}\rightarrow 0$ such that 
$\sup_{(t,x)\in\mathcal{T}\times\mathcal{X}}| \hat\gamma_\ell(t,x) - \bar\gamma(t,x) | = O_p(A_{1n})$ and 
$\sup_{(t,x)\in\mathcal{T}\times\mathcal{X}}| \hat f_\ell(t|x) - \bar f(t,x) | = O_p(A_{2n})$;
(c)~$\hat \gamma_l(t,x)$ and $\hat f_l(t|x)$ are Lipschitz continuous in $t \in \mathcal{T}$ for any $x\in\mathcal{X}$ with the Lipschitz constant independent of  $x$;
(d)~Either $\bar \gamma = \gamma$ or $\bar{f} = f_{T|X}$.
\label{Aunif}
\end{assumption}
Assumptions~\ref{Aunif}(a) is the uniform analog of Assumptions~2.3(b).
Assumption~\ref{Aunif}(b) requires the nuisance estimators to converge uniformly at some rates.
\cite{Fan} make similar assumptions for the conditional average binary treatment effect.
As discussed in Section~3.1, for a specific nuisance estimator to satisfy Assumption~\ref{Aunif}
we may require additional conditions, such as compact $\mathcal{X}$.
We may verify Assumption~\ref{Aunif} for the kernel and series estimators by extending our results in Sections~\ref{SecKernelA} and \ref{SecSeriesA}, respectively,
and using the existing results in the literature (e.g.\ \cite{Newey94ET} and \cite{CFF20AS}).
Verifying Assumption~\ref{Aunif} for the modern ML methods, such as the deep neural networks in \cite{Farrell18}, is more involved and beyond the scope of this paper. 

\begin{theorem}
Let the conditions in Theorem~3.1 and Assumption~\ref{Aunif} hold. 
Then (a) the asymptotically linear representation of $\hat \beta_t$ in (3.5) holds uniformly in $t \in \mathcal{T}$.
(b) Furthermore let the conditions in Theorem~3.3 hold.  
Then the asymptotically linear representation of $\hat \theta_t$ in (3.6) holds uniformly in $t \in \mathcal{T}$.
\label{TIFunif}
\end{theorem}

We consider a multiplier bootstrap method for uniform inference on $\beta_t$ and $\theta_t$ over $t \in\mathcal{T}$.
The method and proof closely follow \cite{SUZ} and Theorem~4.1 in \cite{Fan}, where the moment functions share a similar structure with that of $\hat \beta_t$.
Let $\{\xi_i\}_{i=1}^n$ be a sequence of i.i.d. random variables satisfying Assumption~\ref{AMBS}.
\begin{assumption}[Multiplier bootstrap]
The random variable $\xi_i$ is independent of the sample $\{(Y_i, T_i, X_i)\}_{i=1}^n$, $\mathbb{E}^\ast[\xi_i] = var^\ast(\xi_i) = 1$, and its distribution $P^\ast$ has sub-exponential tails, i.e.\ $P^\ast(|\xi_i| > u) \leq c_1 \exp(-c_2 u)$ for every $u$ and some constants $c_1$ and $c_2$.
\label{AMBS}
\end{assumption}
Assumption~\ref{AMBS} is standard for multiplier bootstrap inference and is satisfied by a normal random variable, for example.
Then we compute
\begin{align*}
\hat \beta_t^\ast = \frac{1}{n}\sum_{\ell=1}^L \sum_{i \in I_\ell} \xi_i \left\{ \hat \gamma_\ell (t, X_i) + \frac{K_h(T_i - t)}{ \hat f_\ell(t|X_i)} \left( Y_i - \hat \gamma_\ell(t,X_i) \right)\right\}.
\end{align*}
We use $\sqrt{nh^{d_t}}(\hat\beta_t^\ast-\hat\beta_t)$ to simulate the limiting process of $\sqrt{nh^{d_t}}(\hat\beta_t-\beta_t)$ indexed by $t\in\mathcal{T}$.
That is, repeat the above procedure for $B$ times and obtain a bootstrap sample of $\{\hat\beta^\ast_{t,b}\}_{b=1}^B$.

For the partial effect, compute the numerical differentiation estimator $\hat\theta_t^\ast$ following Step 3 in the estimation procedure given in Section 2 in the main text using $\hat\beta_t^\ast$.
Then we simulate the limiting process of $\sqrt{nh^{d_t+2}}(\hat\theta_t-\theta_t)$ by $\sqrt{nh^{d_t+2}}(\hat\theta_t^\ast-\hat\theta_t)$. 
Theorem~\ref{TMBS} below 
presents the bootstrap version of Theorems~3.1 and 3.3 and forms the basis of uniform inference.

\begin{theorem}[Multiplier bootstrap]
Let the conditions in Theorem~\ref{TIFunif} and Assumption~\ref{AMBS} hold.
Then uniformly in $t\in\mathcal{T}$,
$\sqrt{nh^{d_t}} \big(\hat \beta_t^\ast - \beta_t\big) = \sqrt{h^{d_t}/n} \sum_{i=1}^n \xi_i\big\{
K_h(T_i-t) (Y_i - \bar\gamma(t, X_i))/\bar f(t,X_i) +  \bar\gamma(t, X_i) - \beta_t  \big\}
+ o_{p^\star}(1)$
and
$\sqrt{nh^{d_t + 2}}( \hat \theta_t^\ast - \theta_{t} ) = \sqrt{h^{d_t+2}/n}  \sum_{i=1}^n$\\$ \xi_i \partial K_h(T_i - t)/\partial t_1 \big(Y_i - \bar\gamma(t, X_i)\big)/\bar f(t,X_i) + o_{p^\star}(1)$, 
where $p^{\star}$ is for the joint distribution of the full sample $\{(Y_i, T_i, X_i)\}_{i=1}^n$ and $\{\xi_i\}_{i=1}^n$.
\label{TMBS}
\end{theorem}

Next we discuss inference using the multiplier bootstrap in \cite{SUZ} and \cite{Fan}.
Theorem~\ref{TMBS} implies that $\sqrt{nh^{d_t}} \big(\hat \beta_t^\ast - \hat\beta_t\big)$
and $\sqrt{nh^{d_t + 2}}( \hat \theta_t^\ast - \hat\theta_{t} )$
converge in distribution to the limiting distribution of 
$\sqrt{nh^{d_t}} \big(\hat \beta_t - \beta_t\big)$
and $\sqrt{nh^{d_t + 2}}( \hat \theta_t - \theta_{t} )$, respectively, conditional on the sample path with probability approaching one (in the sense of Section 2.9 in \cite{VW96}).
Following \cite{SUZ}, obtain $\hat q_t(\alpha)$ as the $\alpha^{th}$ quantile of the sequence $\big\{\hat\beta_{t,b}^\ast-\hat\beta\big\}_{b=1}^B$.
The standard $100(1-\alpha)\%$ percentile bootstrap confidence interval for $\beta_t$ is $\big(\hat\beta_t+\hat q_t(\alpha/2), \hat\beta_t+\hat q_t(1-\alpha/2)\big)$
or $\big(\hat \beta_t -\hat q_t(\alpha/2), \hat \beta_t+\hat q_t(\alpha/2)\big)$.
As shown in Theorem 4.1 in \cite{SUZ}, these pointwise confidence intervals for $t\in\mathcal{T}$ are valid under undersmoothing, i.e.\ using a bandwidth smaller the AMSE optimal bandwidth given in Theorem~3.2.

We can follow the approach in \cite{Fan} to construct uniform confidence bands.
Specifically obtain $\hat Q(\alpha)$ as the $\alpha^{th}$ quantile of the sequence $\big\{\sup_{t\in\mathcal{T}} \sqrt{nh^{d_t}} \big|\hat\beta^\ast_{t,b} - \hat \beta_t\big|/\hat\sigma_t\big\}_{b=1}^B$, where $\hat\sigma_t^2$ is a uniformly consistent estimator of $\mathsf{V}_t$.
The supremum is approximated by the maximum over a chosen fine grid over $\mathcal{T}$.  
Then construct the $100(1-\alpha)\%$  uniform confidence band as $\big(\hat\beta_t - \hat Q(1-\alpha) \hat\sigma_t/\sqrt{nh}, \hat\beta_t + \hat Q(1-\alpha) \hat\sigma_t/\sqrt{nh} \big)$.
For example, we could use $\hat\sigma_t^2 = \hat{\mathsf{V}}_t$ the sample variance estimator described in Section~3.
Following the proof of Theorem 3.2 in \cite{Fan}, one could show 
$\sup_{t\in\mathcal{T} } |\hat{\mathsf{V}}_t - \mathsf{V}_t| = o_p(1)$.
Based on Theorem~\ref{TMBS}, the asymptotic validity of the confidence band could follow the proof of Theorem 4.2 in \cite{Fan}.\footnote{Specifically with more complicated notations,  
we will use the results in \cite{CCK14b} that study the problem of approximating suprema of empirical processes by a sequence of suprema of Gaussian processes.
Then we will use their companion paper \cite{CCK14a} that provide multiplier bootstrap methods for computation of Gaussian approximations.} 
We do not include the formal theoretical details in this paper to conserve space and focus on the new results.

\subsection{Proofs}
\label{Apf}
\textbf{Proof of Theorem~\ref{TIFunif}:}\\[5pt]
\textbf{(a)}
We show the remainder terms (R1-1), (R1-2), (R1-DR), and (R2) are $o_p(1)$ uniformly over $t\in \mathcal{T}$.
Denote the nuisance estimators $\hat \gamma_{i\ell t} = \hat r_\ell(t, X_i)$ and $\hat \lambda_{i\ell t} = 1/\hat f_{\ell}(t|X_i)$ that use $Z_\ell^c$ for $i \in I_\ell$.
Denote $\hat g(t) = n^{-1}\sum_{\ell = 1}^L \sum_{i \in I_\ell} K_h(T_i-t) \Upsilon_{i\ell}(t)$
and $W_{i\ell}(t) = K_h(T_i-t) \Upsilon_{i\ell}(t) - \mathbb{E}\left[K_h(T_i-t) \Upsilon_{i\ell}(t)\right]$, 
where $\Upsilon_{i\ell}(t) = (\hat \lambda_{i\ell t} - \bar\lambda_{it})(\hat \gamma_{i\ell t} - \bar\gamma_{it})$ for (R2). 

We follow similar decompositions in the proof of Theorem 3.1 in \cite{Fan}.
First, to show that $\sup_{t\in \mathcal{T}} \mathbb{E}[\hat g(t)] = o_p(\sqrt{\ln(n)/(nh^{d_t})})$,
the same argument in (A.6) holds uniformly over $t\in\mathcal{T}$ by Assumption~\ref{Aunif}(a).

Since $\mathcal{T}$ is compact, it can be covered by a finite number $M_n$ of cubes $\mathcal{C}_{k,n}$ with centered $t_{k,n}$ and length $m_n$, for $k=1,...,M_n$.
So $M_n \propto 1/m_n^{d_t}$.
Decompose 
\begin{align}
\sup_{t\in \mathcal{T}} \big| \hat g(t) - \mathbb{E}[\hat g(t)] \big|
=&\ \max_{1\leq k\leq M_n }\sup_{t\in \mathcal{T}\cap \mathcal{C}_{k,n}} \big| \hat g(t) - \mathbb{E}[\hat g(t)] \big| 
\notag \\
\leq&\
\max_{1\leq k\leq M_n }\sup_{t\in \mathcal{T}\cap \mathcal{C}_{k,n}} \big| \hat g(t) -\hat g(t_{k,n}) \big| 
\label{Q1}\\
&+ 
\max_{1\leq k\leq M_n } \big| \hat g(t_{k,n}) - \mathbb{E}[\hat g(t_{k,n})] \big| 
\label{Q2}\\
&+ 
\max_{1\leq k\leq M_n }\sup_{t\in \mathcal{T}\cap \mathcal{C}_{k,n}} \big| \mathbb{E}[\hat g(t_{n,k})] - \mathbb{E}[\hat g(t)] \big|.
\label{Q3}
\end{align}

For a positive constant $C$ and positive sequences $A_{1n}$ and $A_{2n}$ given in Assumption~\ref{Aunif}(b), let 
$\mathcal{F}_n(C) \equiv\{( \gamma^\dagger, \lambda^\dagger): 
\sup_{(t,x)\in\mathcal{T}\times\mathcal{X}}| \gamma^\dagger(t,x) - \bar\gamma(t,x) | \leq C A_{1n}, 
\sup_{(t,x)\in\mathcal{T}\times\mathcal{X}}
$\\$| \lambda^\dagger(t,x) - \bar\lambda(t,x) | \leq C A_{2n}.
\}.$ 
Let $\mathcal{A}_{\ell n}(C) \equiv \{(\hat \gamma_{\ell}(\cdot,\cdot), \hat \lambda_{\ell}(\cdot,\cdot)) \in \mathcal{F}_n(C)\}$
and $\mathcal{A}_n(C) \equiv \cap_{\ell=1}^L \mathcal{A}_{\ell n}$.
On $\mathcal{A}_n(C)$, i.e.\ $(\hat \gamma_{i\ell t}, \hat \lambda_{i\ell t}) \in \mathcal{F}_n(C)$ for $\ell=1,...,L$, $\sup_{t\in\mathcal{T}, i\in I_\ell} |\Upsilon_{i\ell}(t)| \leq C^2 A_{1n}A_{2n} \equiv C^2 A_n$.
Assumption~\ref{Aunif}(b) implies that for any $\varepsilon >0$, there exists a positive constant $C$, such that $P(\mathcal{A}_n(C)) \geq 1-\varepsilon$ for $n$ large enough.

Note that the expectation $\mathbb{E}$ is based on the distribution $P$ of the full sample $\{Z_i\}_{i=1}^n$.
Observe that for $\ell\in\{1,...,L\}$
\begin{align}
P\left(  n^{-1}W_{i\ell}(t)  > \eta_n, \mathcal{A}_{\ell n}(C) \right) 
&= \mathbb{E}\left[
P(n^{-1}W_{i\ell}(t)  > \eta_n|Z_\ell^c){\bf 1}\{\mathcal{A}_{\ell n}(C)\} 
\right]\notag \\
&= \mathbb{E}\left[
\int_{\mathcal{Z}}
{\bf 1}\{n^{-1}W_{i\ell}(t)  > \eta_n\} 
f_Z(Z_i) dZ_i{\bf 1}\{\mathcal{A}_{\ell n}(C) \}
\right]
\label{EqFHLZ}
\end{align}
by the law of iterated expectations and cross-fitting with $(\hat\gamma_\ell, \hat \lambda_\ell)$ using $Z_\ell^c$.

We will use the following inequalities. 
By $\exp(w) \leq 1 + w + w^2$ for $|w| \leq 1/2$ and $1+w\leq \exp(w)$ for $w \geq 0$,
we have 
\begin{align}
\mathbb{E}[\exp(W)] \leq 1 + \mathbb{E}[W] + \mathbb{E}[W^2]\leq \exp(\mathbb{E}[W^2])
\label{IE1}
 \end{align}
for a random variable $W$ satisfying $|W| \leq 1/2$ and $\mathbb{E}[W] = 0$.
The Markov's inequality for any positive sequence $a_n$: $P(W > \eta_n) \leq \mathbb{E}[\exp(a_nW)]/\exp(a_n\eta_n)$.


First consider (\ref{Q2}).
For any $\eta_n > 0$, $P\big(\max_{1 \leq k \leq M_n} \big| 
\hat g(t_{k,n}) - \mathbb{E}[\hat g(t_{k,n})] \big| > \eta_n \big)
\leq M_n \sup_{t\in\mathcal{T}} P\big( \big| \hat g(t) - \mathbb{E}[\hat g(t)]\big| > \eta_n, \mathcal{A}_n(C)
 \big) + \varepsilon$.
We show that 
for $t \in\mathcal{T}$,
\begin{align}
&P\big( \big| \hat g(t) - \mathbb{E}[\hat g(t) \big| > \eta_n, \mathcal{A}_n(C) \big) 
\notag\\
& = P\left( \left| n^{-1}\sum_{\ell = 1}^L \sum_{i \in I_\ell} W_{i\ell}(t) \right| > \eta_n, \mathcal{A}_n(C)\right)
\notag\\
& = P\left(  n^{-1}\sum_{\ell = 1}^L \sum_{i \in I_\ell} W_{i\ell}(t)  > \eta_n, \mathcal{A}_n(C)\right)
+ P\left(  n^{-1}\sum_{\ell = 1}^L \sum_{i \in I_\ell} W_{i\ell}(t)  < - \eta_n, \mathcal{A}_n(C)\right)
\notag\\
& \leq \sum_{\ell = 1}^L \sum_{i \in I_\ell} \left\{ P\left(  n^{-1}W_{i\ell}(t)  > \eta_n, \mathcal{A}_{\ell n}(C)\right)
+ P\left(  - n^{-1}  W_{i\ell}(t)  >  \eta_n, \mathcal{A}_{\ell n}(C)\right)\right\}
\notag\\
&\leq \sum_{\ell = 1}^L \sum_{i \in I_\ell} \mathbb{E}\left[ {\bf 1}\{\mathcal{A}_{\ell n}(C)\}  \mathbb{E}\left[\exp\left(a_n n^{-1}W_{i\ell}(t)\right) + \exp( - a_n n^{-1} W_{i\ell}(t))\big| Z_\ell^c\right] \right] \big/\exp(a_n \eta_n)  
\notag\\
&\leq 2 \sum_{\ell = 1}^L n_\ell \exp(-a_n \eta_n)  \mathbb{E}\left[ {\bf 1}\{\mathcal{A}_{\ell n}(C)\}  \mathbb{E}\left[\exp\left(a_n^2 n^{-2}  \mathbb{E}\left[W_{i\ell}(t)^2\big| Z_\ell^c\right]\right)\right]\right].
\label{Ecover}
\end{align}
The second inequality uses (\ref{EqFHLZ}) and the conditional Markov's inequality.
Due to cross-fitting, conditional on $Z_\ell^c$, $\hat \gamma_\ell$ and $\hat\lambda_\ell$ are fixed functions. 
When ${\bf 1}\{\mathcal{A}_{\ell n}(C)\} = 1$, 
we can choose $a_n = \sqrt{\ln(n) nh^{d_t}} /A_n$ such that $\left| a_n n^{-1} W_{i\ell}(t)\right| \leq 1/2$, for all $t, i, \ell$ and for $n$ large enough.
So by (\ref{IE1}), the last inequality holds.

We next choose $\eta_n$ such that $a_n\eta_n \rightarrow \infty$ and $a_n\eta_n \geq a_n^2 n^{-2}  \mathbb{E}\left[W_{i\ell}(t)^2|Z_\ell^c\right]$ for all $t, \ell$, so $\sup_{t\in\mathcal{T}}P\big( \big| \hat g(t) - \mathbb{E}[\hat g(t)] \big| > \eta_n, \mathcal{A}_{\ell n} \big) = o_p(1)$.
When ${\bf 1}\{\mathcal{A}_{\ell n}(C)\} = 1$, for $n$ large enough, there exists a positive constant $c_1$ such that $\mathbb{E}\left[W_{i\ell}(t)^2|Z_\ell^c\right] \leq c_1h^{-d_t} A_n^2$.
We can choose $a_n\eta_n = c_2 \ln(n)$ for some positive constant $c_2$, so we let $\eta_n = c_2 \sqrt{\ln(n)/(nh^{d_t})}A_n$.
Then we can choose $M_n$ such that $P\left(\max_{1\leq k\leq M_n} \big| \hat g(t_{k,n}) - \mathbb{E}[\hat g(t_{k,n})] \big| > \eta_n \right) 
\leq M_n 2n\exp(-c_2$\\ $\ln(n) + c_1\ln(n)/n)+ \varepsilon  \leq 2M_n n^{-(c_2 -c_1n^{-1} -1)} + \varepsilon\leq 2\varepsilon$ for $c_2 \geq 2$ and $n$ large enough.
So $\max_{1\leq k\leq M_n} \big| \hat g(t_{k,n}) - \mathbb{E}[\hat g(t_{k,n})] \big| = O_p(\eta_n) = o_p( \sqrt{\ln(n)/(nh^{d_t})})$.


For (\ref{Q1}), the Lipschitz condition in Assumption~\ref{Aunif}(c) implies
$\sup_{t\in\mathcal{T}\cap \mathcal{C}_{k,n}} \big|
K_h(T_i - t)\Upsilon_{i\ell}(t) - K_h(T_i - t_{k,n})\Upsilon_{i\ell}(t_{k,n}))
\big|
\leq 
c_3 h^{-(d_t+1)} \sup_{t\in\mathcal{T}\cap \mathcal{C}_{k,n}} \| t - t_{k,n}\|
\leq
c_3 h^{-(d_t+1)} m_n$, for some constant $c_3 > 0$ and the Euclidean norm of a vector $\|\cdot\|$.
By choosing $m_n = o(\sqrt{\ln(n)h^{(d_t+2)}/n })$, $\max_{1\leq k\leq M_n }\sup_{t\in \mathcal{T}\cap \mathcal{C}_{k,n}} \big| \hat g(t) -\hat g(t_{k,n}) \big| 
\leq c_3 h^{-(d_t + 1)}m_n $\\$=o_p(\sqrt{\ln(n)/(nh^{d_t})})$.

By the same argument, we can show that for (\ref{Q3}) $\max_{1\leq k\leq M_n }\sup_{t\in \mathcal{T}\cap \mathcal{C}_{k,n}} \big| \mathbb{E}[\hat g(t_{n,k})] - \mathbb{E}[\hat g(t)] \big|  = o_p(\sqrt{\ln(n)/(nh^{d_t})})$.


The same arguments apply to (R1-1) and (R1-2) by defining $\Upsilon_{i\ell}(t)$ accordingly:
let $\Upsilon_{i\ell}(t) = \bar\lambda_{i\ell t}(\hat \gamma_{i\ell t} - \bar\gamma_{i\ell t})$ for (R1-1) and 
$\Upsilon_{i\ell}(t) = (\hat \lambda_{i\ell t} - \bar\lambda_{i\ell t})(Y_t - \bar\gamma_{i\ell t})$ for (R1-2).

For (R1-DR), the argument for the pointwise convergence in the proof of Theorem~3.1 can be extended to uniform convergence by Assumption~\ref{Aunif}(b). 
\\[5pt]
\textbf{(b)} The results in Theorem~3.3 can be extended to uniformity in $t$ by the same arguments in (a).
\hfill $\square$
\\
\\
\textbf{Proof of Theorem~\ref{TMBS}:}\
The proof follows closely the proof of Theorem~\ref{TIFunif}, so we only notice the difference to conserve space.
The derivations proceed conditional on $\xi_i$ and using the law of iterated iterations.  
For notations, let 
the expectation $\mathbb{E}$ based on the distribution $P$ of the full sample $\{(Y_i, T_i, X_i)\}_{i=1}^n$,
the expectation $\mathbb{E}^\ast$ based on the distribution $P^\ast$ of 
$\{\xi_i\}_{i=1}^n$, and
the expectation $\mathbb{E}^\star$ based on the joint distribution $P^\star$ of the full sample $\{(Y_i, T_i, X_i)\}_{i=1}^n$
and $\{\xi_i\}_{i=1}^n$.
Let $\hat g(t) = n^{-1}\sum_{\ell = 1}^L \sum_{i \in I_\ell} \xi_i K_h(T_i-t) \Upsilon_{i\ell}(t)$.
The main difference is in (\ref{Q2}).
We show that 
for $t \in\mathcal{T}$,
\begin{align*}
&P^\star\big( \big| \hat g(t) - \mathbb{E}[\hat g(t) \big| > \eta_n, \mathcal{A}_n(C) \big) 
\\
& = P^\star\left( \left| n^{-1}\sum_{\ell = 1}^L \sum_{i \in I_\ell} \xi_iW_{i\ell}(t) \right| > \eta_n, \mathcal{A}_n(C)\right)
\\
& = P^\star\left(  n^{-1}\sum_{\ell = 1}^L \sum_{i \in I_\ell} \xi_i W_{i\ell}(t)  > \eta_n, \mathcal{A}_n(C)\right)
+ P^\star\left(  n^{-1}\sum_{\ell = 1}^L \sum_{i \in I_\ell}\xi_i W_{i\ell}(t)  < - \eta_n, \mathcal{A}_n(C)\right)
\\
& \leq \sum_{\ell = 1}^L \sum_{i \in I_\ell} \left\{ P^\star\left(  n^{-1}\xi_iW_{i\ell}(t)  > \eta_n, \mathcal{A}_{\ell n}(C)\right)
+ P^\star\left(  - n^{-1} \xi_i W_{i\ell}(t)  >  \eta_n, \mathcal{A}_{\ell n}(C)\right)\right\}
\\
& =  \sum_{\ell = 1}^L \sum_{i \in I_\ell}   \mathbb{E}^\ast\Big[ P\left(n^{-1}W_{i\ell}(t)  > \eta_n/\xi_i, \mathcal{A}_{\ell n}(C)|\xi_i\right){\bf 1}\{\xi_i \geq 0\}\\
&\ \ 
+ P\left(- n^{-1}W_{i\ell}(t)  > - \eta_n/\xi_i, \mathcal{A}_{\ell n}(C)|\xi_i\right){\bf 1}\{\xi_i < 0\} \\
&\ \ + P\left(-n^{-1}W_{i\ell}(t)  > \eta_n/\xi_i, \mathcal{A}_{\ell n}(C)|\xi_i\right){\bf 1}\{\xi_i \geq 0\}
\\
&\ \ + P\left(n^{-1}W_{i\ell}(t)  > - \eta_n/\xi_i, \mathcal{A}_{\ell n}(C)|\xi_i\right){\bf 1}\{\xi_i < 0\}\Big] 
\\
& \leq  \sum_{\ell = 1}^L \sum_{i \in I_\ell}   \mathbb{E}^\star\Big[{\bf 1}\{\mathcal{A}_{\ell n}(C)\}  
 \mathbb{E}\left[
 e^{a_n n^{-1}W_{i\ell}(t)} + e^{ - a_n n^{-1} W_{i\ell}(t)}\Big| Z_\ell^c, \xi_i\right] 
e^{-a_n\eta_n/|\xi_i|}
\Big]
\\
&\leq 2\sum_{\ell = 1}^L  n_\ell \mathbb{E}^\star\left[\exp(-a_n \eta_n /|\xi_i|) \right] {\bf 1}\{\mathcal{A}_{\ell n}(C)\} \mathbb{E}\left[\exp\left(a_n^2 n^{-2}  \mathbb{E}\left[W_{i\ell}(t)^2\big| Z_\ell^c\right]\right)\right].
\end{align*}

The same arguments following (\ref{Ecover}) are valid conditional on $\xi_i$.
Due to $\xi_i$, we choose $a_n\eta_n = c_2 \ln(n) \ln(n)$ for some positive constant $c_2$. 
So $\eta_n = c_2\sqrt{\ln(n)/(nh^{d_t})}A_n \ln(n)$.
Next we show that we can choose $M_n$ and $c_2$ such that 
 for $n$ large enough, $P^\star\big(\max_{1\leq k\leq M_n}
 \big| \hat g(t_{k,n}) - \mathbb{E}^\star[\hat g(t_{k,n})] \big| > \eta_n \big) 
\leq M_n 2n\mathbb{E}^\ast\big[ \exp(-c_2 \ln(n)\ln(n)/|\xi_i|  + c_1\ln(n)/n)\big]+ \varepsilon  \leq 2M_n \mathbb{E}^\ast\big[n^{-c_2\ln(n)/|\xi_i|+c_1/n + 1}\big] + \varepsilon\leq 2\varepsilon$.

By Assumption~\ref{AMBS}, there exist some constants $c_4,c_5$ such that
$P^\ast\big(-c_2\ln(n)/|\xi_i| + c_1/n + 1 \geq 0\big) 
\leq P^\ast\big( |\xi_i| \geq c_2\ln(n)/(c_1+1)\big)
\leq c_4 \exp(-c_5 c_2\ln(n)/(c_1+1)) 
\leq c_4 n^{-c_5c_2/(c_1+1)}$.
So $\mathbb{E}^\ast\big[n^{-c_2\ln(n)/|\xi_i|+c_1/n + 1} \big] \leq  \mathbb{E}^\ast\big[n^{-c_2\ln(n)/|\xi_i|+c_1/n + 1} {\bf 1}\big\{ -c_2\ln(n)/|\xi_i| + c_1/n+ 1 \leq 0 \big\} \big] 
+ 
n^{c_1 + 1} P^\ast\big(-c_2\ln(n)/|\xi_i| + c_1/n + 1 \geq 0\big) 
$, where the second term $\leq c_4n^{c_1 +1 -c_5c_2/(c_1+1)} = o(1)$ by choosing $c_2$.
Therefore we show that we can choose $M_n$ and $c_2$ such that $P^\star\big(\max_{1\leq k\leq M_n} \big| \hat g(t_{k,n}) - \mathbb{E}[\hat g(t_{k,n})] \big| > \eta_n \big) \leq 2\varepsilon$ for $n$ large enough.


For (\ref{Q1}), $\sup_{t\in\mathcal{T}\cap \mathcal{C}_{k,n}} \big|
\xi_iK_h(T_i - t)\Upsilon_{i\ell}(t) - \xi_iK_h(T_i - t_{k,n})\Upsilon_{i\ell}(t_{k,n}))
\big|
\leq 
c_3 h^{-(d_t+1)}
$\\$
\times \sup_{t\in\mathcal{T}\cap \mathcal{C}_{k,n}} \| t - t_{k,n}\||\xi_i|
\leq
c_3 h^{-(d_t+1)} m_n|\xi_i|$, for some constant $c_3 > 0$.
We can choose $m_n = o(\sqrt{\ln(n)h^{(d_t+2)}/n })$ such that 
$\max_{1\leq k\leq M_n }\sup_{t\in \mathcal{T}\cap \mathcal{C}_{k,n}} \big| \hat g(t) -\hat g(t_{k,n}) \big| 
\leq c_3 h^{-(d_t + 1)}m_n n^{-1}\sum_{i=1}^n |\xi_i| = o_{p^\star}(\sqrt{\ln(n)/(nh^{d_t})})$, 
since $n^{-1}\sum_{i=1}^n |\xi_i| = O_{p^\ast}(1)$.


Putting together, the remainder term $\sqrt{n_\ell h^{d_t}} n_\ell^{-1}\sum_{i \in I_\ell} \xi_i \big\{ \psi(Z_i, \beta_t, \hat\gamma_{i\ell}, \hat\lambda_{i\ell}) - \psi(Z_i, \beta_t, $
\\
$\bar\gamma_{i\ell}, \bar\lambda_{i\ell})\big\} = o_{p^\star}(1)$ uniformly in $t\in\mathcal{T}$.
Therefore $\sqrt{nh^{d_t}} \big(\hat \beta_t^\ast - \beta_t\big) = \sqrt{h^{d_t}/n} \sum_{i=1}^n \xi_i \psi_i + o_{p^\star}(1)$, where 
$\psi_i \equiv K_h(T_i-t) (Y_i - \bar\gamma(t, X_i)/\bar f(t,X_i) +  \bar\gamma(t, X_i) - \beta_t$.

It follows that  
$\sqrt{nh^{d_t}} \big(\hat \beta_t^\ast - \hat\beta_t\big) = - \sqrt{nh^{d_t}} \big(\hat \beta_t - \beta_t\big)  + \sqrt{h^{d_t}/n} \sum_{i=1}^n \xi_i \psi_i + o_{p^\star}(1)
=  \sqrt{h^{d_t}/n} \sum_{i=1}^n (\xi_i-1) \psi_i + o_{p^\star}(1).$
Conditional on the sample path, 
$\sqrt{h^{d_t}/n} \sum_{i=1}^n (\xi_i-1) \psi_i $ converges in distribution to the limiting distribution of $\sqrt{nh^{d_t}} \big(\hat \beta_t - \beta_t\big)$ for any $t\in\mathcal{T}$.


The result for the partial effect follows the same arguments.
\hfill $\square$




\end{document}